\documentclass[a4paper,11pt]{article}
\usepackage{jheppub} 
\usepackage{lineno}
\usepackage{graphicx}
\usepackage[utf8,latin1]{inputenc}
\usepackage{dcolumn}
\usepackage{mathrsfs}
\usepackage{subcaption}
\usepackage{caption}
\usepackage{comment}
\usepackage{academicons}
\usepackage{mathtools, nccmath}
\usepackage[overload]{empheq}
\usepackage{tikz,xcolor,color}
\usepackage{url}
\usepackage{cases}
\usepackage{setspace}
\usepackage{lipsum}
\usepackage[normalem]{ulem}
\usepackage{bm}
\usepackage{soul}
\usepackage{hhline}
\usepackage{bigints}
\usepackage{hyperref}
\usepackage{amsmath}
\usepackage[T1]{fontenc}
\usepackage{multirow}
\usepackage{adjustbox}
\usepackage{fancyhdr}

\usepackage{hyperref}
\hypersetup{colorlinks, linkcolor={red},citecolor={blue},urlcolor={blue}}  

\newcommand{\bsubeqs}{\begin{subequations}}
\newcommand{\esubeqs}{\end{subequations}}
\definecolor{lime}{HTML}{A6CE39}
\DeclareRobustCommand{\orcidicon}{
	\begin{tikzpicture}
	\draw[lime, fill=lime] (0,0) 
	circle [radius=0.16] 
	node[white] {{\fontfamily{qag}\selectfont \tiny ID}};
	\draw[white, fill=white] (-0.0625,0.095) 
	circle [radius=0.007];
	\end{tikzpicture}
	\hspace{-2mm}
}

\newcommand{\dd}{{\rm d}}

\newcommand{\ee}{{\rm e}}

\fancypagestyle{plain}{%
  \fancyhf{}
  \fancyfoot[C]{\iffloatpage{}{\thepage}}
  }
\foreach \x in {A, ..., Z}{%
	\expandafter\xdef\csname orcid\x\endcsname{\noexpand\href{https://orcid.org/\csname orcidauthor\x\endcsname}{\noexpand\orcidicon}}
}

\arxivnumber{2506.21148} % if you have one
\codelink{https://github.com/ziliang-wang0/SHADOW-TD}

\title{\boldmath Exploring the role of accretion disk geometry in shaping black hole shadows}

\author[a]{Zi-Liang Wang}

\affiliation[a]{Department of Physics, School of Science, Jiangsu University of Science and Technology, Zhenjiang, 212003, China}

\emailAdd{ziliang.wang@just.edu.cn}

\abstract{We study black hole imaging in the context of geometrically thick accretion disks in Schwarzschild spacetime. By decomposing the emitting region into a set of one-dimensional luminous segments, each characterized by its inclination angle and inner radius, we construct transfer functions that capture key image features---namely, the direct image, lensing ring, and photon ring. This approach allows a unified treatment of disk geometry and viewing angle. We explore three regimes: optically thin, optically thick, and partially optically thick disks. For optically thin flows, increasing the disk thickness (characterized by the half-opening angle $\psi_0$) broadens the lensing ring, gradually bridging the photon ring and the direct image. The photon ring remains narrow, but its position robustly defines the innermost edge of the lensing structure. In the optically thick case, image features are primarily determined by the first intersection of traced light rays with the disk, and we provide analytical criteria for the presence of lensing and photon rings based on the critical deflection angles. For partially optically thick disks, we adopt a simplified radiative transport model and find a critical absorption coefficient $\chi \sim (6M\psi_0)^{-1}$ beyond which the image rapidly transitions from an optically thin- to thick-disk appearance. These results help clarify the respective roles of the photon and lensing rings across different disk configurations, and may offer a useful framework for interpreting future high-resolution black hole observations.}

\begin{document}
\maketitle
\flushbottom

\section{Introduction}
\label{sec:intro}

Since the Event Horizon Telescope (EHT) successfully imaged the shadows of the supermassive black holes in M87*~\cite{EventHorizonTelescope:2019dse,EventHorizonTelescope:2019ggy} and Sagittarius A* (Sgr A*)~\cite{EventHorizonTelescope:2022wkp}, black hole shadow observations have become a cornerstone of modern astrophysics, offering a unique opportunity to test general relativity in the strong-field regime and to probe the nature of black holes~\cite{Vagnozzi:2022moj,Virbhadra:1999nm,Adler:2022qtb,Zhang:2024lsf,Sokoliuk:2022owk,Virbhadra:2022iiy,Hou:2023bep,Li:2025ixk,Sui:2023yay,Li:2024oyc}. Interpreting such images critically depends not only on the background spacetime geometry but also on the physical processes occurring near the black hole---particularly the structure and geometry of the surrounding accretion flow. 

It is well understood that the spacetime geometry, which governs the trajectories of photons via gravitational curvature, plays a crucial role in shaping the black hole shadow. However, for a fixed spacetime background, the optical appearance of the black hole can be substantially affected by the structure and dynamics of the surrounding accretion flow. Different accretion geometries---such as geometrically thin disks, thick disks, and spherically symmetric flows---can produce distinct observational signatures. This naturally raises the question: between the spacetime geometry and the accretion flow, which plays the dominant role in shaping the black hole shadow? 

Recent studies focusing on Schwarzschild black holes have presented differing perspectives on this issue.  Gralla et al.~\cite{Gralla:2019xty} argue that the structure of the accretion flow plays a more dominant role in shaping the observed black hole shadow than intrinsic features such as the photon ring. In contrast, Narayan et al.~\cite{Narayan:2019imo} demonstrate that the shadow edge is largely governed by the underlying spacetime geometry, with minimal dependence on the specific structure of the accreting material. These differing conclusions stem from the distinct accretion models considered: Gralla et al. mainly investigate a thin disk configuration, whereas Narayan et al. focus on  spherically symmetric, optically thin flows---specifically, both static gas and infalling gas, with the latter modeled as a Bondi---type inflow~\cite{Bondi:1952,Abramowicz:2011xu}.

The contrast between these findings underscores the importance of systematically studying how accretion geometry affects black hole shadows. In this work, we examine this question by considering the Schwarzschild black hole and exploring different accretion geometries. We analyze how different accretion geometries influence the structure of the observed black hole shadow, including its brightness distribution and ring features. To this end, we begin with a brief overview of the primary accretion disk models commonly encountered in astrophysics.

A key quantity in the study of luminous astrophysical objects is the Eddington luminosity $L_{\mathrm{Edd}}$, which represents the maximum luminosity an object can sustain when the outward force from radiation pressure balances the inward pull of gravity on infalling matter~\cite{kato2008black}.  The corresponding Eddington accretion rate is denoted by $\dot{M}_{\mathrm{Edd}}\equiv 10 L_{\mathrm{Edd}}/c^2 $~\cite{Yuan:2014gma}. Accretion disks are inherently complex and cannot be fully characterized by a single parameter. Nevertheless, an useful approach is to classify them approximately based on their mass accretion rate relative to the Eddington accretion rate~\cite{Abramowicz:2011xu}.

The most widely studied model is the geometrically thin, optically thick disk, initially developed in the early 1970s by Shakura and Sunyaev~\cite{Shakura:1972te} and others~\cite{novikov1973astrophysics,lynden1974evolution}. In this model, the gas temperature is low compared to the local virial temperature, and the mass accretion rate remains sub-Eddington~\cite{Yuan:2014gma}. The thin disk paradigm has become a standard model for a variety of accreting systems.

When the mass accretion rate approaches or exceeds $\dot{M}_{\mathrm{Edd}}$, however, the thin disk approximation breaks down. In these cases, the accretion flow is better described by the slim disk model, which incorporates advective energy transport and reduced radiative efficiency. For even higher mass accretion rates, the flow becomes geometrically thick, forming structures often referred to as thick disks or Polish doughnut (thick torus)~\cite{Abramowicz:2011xu}.

At the other extreme lies the {advection-dominated accretion flows} (ADAF)~\cite{Narayan:1994xi}, a radiatively inefficient solution characterized by low mass accretion rates, high gas temperatures (nearly at virial values), and significant energy advection. This model has been successfully applied to low-luminosity systems such as the Galactic center black hole, Sgr A*, and several low-luminosity active galactic nuclei (AGNs)~\cite{kato2008black,Liu:2022cph}.

While the above classification scheme offers a useful framework, it is necessarily simplified. Overlaps in mass accretion rates between models do occur. For example, ADAF and thin disks can both exist at similar mass accretion rates under different physical conditions~\cite{Abramowicz:2011xu}. A more complete characterization of an accretion flow involves several additional physical parameters, including opacity, geometric thickness, radiative efficiency, the inner edge, and the radiation pressure, among others.  In this work, our primary focus is on how the geometry of the accretion structure influences the observable features of the black hole shadow. Accordingly, the relative thickness and the inner edge of the accretion disk play a central role in our analysis. In addition, we also incorporate opacity as a key parameter to account for different radiative transfer regimes.

\begin{figure}[htbp]
\centering
\includegraphics[width=.5\textwidth]{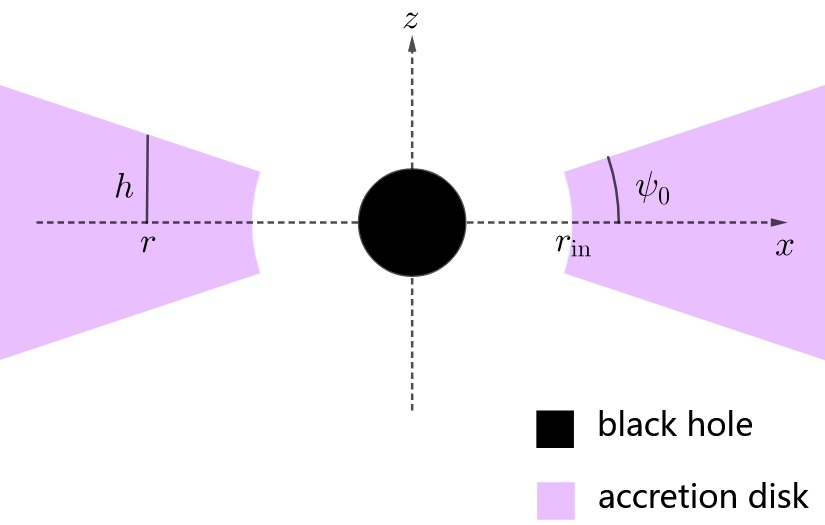}
\caption{Schematic meridional cross-sectional view ($x-z$ plane) of a geometrically thick accretion disk. The disk is symmetric about the equatorial plane and extends outward from an inner radius $r_{\mathrm{in}}$. In general, the relative thickness is characterized by $h/r$. In this paper, we consider a simplified scenario where the thickness is described by an half-opening angle $\psi_0$ at $r_{\mathrm{in}}$. \label{fig:1}}
\end{figure}

The relative thickness is defined as the ratio of the half thickness of the disk $h$~\cite{kato2008black} to the radial distance $ r $ from the black hole, i.e., $h/r$. This dimensionless quantity provides a convenient measure of the disk's aspect ratio, see Fig.~\ref{fig:1}. In geometrically thin disks, $ h/r \ll 1 $, indicating that the disk is confined to the equatorial plane.  Slim disks are primarily supported by radiation pressure, which leads to a larger vertical scale height compared to standard thin disks, with a typical relative thickness of $ h/r \lesssim 1 $~\cite{kato2008black}. Geometrically thick disks, on the other hand, can exhibit even larger relative thickness.   For the ADAF, the gas temperature is extremely high, resulting in a significant pressure-gradient force. This force becomes comparable to the centrifugal and gravitational forces in the radial direction. As a consequence, the disk is no longer geometrically thin but becomes moderately thick, with a relative scale height $ h/r \lesssim 1 $, supported primarily by thermal pressure~\cite{kato2008black}. 

Opacity describes how easily radiation passes through a medium and is a key factor in shaping the appearance of black hole accretion flows. In standard thin disks, the opacity is high, mainly due to electron scattering and absorption, making them optically thick and bright. ADAF, on the other hand, are optically thin with low opacity, allowing radiation to escape more easily but leading to much fainter images. Slim disks lie in between, with moderate opacity.

Another relevant quantity is the inner edge $r_{\mathrm{in}}$ of the accretion disk. In the standard geometrically thin disk model, the gas orbits the black hole in nearly circular trajectories with the local Keplerian velocity. For a Schwarzschild black hole, the inner edge of the thin disk coincides with the innermost stable circular orbit (ISCO), located at $r_{\mathrm{ISCO}} = 6M$ (throughout this paper, we adopt geometric units with $G_N = c = 1$).
 In contrast, thick disks and slim disks, which have higher accretion rates and incorporate pressure support and advective effects, can have inner edges that extend inside the ISCO. The inner edge can reach the marginally bound orbit (MBO), $r_{\mathrm{MBO}} = 4M$~\cite{Abramowicz:2011xu}. A similar inward extension of the inner edge is also expected for ADAF. However, unlike the standard thin disk, these models do not have a single, well-defined inner edge. As pointed out in Ref.~\cite{Krolik_2002},  there are at least four different definitions of inner edge, depending on the physical processes involved: turbulence edge, stress edge, reflection edge and radiation edge. Ref.~\cite{kato2008black} notes that as the accretion rate increases, the inner edge of the radiation can move inward from $r_{\mathrm{ISCO}}$ toward the event horizon $r_{\rm h}=2M$. 

This paper is organized as follows. In Sec.~\ref{sec:Coordinate systems and accretion disk models}, we introduce the coordinate systems used for imaging formation, and define the key parameters characterizing the simplified accretion disk models considered in this work. In Sec.~\ref{sec:imaging of one-dimensional segments}, we develop a segment-based formalism to analyze photon trajectories and image structures, introducing transfer functions associated with one-dimensional luminous segments. In Sec.~\ref{sec:Image of the accretion disk}, we apply this framework to study black hole images arising from geometrically thick accretion flows under different optical conditions, including optically thin, optically thick, and partially optically thick regimes. Sec.~\ref{sec:Conclusions and Discussion} provides concluding remarks and discusses potential extensions and limitations of the present study. Appendix~\ref{appA} contains a brief introduction to the local proper reference frame used to describe the emitter and observer kinematics. Appendix~\ref{appB} reviews the classical model of a geometrically thin disk and derives several radiation flux formulas referenced in the main text.

nai

\section{Coordinate Systems and Accretion Disk Models}
\label{sec:Coordinate systems and accretion disk models}

We focus on the Schwarzschild black hole, whose metric in spherical form with coordinates $(t, r, \theta, \phi)$ is given by
\begin{align}\label{eq:metric}
\dd s^2 = -B(r)\dd t^2 + B(r)^{-1}\dd r^2 + r^2\left(\dd \theta^2 + \sin^2\theta \, \dd \phi^2\right)\,,
\end{align}
with 
\begin{align}
B(r) = 1-\frac{2M}{r}\,.
\end{align}
One can also define the Cartesian coordinates $(x, y, z)$ which are related to the spatial spherical coordinates $(r, \theta, \phi)$ via the standard coordinate transformation. Throughout this paper, we assume that the accretion flow is symmetric with respect to the equatorial plane ($\theta = \pi/2$) and axisymmetric about the $z$-axis.

\subsection{Coordinate systems}

\begin{figure}[htbp]
\centering
\includegraphics[width=.8\textwidth]{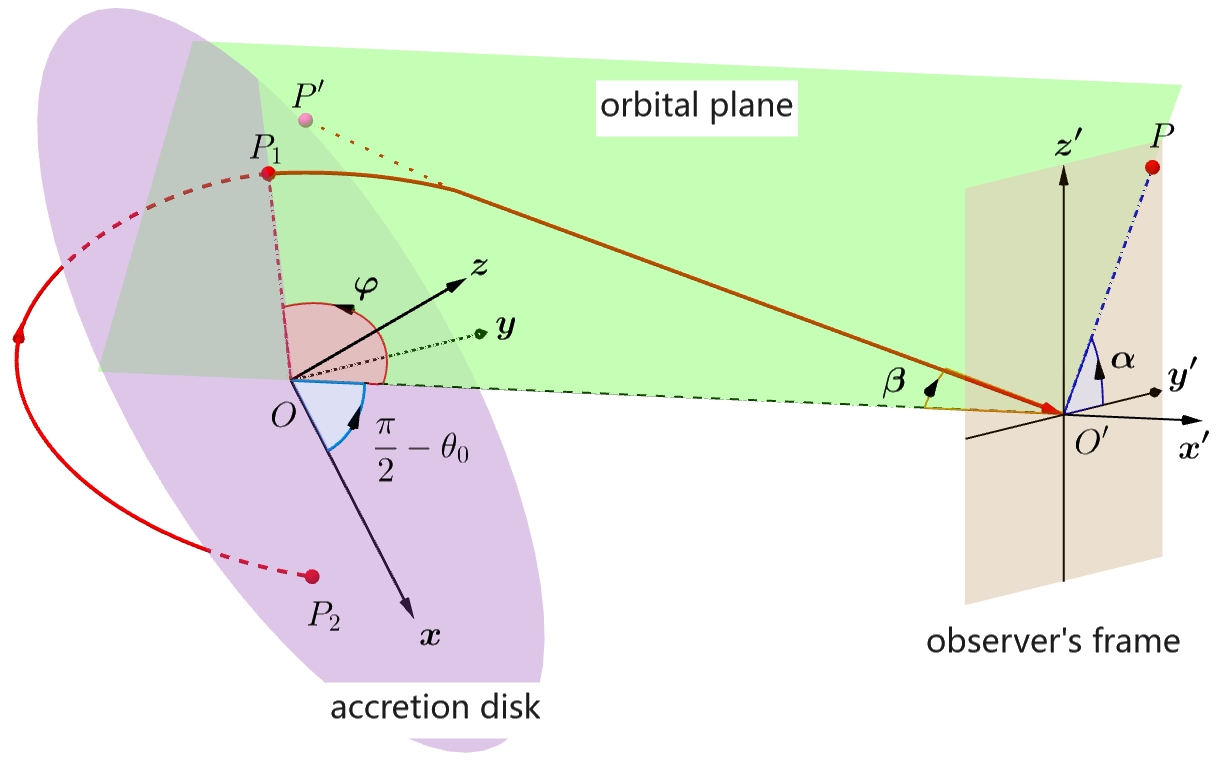}
\caption{Coordinate systems for a photon trajectory (red solid curve) intersecting with a thin accretion disk. The Cartesian coordinates $(x, y, z)$ are related to the spatial spherical coordinates $(r, \theta, \phi)$ via the standard coordinate transformation.  The observer $O'$ is located at a radius $r_{\rm obs} \gg M$ (in numerical calculations, we set $r_{\rm obs} = 10^5 M$). The accretion disk lies in the $z=0$ plane with an inclination angle $\theta_0$. The observer's local frame $(x', y', z')$ is obtained by first rotating the $(x, y, z)$ frame clockwise around the $y$-axis by an angle $\pi/2-\theta_0$, and then translating the origin to $O'$. From the observer's perspective, light rays emitted from points $P_1$ and $P_2$ appear to originate from a single point $P'$, whose projection onto the observer's image plane $y'$-$z'$ is denoted by $P$.  The position of $P$ on the image plane is characterized by  polar coordinates $(b, \alpha)$, where $b$ is the impact parameter and $\alpha$ is the angle between the projection of the photon trajectory onto the image plane and the $y'$-axis. $\beta$ is defined as the angle between  the observer-black hole axis ($\overline{OO'}$) and the incident photon trajectory.  $\varphi$ denotes the angle between the observer-black hole axis and the line connecting the black hole center to the photon emission point $P_1$. \label{fig:2}}
\end{figure}

Our primary interest is to understand how the region near the black hole appears to a distant observer in the presence of different types of surrounding accretion flows. To investigate this, we trace the trajectories of light rays (null geodesics) backward from the observer's viewpoint toward the vicinity of the black hole. Throughout this paper, we assume that the observer is located at $r_{\rm obs} = 10^5 M$ with $\phi = 0$ and $\theta = \theta_0$. For a thin accretion disk lying in the $z = 0$ plane, as illustrated in Fig.~\ref{fig:2}, $\theta_0$ corresponds to the inclination angle. In the case of a geometrically thick disk, which is symmetric about the $z = 0$ plane, the inclination angle is defined as the angle between the observer's direction and the $z$-axis.   To understand the observed appearance of black holes, it is convenient to work in the observer's local proper reference frame $(t',x', y', z')$. The local spatial coordinates $(x', y', z')$ are related to the global coordinates $(x, y, z)$ by first rotating the $(x, y, z)$ system clockwise around the $y$-axis by an angle $\pi/2-\theta_0$, and then translating the origin to the observer's position. As we will see below, the $y'$-$z'$ plane can be interpreted as the image plane.

Fig.~\ref{fig:2} displays a representative trajectory of a photon observed by an observer located at $O'$. From the observer's perspective, light rays emitted from point $P_1$ or $P_2$ appear to originate from point $P'$. Consequently, the apparent distance between the black hole center and $P'$ can be expressed as
\begin{align}\label{eq:dP}
 d_{P'}\simeq r_{\rm obs} \tan \beta\,,
\end{align}
where $\beta$ is defined as the angle between the observer-black hole axis and the incident photon trajectory. {Eq.~\eqref{eq:dP} holds approximately under the assumption that the observer is located far from the black hole, i.e., \( r_{\rm obs} \gg M \), so that the local geometry can be treated as nearly flat and the angle \( \beta \) remains sufficiently small. In this regime, \( \beta \) can be identified with the angular radius commonly used in astronomy.}
For a distant observer, it can be shown that 
\begin{align}\label{eq:apparent_size}
d_{P'} \simeq b \equiv L/E\,,
\end{align}
where $L$ and $E$ are conserved quantities along the geodesic, representing the photon's angular momentum and energy, respectively. The detailed proof of this statement is provided in Appendix~\ref{appA}. By denoting the four-momentum of the photon as $k^{\mu} =\dd x^{\mu}/\dd \lambda$ with $\lambda$ being the affine parameter along the geodesic, we also show in Appendix~\ref{appA} that 
\begin{align}\label{eq:kphi/kt}
\frac{k_{\phi}}{k_{t}}=b \cos \alpha \sin \theta_0 \,, 
\end{align}
 where $\alpha$ denotes the angle between the projection of the photon trajectory onto the $y'$-$z'$ plane and the $y'$-axis (see Fig.~\ref{fig:2}). Therefore, letting $P$ denote the projection of the apparent light source $P'$ onto the observer's image plane $y'$-$z'$, the position of $P$ is characterized by the impact parameter $b$ and the angle $\alpha$.

\subsection{Accretion disk models}
In Sec.~\ref{sec:intro}, we briefly reviewed various accretion disk models in astrophysics. Given the complexity of realistic disks, in this paper we adopt simplified models. The relevant quantities of the simplified accretion disk models are discussed below.
\subsubsection{Disk geometry and dynamics of the flow} 
\label{sec.Disk geometry and dynamics of the flow}
The disk geometry and dynamics of the flow are characterized by the following parameters:
\begin{itemize}
\item Inner edge $r_{\mathrm{in}}$ and half-opening angle $\psi_0$~\footnote{The half-opening angle $\psi_0$ (or the opening angle $2\psi_0$) is a commonly used parameter in the literature~\cite{Abramowicz:2011xu,Arshakian:2004kh} to characterize the thickness of accretion flows.}. The accretion disk is symmetric about the equatorial plane and axisymmetric around the $z$-axis, extending outward from an inner radius $r_{\mathrm{in}}$. The vertical thickness of the disk is characterized by an half-opening angle $\psi_0$ at $r_{\mathrm{in}}$, see Fig.~\ref{fig:1}. We consider the following parameter ranges: $r_{\rm h} < r_{\mathrm{in}}\leq r_{\mathrm{ISCO}}$ and $0< \psi_0 \leq \pi/2$. 
\item Opacity. The optically thin model neglects all absorption and scattering processes, allowing photons to propagate freely. In contrast, the optically thick model includes significant absorption and scattering, leading to strong coupling between radiation and matter, which prevents photons from escaping freely within the disk. In an optically thick disk, only photons emitted from its surface can escape and potentially be detected. In this work, we consider both optically thin and optically thick scenarios. Additionally, we explore an intermediate regime characterized by a finite, effective absorption coefficient.
\item Uniform motion of the disk matter. Denoting the four-velocity of the matter by $U^{\mu}$,  we consider three types of motion:
\begin{itemize}
    \item {Rotation-Dominated Flows}: Matter moves primarily along circular orbits around the black hole. For nearly Keplerian motion, the local Keplerian angular velocity is given by
    \begin{align}
    \Omega_{\rm K} = \frac{U_{\rm K}^{\phi}}{U_{\rm K}^{t}}= \left(\frac{M}{r^3}\right)^{1/2}\,.
    \end{align}
    Sub-Keplerian rotation is also considered, characterized by an angular velocity $\Omega = \kappa_{\mathrm{K}} \Omega_{\mathrm{K}}$ with $0< \kappa_{\mathrm{K}} \leq 1$. This type of motion is typical for geometrically thin disks and some slim disk configurations.

    \item {Infall-Dominated Flows}: Matter moves predominantly in a radial direction toward the black hole. For purely radial free-fall motion, the local radial velocity is
    \begin{align}
    v^r _{\rm ff}=\frac{U^r_{\rm ff}}{U^t _{\rm ff}}=-B(r)\left(\frac{2M}{r}\right)^{1/2}\,.
    \end{align}
    We consider deviations from pure free fall which are modeled by $v^r = \kappa_{\mathrm{ff}} v^r_{\mathrm{ff}}$ with $0\leq\kappa_{\mathrm{ff}} < 1$. A representative example is the ADAF, where the flow is described by a specific heat ratio of 5/3~\cite{Narayan:1994xi,kato2008black,Abramowicz:2011xu}. 

    \item {Mixed Flows}: The motion involves a combination of substantial rotation and radial infall, representing an intermediate regime between the two limiting cases above. This type of motion is typical for thick disks at high accretion rates.
\end{itemize}

\end{itemize}

\subsubsection{Emission profile and absorption}
The primary observable of our interest is the radiation flux emitted from the accretion disk. Using the standard ray-tracing method, light rays are traced backward in time from the observer toward the black hole. If a light ray intersects the accretion disk, it may contribute to the observed flux.

The simplest scenario arises for an optically thick disk. In this case, the observed flux is determined solely by the radiation emitted at the \emph{first} intersection point of the ray with the disk surface during the backward tracing. In contrast, for an {optically thin} disk, all regions traversed by the light ray within the disk can contribute to the observed flux, as radiation can be emitted throughout the volume.

For an optically thick disk, the {local emitted flux}---that is, the radiative energy emitted per unit time per unit area of the disk surface---is modeled as
\begin{align}\label{eq:deltaF}
F_{\rm em}(r) \approx \frac{\epsilon \dot{M}r}{4\pi} \left( -\Omega + \frac{r_{\rm in}^2 \, \Omega(r_{\rm in})}{r^2} \right) \frac{\mathrm{d}\Omega}{\mathrm{d}r},
\end{align}
where:
\begin{itemize}
    \item $\dot{M}$ is the mass accretion rate,  assumed constant for simplicity.
    \item $\epsilon$ is a dimensionless efficiency parameter that characterizes the conversion of rotational (orbital) energy into radiation due to viscous dissipation in the differentially rotating disk. In our numerical calculations, we simply set $\epsilon = 1$.
    \item $\Omega$ is the local angular velocity and $r_{\rm in}$ denotes the inner radius of the accretion disk.
\end{itemize}

A derivation of Eq.~\eqref{eq:deltaF} based on the classical Shakura-Sunyaev thin-disk model is provided in Appendix~\ref{appB}.

For an optically thin disk, radiation may be emitted continuously along the photon path. In this case, the contribution to the emitted flux over a small path length interval $\delta l$ is modeled as
\begin{align}\label{eq:deltaF2}
\delta F_{\rm em} =q^-\,\sqrt{B}\,  \delta l \approx \frac{\epsilon \dot{M}}{4\pi \psi_0} \left( -\Omega + \frac{r_{\rm in}^2 \, \Omega(r_{\rm in})}{r^2} \right) \frac{\mathrm{d}\Omega}{\mathrm{d}r} \sqrt{B} \, \delta l,
\end{align}
{where \( q^- \) denotes the rate of radiative energy loss per coordinate volume element (a derivation of \( q^- \) is provided in Appendix~\ref{appB})\footnote{The derivation of \( q^- \) in Appendix~\ref{appB} is purely classical, where the volume element is implicitly assumed to be $r^2\sin \theta \dd r \dd \theta \dd \phi$. To convert this coordinate volume element into the proper volume element, a factor  $\sqrt{B}$ should be included. For a more relativistic treatment, one may adopt the Novikov-Thorne model~\cite{novikov1973astrophysics}, which, however, involves complicated calculations, especially when generalized to geometrically thick disks. Given the goals of this paper, our simplified approach suffices to capture the essential physics.}, and \( \psi_0 \) is the half-opening angle defined in Sec.~\ref{sec.Disk geometry and dynamics of the flow} The factor $\sqrt{B}$ in \eqref{eq:deltaF2} accounts for the proper radial distance, converting the coordinate volume element into the proper volume element. The interval $\delta l$ represents the spatial distance travelled by a photon with four-momentum $k^{\mu}$, as measured by a local observer with four-velocity $V^\mu$. It can be expressed as  
\begin{align}
\delta l =\sqrt{(g_{\mu \nu}+V_{\mu}V_{\nu})k^{\mu}k^{\nu}}\, \delta \lambda =-k_{\mu}V^{\mu}\,\delta \lambda\,. 
\end{align}
In numerical computations, the affine parameter interval $\delta \lambda$ can be related to the $r$-component of the photon's four-momentum, yielding
\begin{align}
\delta l =\left(B-B^2 b^2 /r^2 \right)^{-1/2} {\delta r}\,,
\end{align}
 where the four-velocity $V^{\mu}$ of a local static observer has been used. (Note that this local static observer is hypothetical and is assumed to reside on (or within) the accretion disk for illustrative purposes. It should not be confused with the distant observer.) At locations where $\delta r$ vanishes (e.g., near turning points), an alternative expression involving the $\phi$-component of the photon's four-momentum may be used.}

To account for absorption within the accretion flow, we adopt an approximate recurrence relation for the total local flux at a given radius $r_2$,
\begin{align}\label{eq:F_lambda}
F(r_2) = F(r_1) \, e^{-\chi \, \delta l} \left[ \frac{g(r_1)}{g(r_2)} \right]^4 + \delta F_{\rm em}(r_2),
\end{align}
where:
\begin{itemize}
    \item $F(r_1)$ is the accumulated flux at the previous point along the ray path, effectively corresponding to the last interaction with the disk in the numerical integration,
    \item $g(r) = \nu_{\rm obs}/\nu_{\rm em}(r)$ is the redshift factor,
    \item $\chi$ denotes the effective absorption coefficient, treated as constant in this work for simplicity. An optically thick medium corresponds to the limit $\chi \to \infty$, whereas an optically thin medium is represented by $\chi \to 0$.
\end{itemize}

Note that our model assumes that the radiation originates solely from the dissipation of rotational energy through differential rotation.\footnote{One could, of course, consider more general relativistic scenarios in which radiation arises from processes other than viscous dissipation, such as magnetic reconnection or shocks. However, the approximation of considering only radiation from viscous dissipation due to differential rotation is sufficient for our purposes.}
 Consequently, in the case of purely radial infall, where $\dd \Omega / \dd r = 0$, no radiation is produced within this framework. To avoid this unphysical outcome, we do not consider purely infalling motion. Instead, we model infall-dominated flows that retain a small but nonzero rotational component.

In the case of an optically thick disk, the observed bolometric flux is given by
\begin{align}
F_{\rm obs} = \left[ g(r) \right]^4 F_{\rm em}(r)\,.
\end{align}
In the case of optically thin or partially optically thick disks, the observed bolometric flux is computed using Eq.~\eqref{eq:F_lambda}, by integrating the local flux along the entire portion of the light ray that traverses the disk on its way to the observer. In the subsequent numerical calculations, we set the outer radius of the accretion disk to be $r_{\rm out} = 50M $, unless otherwise specified.

At the end of this section, we summarize the key quantities relevant for computing bolometric flux measured by a static distant observer:\footnote{It should be noted that for given values of $\kappa_{\rm ff}$ and $\kappa_{\rm K}$, the requirement for $U^{t}$ to be real imposes a lower limit on the radius $r$. Consequently, the inner radius $r_{\rm in}$ must be larger than this minimum value.}
\begin{subequations} 
\begin{align}
g(r) &= \frac{\nu_{\rm obs}}{\nu_{\rm em}(r)} = \left[U^t \left(1 + v^r \frac{k_r}{k_t} + \Omega \frac{k_\phi}{k_t} \right) \right]^{-1}, \\
U^t &= \left[B - \frac{(v^r)^2}{B} - r^2 \sin^2 \theta\, \Omega^2 \right]^{-1/2}, \\
v^r &= -\kappa_{\rm ff}\, B\, \left(\frac{2M}{r}\right)^{1/2}, \\
\Omega &= \kappa_{\rm K} \left(\frac{M}{r^3}\right)^{1/2}, \\
\frac{k_r}{k_t} &= \pm \frac{\sqrt{1 - B\, b^2 / r^2}}{B}, \\
\frac{k_\phi}{k_t} &= b \cos \alpha \sin \theta_0.
\end{align}
\end{subequations}

\section{Imaging of One-dimensional Segments}
\label{sec:imaging of one-dimensional segments}
To develop a clearer understanding of the observational features associated with geometrically thick accretion flows, in this section we investigate the relativistic imaging of one-dimensional luminous segments that extend radially from the vicinity of a Schwarzschild black hole. These segments, defined by their inner radius and inclination angle relative to the observer-black hole axis, serve as simplified geometric emitters that allow us to isolate and examine key general relativistic effects, including light bending, Doppler boosting, and gravitational redshift.  By systematically varying both the orientation and the inner radius of the segments, we examine how these parameters influence the observed flux profile and the emergence of peak-like features along the projected image line on the observer's screen. This study establishes a conceptual and computational foundation for the more complex modeling of thick accretion disk imaging presented in Sec.~\ref{sec:Image of the accretion disk}.

\begin{figure}[htbp]
\centering
\includegraphics[width=.5\textwidth]{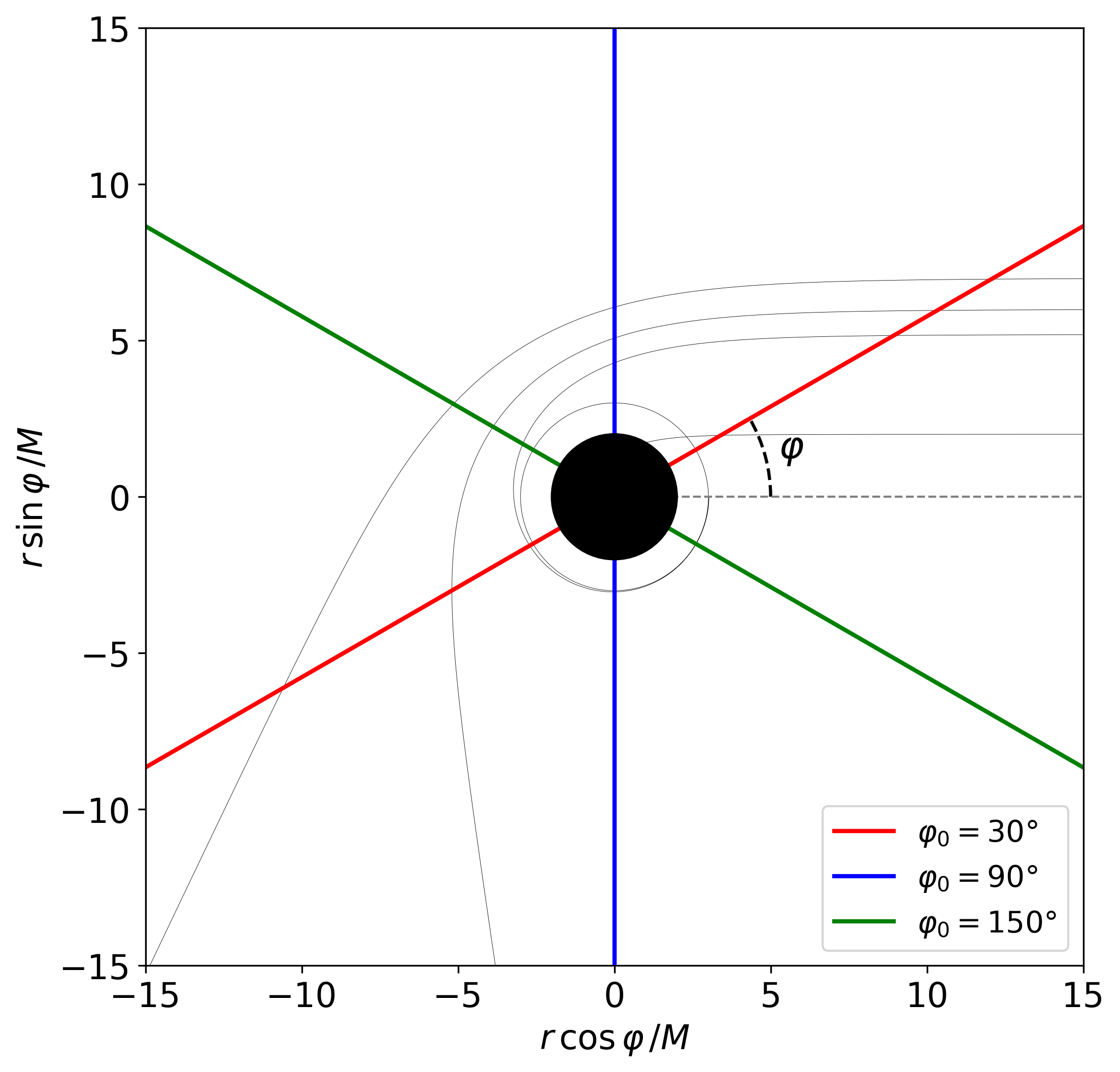}
\caption{Sketch of one-dimensional luminous segments lying in the photon's orbital plane. {The colorful thick solid lines depict representative emitting segments.} Thin black lines illustrate representative photon trajectories, while the dashed line indicates the observer-black hole axis.  \label{fig:1d_sketch}}
\end{figure}

A schematic illustration of luminous segments with different inclination angles is presented in Fig.~\ref{fig:1d_sketch}. Here, $\varphi_0 \in (0, \pi]$ denotes the inclination angle of an individual luminous segment, which should not be confused with $\theta_0$, the inclination angle of the entire accretion disk as introduced in Sec.~\ref{sec:Coordinate systems and accretion disk models}. The latter follows the standard convention commonly used in astrophysics. {As illustrated in Fig.~\ref{fig:1d_sketch}, regardless of the inclination angle of the accretion disk, we assume that the luminous segments lie in the photon's orbital plane rather than the physical disk plane. It should be emphasized that a two- or three-dimensional accretion disk can always be viewed as a composition of such luminous segments, regardless of its inclination angle.} Without loss of generality, we restrict our analysis in this section to the image line corresponding to photons with positive impact parameters. Under this setup, a thin accretion disk with an inclination angle $\theta_0$ can be mathematically modeled as a continuous distribution of radial luminous segments, each characterized by an inclination angle $\varphi_0$ ranging from $[\pi/2 - \theta_0,\, \pi/2 + \theta_0]$.

\subsection{Transfer functions for one-dimensional segments}

\begin{figure}[htbp]
    \centering
    \begin{subfigure}{0.495\textwidth}
        \includegraphics[width=\linewidth]{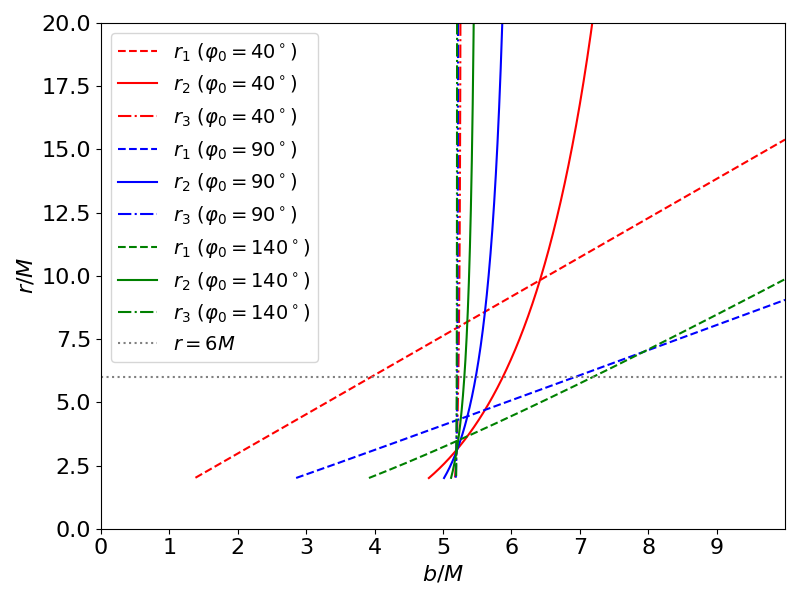}
        %\caption{1}
    \end{subfigure}
    \hfill
    \begin{subfigure}{0.495\textwidth}
        \includegraphics[width=\linewidth]{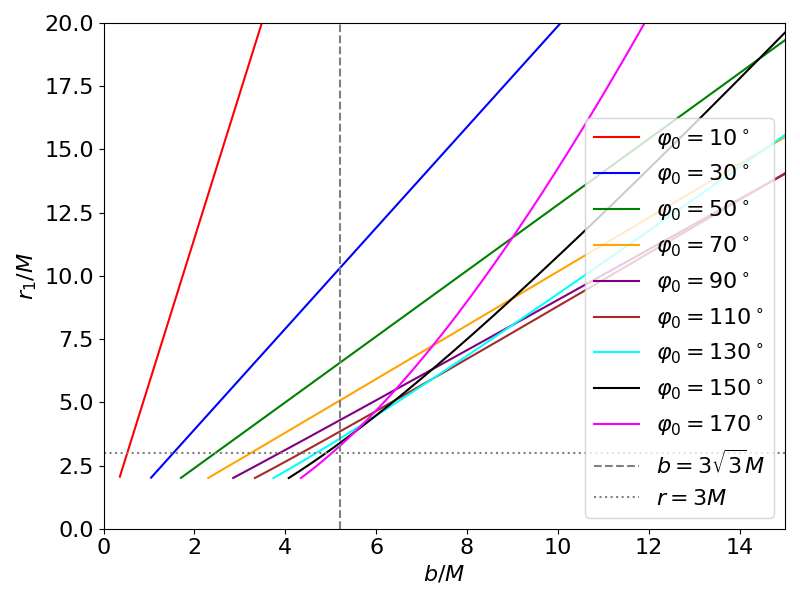}
        %\caption{2}
    \end{subfigure}

    \vspace{0.5cm}

    \begin{subfigure}{0.495\textwidth}
        \includegraphics[width=\linewidth]{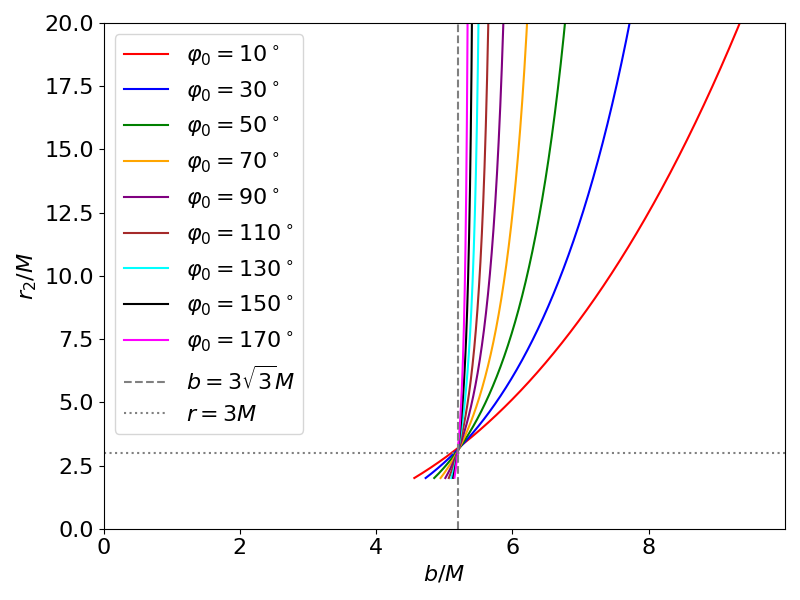}
        %\caption{3}
    \end{subfigure}
    \hfill
    \begin{subfigure}{0.495\textwidth}
        \includegraphics[width=\linewidth]{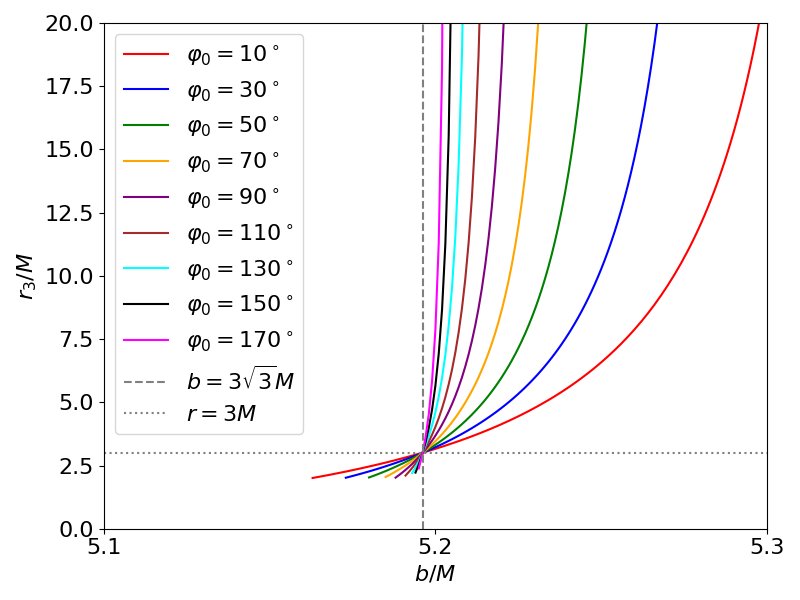}
       % \caption{4}
    \end{subfigure}

    \caption{The first three transfer functions for luminous segments at different inclination angles.  
Top left: Selected inclination angles.  Top right, bottom left, and bottom right: First, second, and third transfer functions over a broad range of inclination angles, respectively.}
    \label{fig:transfer function}
\end{figure}

An useful quantity in the study of disk imaging is the transfer function, which was first introduced in Ref.~\cite{Gralla:2019xty} in the context of imaging a face-on accretion disk. In this work, we generalize the concept of the transfer function to configurations with arbitrary inclination angles. We denote the transfer function by $r_m(b)$ ($m = 1, 2, 3, \dots$), which specifies the radial coordinate of the $m$-th intersection between a photon trajectory and a radial segment with inner edge located at $r_{\rm in}=r_{\rm h}$.  

Fig.~\ref{fig:transfer function} illustrates the first three transfer functions for luminous segments at different inclination angles. An important features can be identified from the figure is that, for the second and third transfer functions, a critical transition occurs near $r \approx r_{\rm{sp}} = 3M$, where the impact parameter $b$ attains $3\sqrt{3}M$ and is nearly independent of the inclination angle $\varphi_0$. This radius corresponds to the photon sphere and plays a key role in shaping the image structure. For radii larger than $r_{\rm{sp}}$, the transfer functions exhibit approximately monotonic behavior: the impact parameter $b$ tends to decrease as the inclination angle $\varphi_0$ increases. In contrast, for $r < r_{\rm{sp}}$, the trend reverses: $b$ increases with increasing $\varphi_0$.  The first transfer function does not exhibit this monotonic behavior for radii greater than $r_{\rm{sp}}$.  In the notation of Ref.~\cite{Gralla:2019xty}, the image components associated with the first three transfer functions are referred to as the ``direct image,'' ``lensing ring,'' and ``photon ring,'' respectively.  It should also be noted that for segments with inner edge $r_{\rm in}>r_{\rm sp}$, the appearance of a lensing ring or photon ring does not necessarily require multiple intersections with the segments.  This distinction becomes particularly important in the optically thick case, where light rays typically intersect the emitting surface only once. Therefore, the classical definitions---based on multiple intersections with the disk---no longer strictly apply. In this work, we adopt the following generalized definitions:
\begin{itemize}
    \item Lensing ring: the set of image points formed by light rays whose total azimuthal angle change (in the orbital plane) outside the horizon exceeds $ \pi $.
    \item Photon ring: the set of image points formed by light rays whose total azimuthal angle change (in the orbital plane) outside the horizon exceeds $ 2\pi $.
\end{itemize}
These definitions are applicable even in scenarios where only a single intersection with the disk occurs, thereby allowing consistent terminology across both optically thin and optically thick cases. 

It should be noted that in the case of one-dimensional emission, these ``rings'' manifest as distinct peaks along the image line, rather than as extended ring-like structures.

\subsection{Image of one-dimensional segments}
\begin{figure}[htbp]
    \centering
    \begin{subfigure}{0.49\textwidth}
        \includegraphics[width=\linewidth]{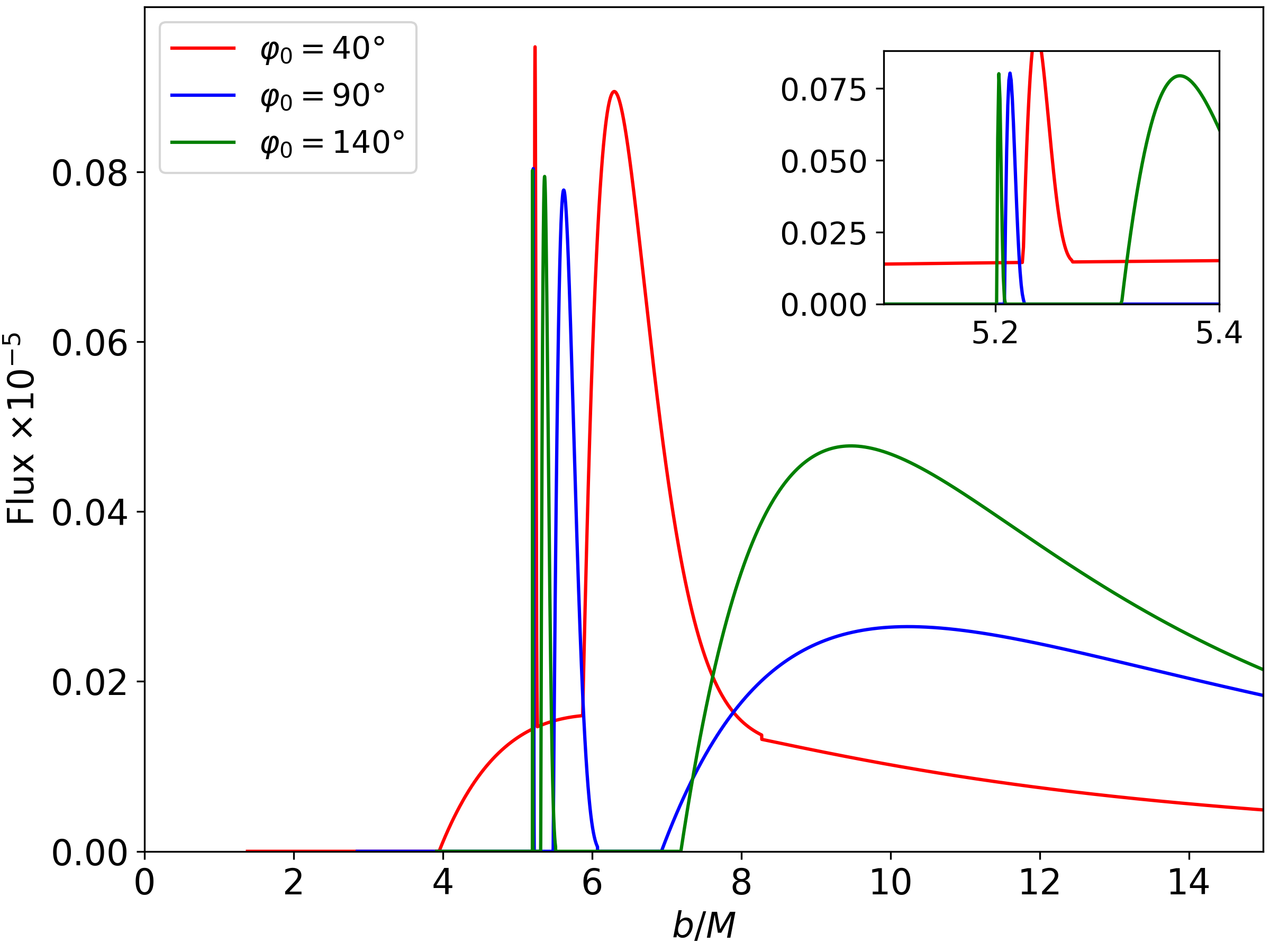}
        %\caption{1}
    \end{subfigure}
    \hfill
    \begin{subfigure}{0.49\textwidth}
        \includegraphics[width=\linewidth]{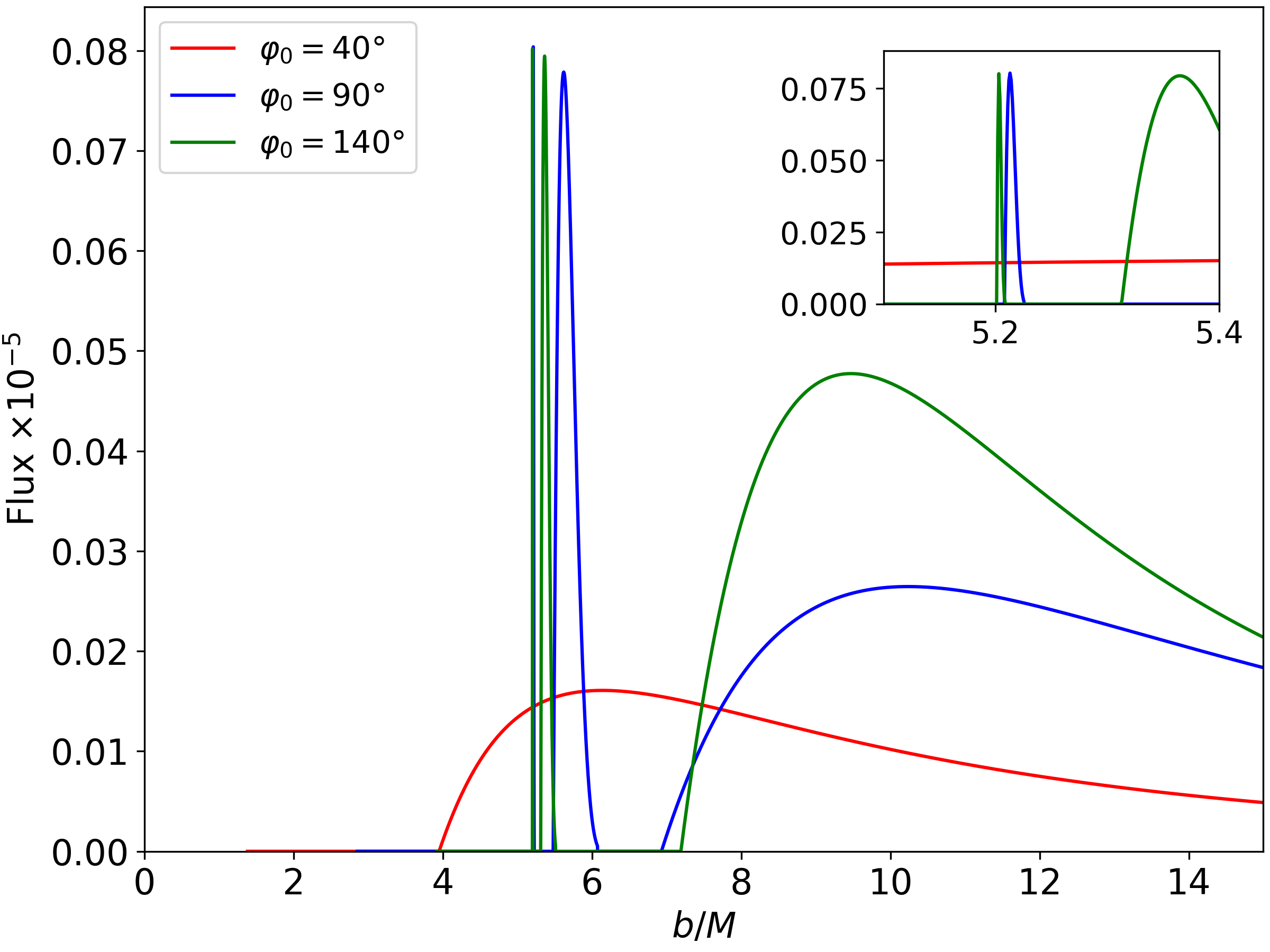}
        %\caption{2}
    \end{subfigure}
    \caption{Observed flux from one-dimensional luminous segments with an inner radius of $r_{\rm in} = 6M$. The left panel corresponds to the optically thin case, while the right panel shows the optically thick case. These image lines correspond to projections of certain thin accretion disk structures onto the observer's image plane. Specifically, $\varphi_0 = 90^\circ$ corresponds to the $\alpha = {\pi}/{2}$ line (positive $z'-$axis) in the image of a face-on disk with inclination angle $\theta_0 = 0$; $\varphi_0 = 40^\circ$ and $\varphi_0 = 140^\circ$ correspond to the $\alpha = {3\pi}/{2}$(negative $z'-$axis) and $\alpha = {\pi}/{2}$ lines, respectively, in the image of a disk with inclination angle $\theta_0 = 50^\circ$. All disks assume $\kappa_{\rm ff} = 0.5$ and $\kappa_{\rm K} = 0.1$. \label{fig:1d_emission1}}
\end{figure}

One important application of the transfer function is to determine the inner boundary of the image line, as well as the minimum radius at which potential peaks---such as those corresponding to the lensing ring or photon ring---can appear.  An explicit example of this can be seen by comparing the transfer functions in the top-right panel of Fig.~\ref{fig:transfer function} with the flux profiles shown in Fig.~\ref{fig:1d_emission1}.\footnote{In calculating the flux contribution from each luminous segment, we continue to model it as a one-dimensional object, but assign it an infinitesimal area element to ensure a physically meaningful emission flux.}

Specifically, in the top-right panel of Fig.~\ref{fig:transfer function}, the intersection points between the transfer functions and the horizontal line $r = 6M$ indicate the inner edge of the image line or the minimum radius at which peaks may appear. These features are directly reflected in the flux distribution shown in Fig.~\ref{fig:1d_emission1}. In the optically thin regime, the correspondence is typically one-to-one, with each intersection point mapping directly to the inner edge either of the image line or a distinct brightness peak. In contrast, for optically thick emission, the presence of peaks associated with the lensing ring and photon ring depends sensitively on the inclination angle $\varphi_0$ and the segment's inner radius $r_{\rm in}$; such peaks may be absent in some configurations---for instance, see the case with $\varphi_0 = 40^\circ$ in the right panel of Fig.~\ref{fig:1d_emission1}.

\begin{figure}[htbp]
\centering
\includegraphics[width=.5\textwidth]{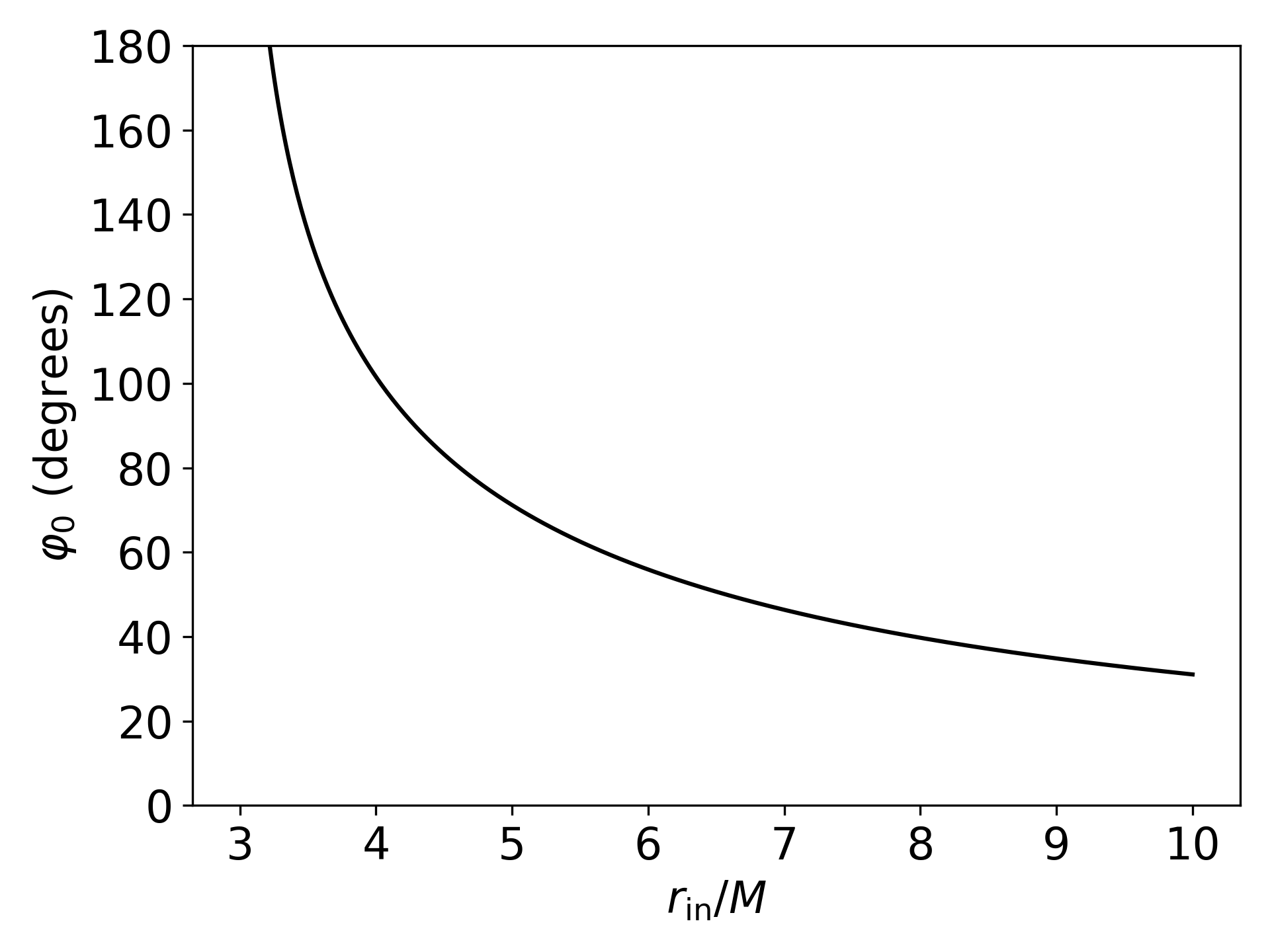}
\caption{Minimum inclination angle $\varphi_0$ required for the appearance of the photon ring peak in the image line, given an optically thick luminous segment with inner radius $r_{\rm in}$.\label{fig:critical photon orbit}}
\end{figure}

In fact, an analytical condition can be derived for the appearance of the photon ring in the optically thick case. The critical photon orbit---i.e., a null geodesic arriving from infinity with the critical impact parameter $b = 3\sqrt{3}M$---asymptotically spirals toward the photon sphere. The radial trajectory of this orbit as a function of the deflection angle $\varphi$ is given by the following expression~\cite{Chandrasekhar1985}:
\begin{subequations}\label{eq:critical photon orbit1}
\begin{align}
\frac{1}{r(\varphi)} = -\frac{1}{6M} + \frac{1}{2M} \tanh^2\left(\frac{\varphi - \bar{\varphi}}{2} \right)\,,
\end{align}
with
\begin{align}
\bar{\varphi} = -2\,\mathrm{arctanh}\left(\frac{1}{\sqrt{3}}\right)\,,
\end{align}
\end{subequations}
in our setup. A photon ring peak appears in the image line of a luminous segment if the critical photon orbit does not intersect the segment when $0 < \varphi \leq \pi$. More precisely, for a luminous segment with inner radius $r_{\rm in}$, the photon ring peak will be present in its image if and only if its inclination angle $\varphi_0$ satisfies
\begin{align}\label{eq:critical photon orbit2}
\varphi_0 > \bar{\varphi} + 2\, \mathrm{arctanh} \left( \sqrt{\frac{2M}{r_{\rm in}} + \frac{1}{3}} \right)\,.
\end{align}
This provides a criterion for the visibility of photon ring features in the optically thick regime. In Fig.~\ref{fig:critical photon orbit}, we plot the right-hand side of Eq.\eqref{eq:critical photon orbit2} as a function of $r_{\rm in}$. This curve indicates the minimum inclination angle $\varphi_0$ required for the appearance of the photon ring peak in the image line, given a luminous segment with inner radius $r_{\rm in}$.

\begin{figure}[htbp]
    \centering
    \begin{subfigure}{0.49\textwidth}
        \includegraphics[width=\linewidth]{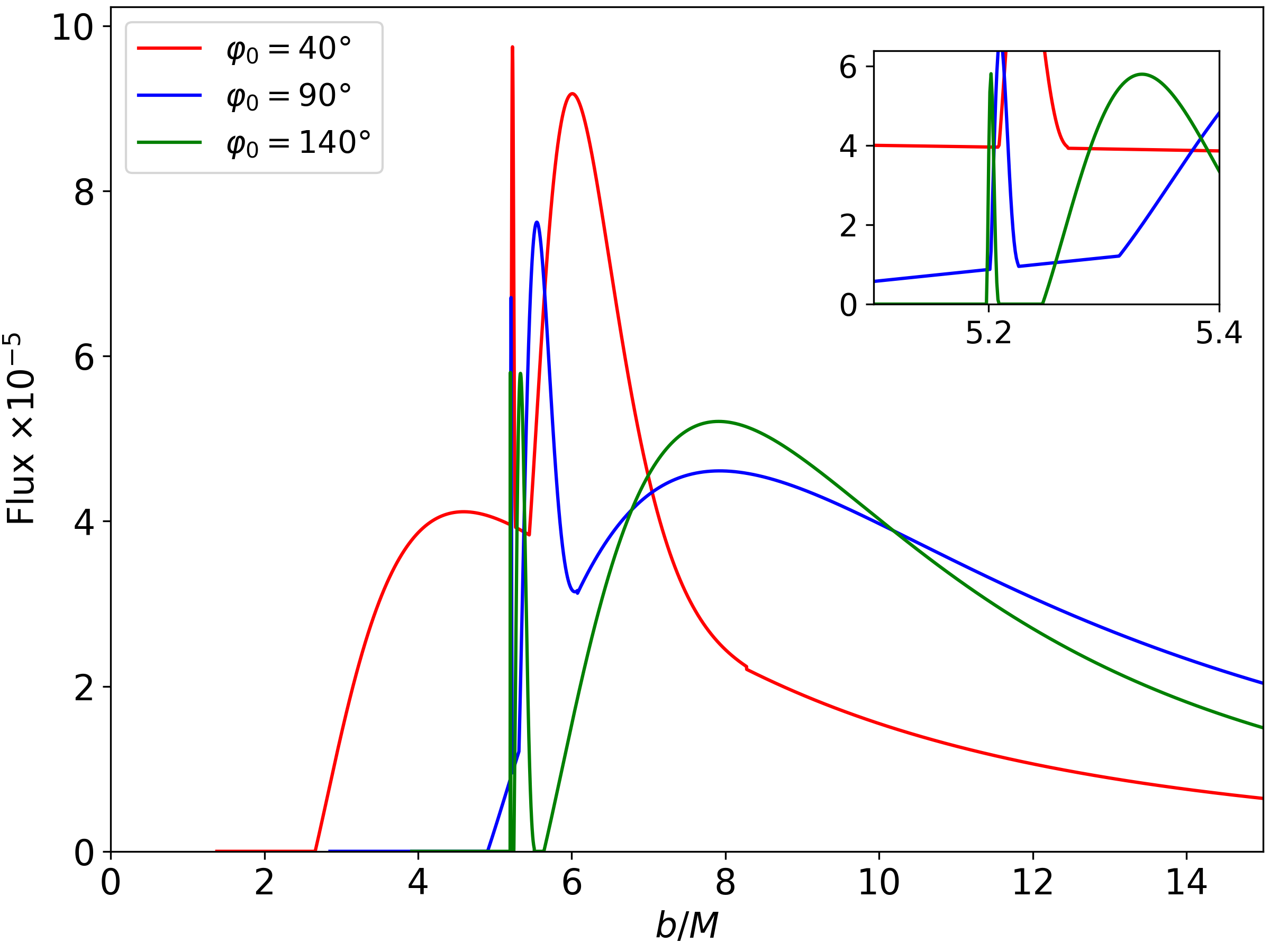}
        %\caption{1}
    \end{subfigure}
    \hfill
    \begin{subfigure}{0.49\textwidth}
        \includegraphics[width=\linewidth]{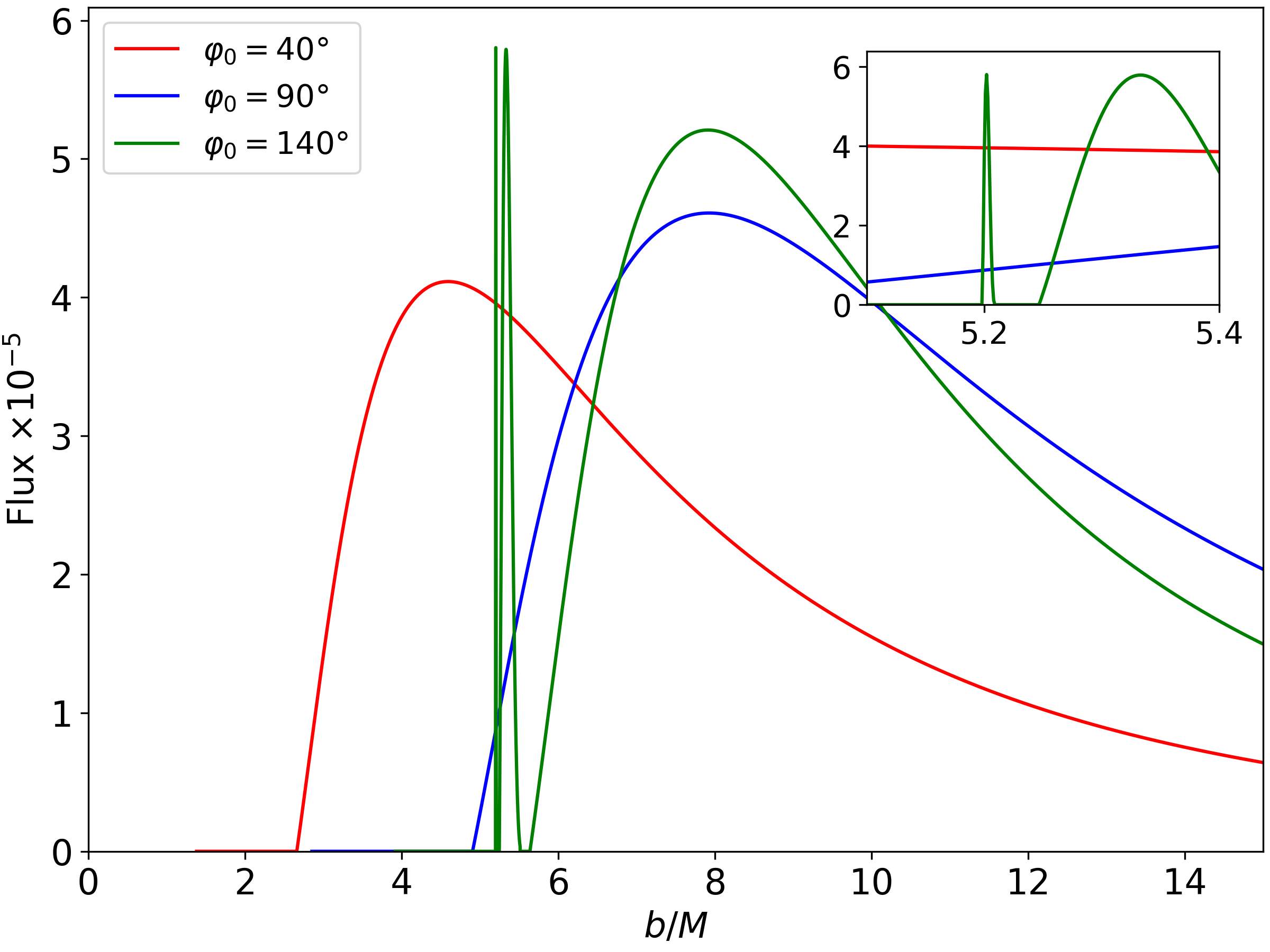}
        %\caption{2}
    \end{subfigure}
    \caption{Same as Fig.~\ref{fig:1d_emission1}, but with $r_{\rm in} = 4M$, $\kappa_{\rm ff} = 0.1$ and $\kappa_{\rm K} = 0.9$. \label{fig:1d_emission2}}
\end{figure}

In Fig.~\ref{fig:1d_emission2}, we present the observed flux from one-dimensional luminous segments with an inner radius of $r_{\rm in} = 4M$. According to Eq.~\eqref{eq:critical photon orbit2}, the minimum inclination angle required for the appearance of the photon ring peak is $\varphi = 101.5^\circ$. Consequently, as shown in the right panel of Fig.~\ref{fig:1d_emission2}, the photon ring peak is absent for both $\varphi_0 = 90^\circ$ and $40^\circ$. Except for this, there are several prominent differences between Fig.~\ref{fig:1d_emission1} and Fig.~\ref{fig:1d_emission2}. First, the images in Fig.~\ref{fig:1d_emission2} are overall brighter than those in Fig.~\ref{fig:1d_emission1}, due to the larger angular velocity characterized by $\kappa_{\rm K}$. (Recall that the emitted flux is proportional to $\dd \Omega / \dd r$.)  Even after removing this overall brightness difference, the flux associated with $\varphi_0 = 40^\circ$ inside the critical impact parameter in Fig.~\ref{fig:1d_emission2} remains significantly higher than its counterpart in Fig.~\ref{fig:1d_emission1}. This is because, for $\varphi_0 = 40^\circ$, the dominant contribution comes from the near side of the segment, where emitters are moving away from the observer. A larger infall velocity in this case leads to a stronger Doppler effect, reducing the observed intensity.  Another notable distinction lies in the dependence of image structure on the inner radius of the segment. As $r_{\rm in}$ decreases, the inner boundaries of both the direct image and the rings features systematically shift inward. This behavior is fully consistent with the trends predicted by the transfer function analysis in Fig.~\ref{fig:transfer function}. 

Let us return to the optically thin case. As shown in both Fig.~\ref{fig:1d_emission1} and Fig.~\ref{fig:1d_emission2}, the lensing ring exhibits a broader width when the inclination angle $\varphi_0$ becomes smaller. This is because the second transfer function has a shallower slope for smaller values of $\varphi_0$, leading to a more extended radial range contributing to the image.

Another observation from the left panel of Fig.~\ref{fig:1d_emission1} is that, for $\varphi_0 = 90^\circ$ and $140^\circ$, the observed flux profile contains three distinct peaks---corresponding to the photon ring, lensing ring, and direct image---though the visibility of the photon ring peak may depend on the experimental resolution. In contrast, for $\varphi_0 = 40^\circ$, only two peaks are present: the photon ring and the lensing ring. This is because the inner edge of the direct image extends into the region inside the critical impact parameter, and its flux contribution is spread over a wider area, resulting in a smoother profile that lacks a clearly defined peak due to the overlapping influence of the photon and lensing rings. In the right panel of Fig.~\ref{fig:1d_emission1}, which shows the optically thick scenario, we observe a broad and smooth flux enhancement for $\varphi_0 = 40^\circ$ associated with the direct image. Since the photon ring and lensing ring are absent in this case, the direct image becomes more apparent, albeit without a sharply defined peak.

\section{Imaging of Accretion Disks}
\label{sec:Image of the accretion disk}

Now we move on to study the full imaging of geometrically thick accretion disks. As we will see, the insights obtained from the segment-based transfer functions provide a theoretical foundation for interpreting the key features observed in the complete images, including the behavior and formation of the photon ring, lensing ring, and direct image under different geometric configurations.

Specifically, this section systematically explores how the observed intensity profiles depend on the disk's inner radius $r_{\rm in}$ and half-opening angle $\psi_0$, which jointly characterize the size and thickness of the emitting region. We focus on two representative inclination angles of the observer: $\theta_0 = 20^\circ$ and $\theta_0 = 50^\circ$, which correspond roughly to the inferred viewing angles of M87*~\cite{EventHorizonTelescope:2019dse} and Sgr~A*~\citep{EventHorizonTelescope:2022wkp}, respectively. To further capture the physical nature of different accretion regimes, we adopt different motion of the flow for these two cases. For $\theta_0 = 20^\circ$, we choose $\kappa_{\rm ff} = 0.1$ and $\kappa_{\rm K} = 0.9$, representing a disk dominated by Keplerian rotation. For $\theta_0 = 50^\circ$, we take $\kappa_{\rm ff} = 0.5$ and $\kappa_{\rm K} = 0.1$, modeling a disk with a more significant contribution from advection-dominated accretion flows (ADAF). 

Moreover, we classify our results by opacity---into optically thin, optically thick, and partially optically thick regimes---to highlight the distinct radiative features of each. In summary, within the scope allowed by our computational resources, we have incorporated a wide range of parameters in an effort to gain a more comprehensive understanding of black hole imaging.

\subsection{Optically thin disks}
\label{sec:Optically thin disks}

\begin{figure}[htbp]
\centering
\includegraphics[width=.85\textwidth]{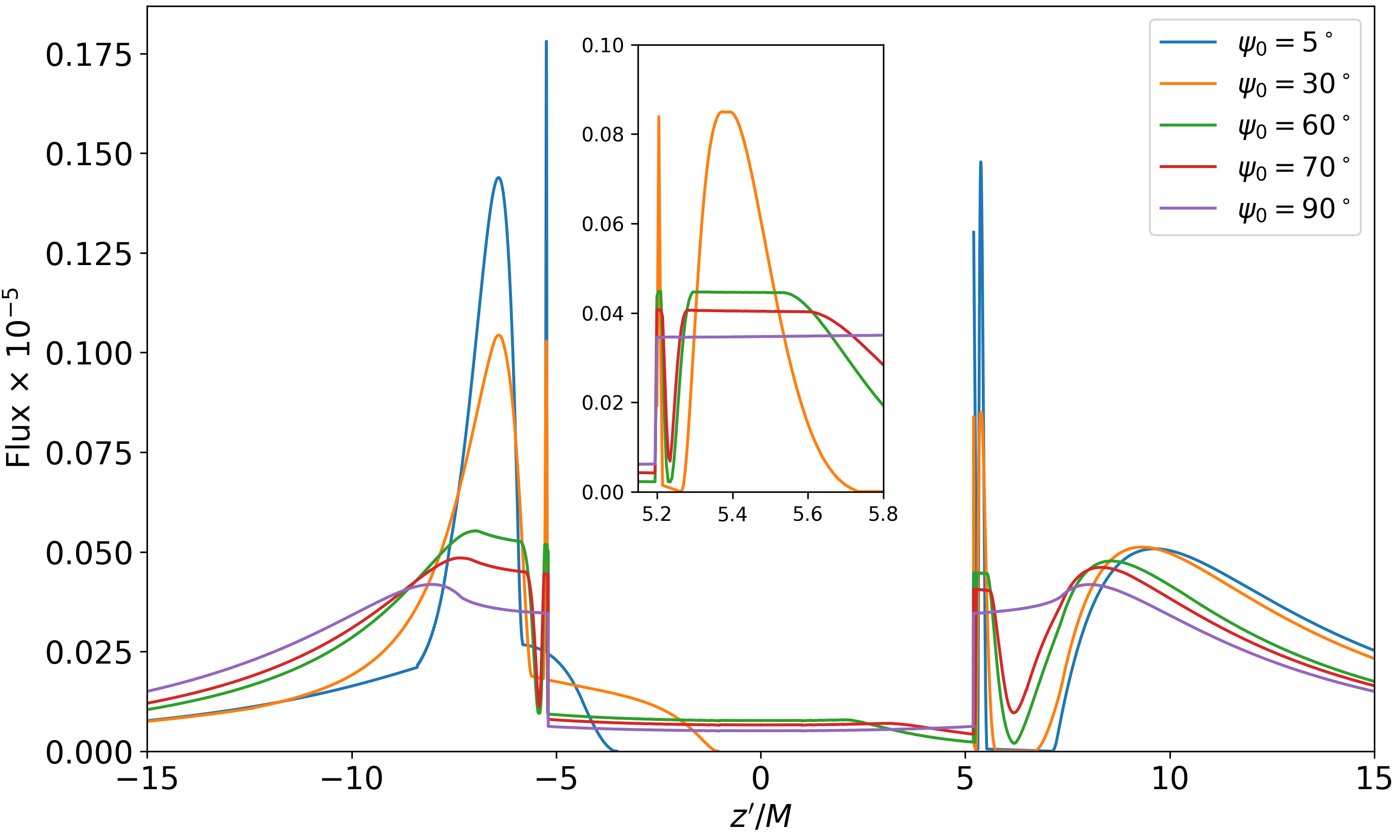}
\caption{Observed flux along the $z'$ axis of the image plane for optically thin disks with an inclination angle $\theta_0 = 50^\circ$, an inner radius of $r_{\rm in} = 6M$, and varying half-opening angles $\psi_0$. The inset highlights a zoomed-in region around the lensing ring for thick disks, aiming to illustrate the trend of its evolution with increasing disk thickness. All disks assume $\kappa_{\rm ff} = 0.5$ and $\kappa_{\rm K} = 0.1$. \label{fig:fig8}}
\end{figure}

\begin{figure}[htbp]
\centering
\includegraphics[width=.85\textwidth]{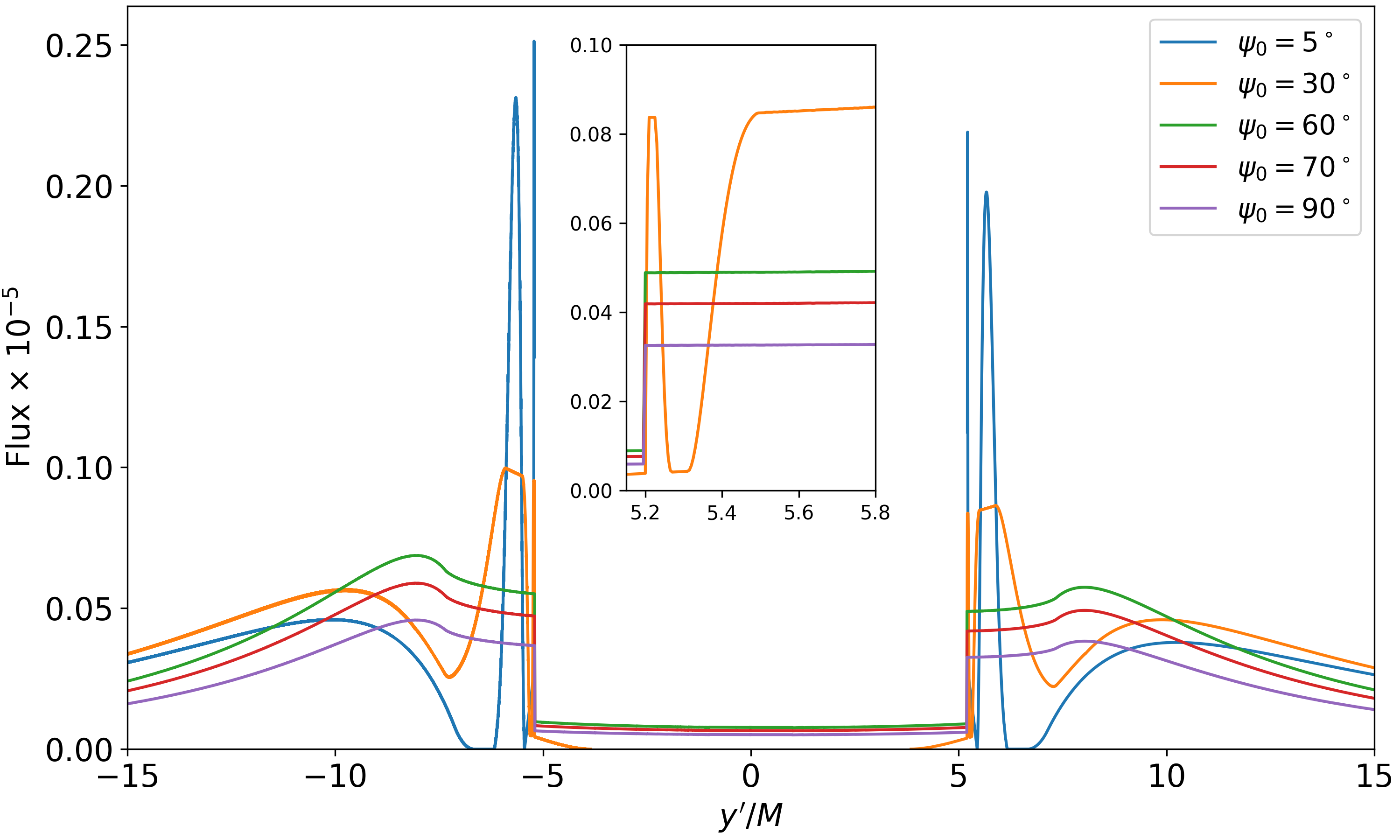}
\caption{Observed flux along the $y'$ axis of the image plane for optically thin disks with an inclination angle of $\theta_0 = 50^\circ$, an inner radius of $r_{\rm in} = 6M$, and varying half-opening angles $\psi_0$. All disks assume $\kappa_{\rm ff} = 0.5$ and $\kappa_{\rm K} = 0.1$. \label{fig:fig9}}
\end{figure}

The observed flux along the $z'$ and $y'$ axes of the image plane for optically thin disks with an inclination angle of $\theta_0 = 50^\circ$ and inner radius of $r_{\rm in} =6M$ are presented in Figs.~\ref{fig:fig8} and \ref{fig:fig9}. Different colors of the curves correspond to different values of the half-opening angle $\psi_0$, and thus represent disks with varying thickness. As we will show below, these flux profiles can be well understood in terms of the one-dimensional luminous segments discussed in Sec.~\ref{sec:imaging of one-dimensional segments}.

First, we focus on the positive $z'$ axis in Fig.~\ref{fig:fig8}. The envelope of the $\psi_0 = 5^\circ$ flux profile, corresponding to a relatively thin disk, closely resembles that of the $\varphi_0 = 140^\circ$ profile shown in the left panel of Fig.~\ref{fig:1d_emission1}. This similarity arises because, in the line of sight, a disk with inclination angle $\theta_0 = 50^\circ$ and half-opening angle $\psi_0 = 5^\circ$ is composed of luminous segments with inclination angles $\varphi_0$ ranging from $135^\circ$ to $145^\circ$. 

%As $\psi_0$ increases, the corresponding luminous segments span a broader range of inclination angles. \footnote{Note that a segment with $\varphi_0 = 200^\circ$ is equivalent to a segment with $\varphi_0 = 20^\circ$, due to periodicity.}
For the case of $\psi_0 = 30^\circ$, the relevant luminous segments span inclination angles from $\varphi_0 = 110^\circ$ to $170^\circ$. As shown from the second transfer function in Fig.~\ref{fig:transfer function}, decreasing the lower bound of $\varphi_0$  extends the outer edge of the lensing ring, while increasing the upper bound pushes the inner edge inward. Consequently, increasing the disk's half-opening angle $\psi_0$ incorporates a broader range of inclined segments, thereby widening the lensing ring in the observed image. This trend is clearly reflected in Fig.~\ref{fig:fig8}, particularly in the zoomed-in region around the lensing ring. Moreover, as $\psi_0$ increases in this situation, the inner edge of the direct image moves inward, as indicated by the evolution of the first transfer function. In the limiting case of $\psi_0 = 90^\circ$, which corresponds to a geometrically spherical accretion flow, the contributing inclination angles span the full range from $\varphi_0 = 0$ to $180^\circ$. In this regime, the inner edge of the lensing ring coincides with the photon ring, while the outer edge merges smoothly into the direct image. As a result, the image exhibits a continuous and broadened flux profile, characterized by a single inner edge located near the critical impact parameter.

Note that for $\psi_0 > 40^\circ$, the observer-black hole axis penetrates the disk, so there is no strictly dark region with zero flux inside the critical impact parameter. Nevertheless, a relatively dim area remains visible, which still corresponds to the black hole shadow in a generalized sense. It is also worth noting that the maximum observed flux decreases as $\psi_0$ increases. This behavior results from our assumption of a fixed mass accretion rate $\dot{M}$ across all disk models. With increasing $\psi_0$, the disk becomes geometrically thicker, and the emitted radiation is distributed over a larger volume. As a consequence, the energy output per unit volume---and thus the peak intensity detected by the observer---becomes smaller.

Next, let us focus on the negative $z'$ axis in Fig.~\ref{fig:fig8}. In this direction, the disk can be effectively interpreted as a superposition of luminous segments with relatively small inclination angles, i.e., $\varphi_0 < 90^\circ$. For instance, in the case of $\psi_0 = 30^\circ$, the contributing segments correspond to inclination angles ranging from $\varphi_0 = 20^\circ$ to $80^\circ$. Consequently, the inner edge of the direct image lies within the critical impact parameter,  as indicated by the first transfer function. As a result, only the peaks associated with the lensing ring and the photon ring are present. Unlike the case on the positive $z'$ axis, there is no distinct gap between the lensing ring and the direct image regions; instead, the lensing ring exhibits a finite width even for small $\psi_0$.  Similarly, the inner edge of the lensing ring decreases with increasing $\psi_0$, gradually approaching the photon ring.

Now, we focus on the $y'$ axis in Fig.~\ref{fig:fig9}. The flux observed from the negative $y'$ axis is slightly brighter than that from the positive $y'$ axis. This arises from the Doppler effect due to the presence of a slow rotation in the disks. For purely infalling flows (with the emission profile depending only on $r$), the observed flux should be strictly symmetric about $y' = 0$, since both sides can be understood as composed of segments with the same ranges of inclination angles. However, in this direction, the range of contributing inclination angles does not increase linearly with $\psi_0$, unlike the case along the $z'$ axis. For instance, once $\psi_0 > 40^\circ$, the disk covers the observer-black hole axis, and the range of contributing inclination angles for the segments spans the full interval $0 < \varphi_0 \leq 180^\circ$. This explains why in Fig.~\ref{fig:fig9} the flux profiles for $\psi_0 = 60^\circ$ and $\psi_0 = 70^\circ$ share the same envelope as that for $\psi_0 = 90^\circ$.  Again, the averaged flux of the lensing ring becomes smaller as $\psi_0$ increases, due to the lower radiation energy density per unit volume in thicker disks.

\begin{figure}[htbp]
\centering
\includegraphics[width=.85\textwidth]{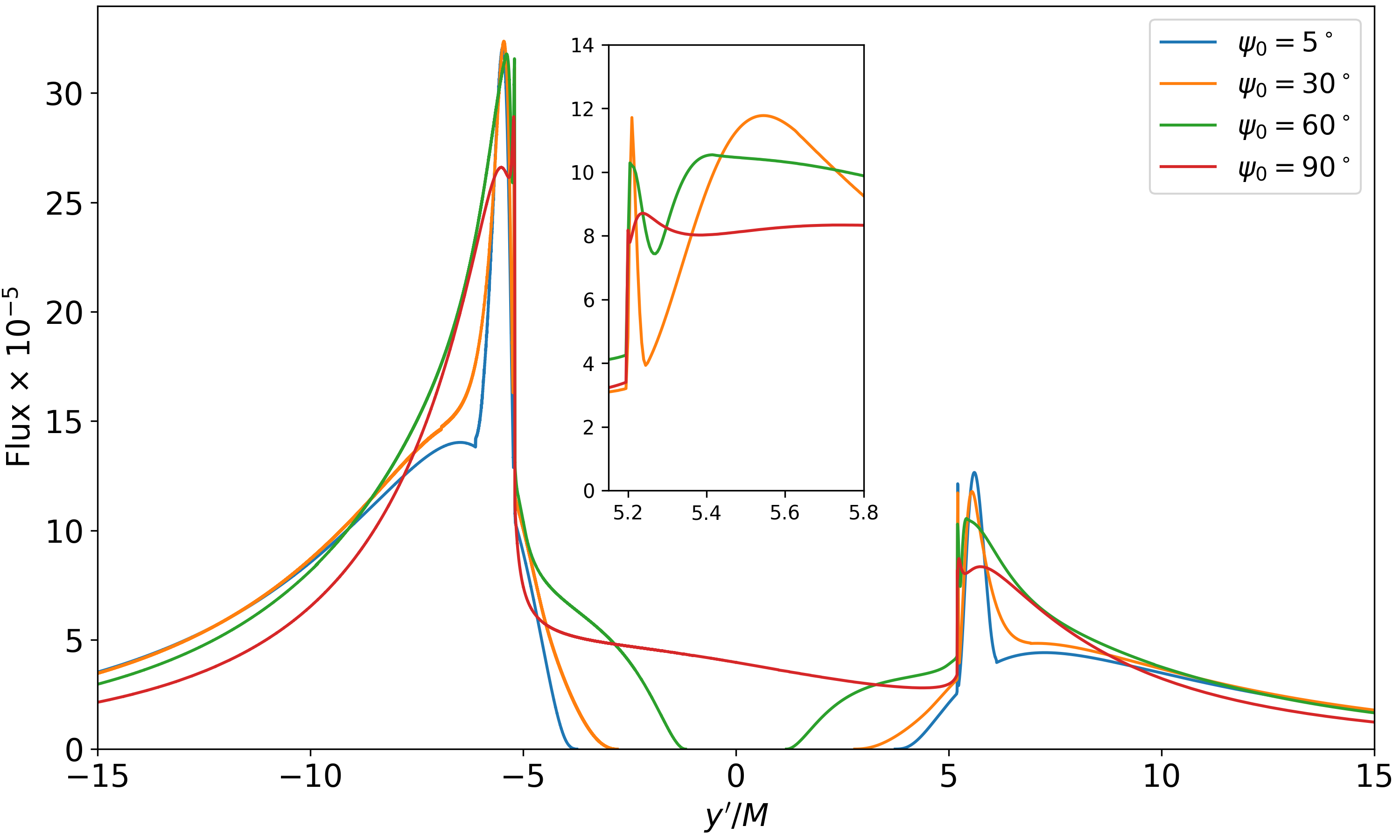}
\caption{Observed flux along the $y'$ axis of the image plane for optically thin disks with an inclination angle of $\theta_0 = 20^\circ$, an inner radius of $r_{\rm in} = 3M$, and varying half-opening angles $\psi_0$. All disks assume $\kappa_{\rm ff} = 0.1$ and $\kappa_{\rm K} = 0.9$. \label{fig:fig10}}
\end{figure}

The flux profile along the $y'$ axis of the image plane for optically thin disks with an inclination angle of $\theta_0 = 20^\circ$ and an inner radius of $r_{\rm in} = 3M$ is shown in Fig.~\ref{fig:fig10}. In this configuration, the disk is assumed to rotate with a nearly Keplerian angular velocity, characterized by $\Omega_{\rm K} = 0.9$. A pronounced asymmetry is observed between the negative and positive $y'$ directions, with significantly higher flux on the approaching side. This difference arises from the strong Doppler effect due to the high orbital speed of the emitting material. Another notable difference lies in the flux distribution inside the critical impact parameter when compared to Fig.~\ref{fig:fig9}. For the case of $\theta_0 = 20^\circ$, the observer--black hole axis penetrates the disk only when $\psi_0 > 70^\circ$. As a result, a region of vanishing flux persists for $\psi_0 < 70^\circ$. It is important to emphasize that, in principle, the flux is expected to diverge at the critical impact parameter when the inner edge of the emitting region approaches the photon ring. This theoretical divergence is regulated in our numerical simulations due to the finite step size in ray-tracing. Nevertheless, a sharp cusp or pronounced local maximum consistently emerges near the critical impact parameter, clearly indicating the presence of the photon ring.

\begin{figure}[htbp]
    \centering
    \renewcommand{\thesubfigure}{} % 取消 subcaption 编号
    \setlength{\tabcolsep}{4pt} % 图间距微调
    \begin{tabular}{c *{3}{>{\centering\arraybackslash}m{0.28\textwidth}}}
        % Top header row
        & $r_{\mathrm{in}}=3$ & $r_{\mathrm{in}}=4$ & $r_{\mathrm{in}}=6$ \\

        % psi = 5
        \raisebox{-0.3\height}{\rotatebox{90}{$\psi_0=5^\circ$}} &
        \includegraphics[width=\linewidth]{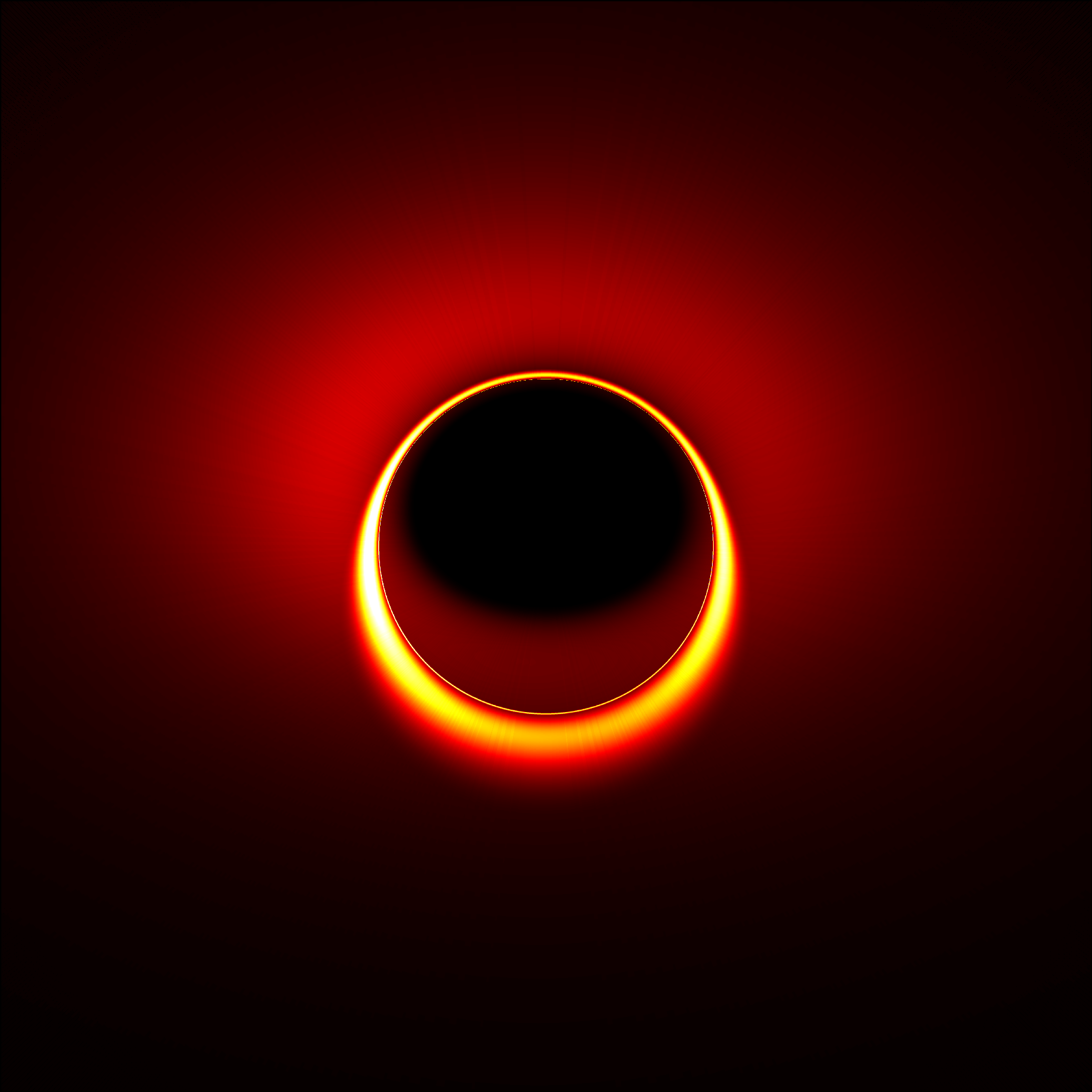} &
        \includegraphics[width=\linewidth]{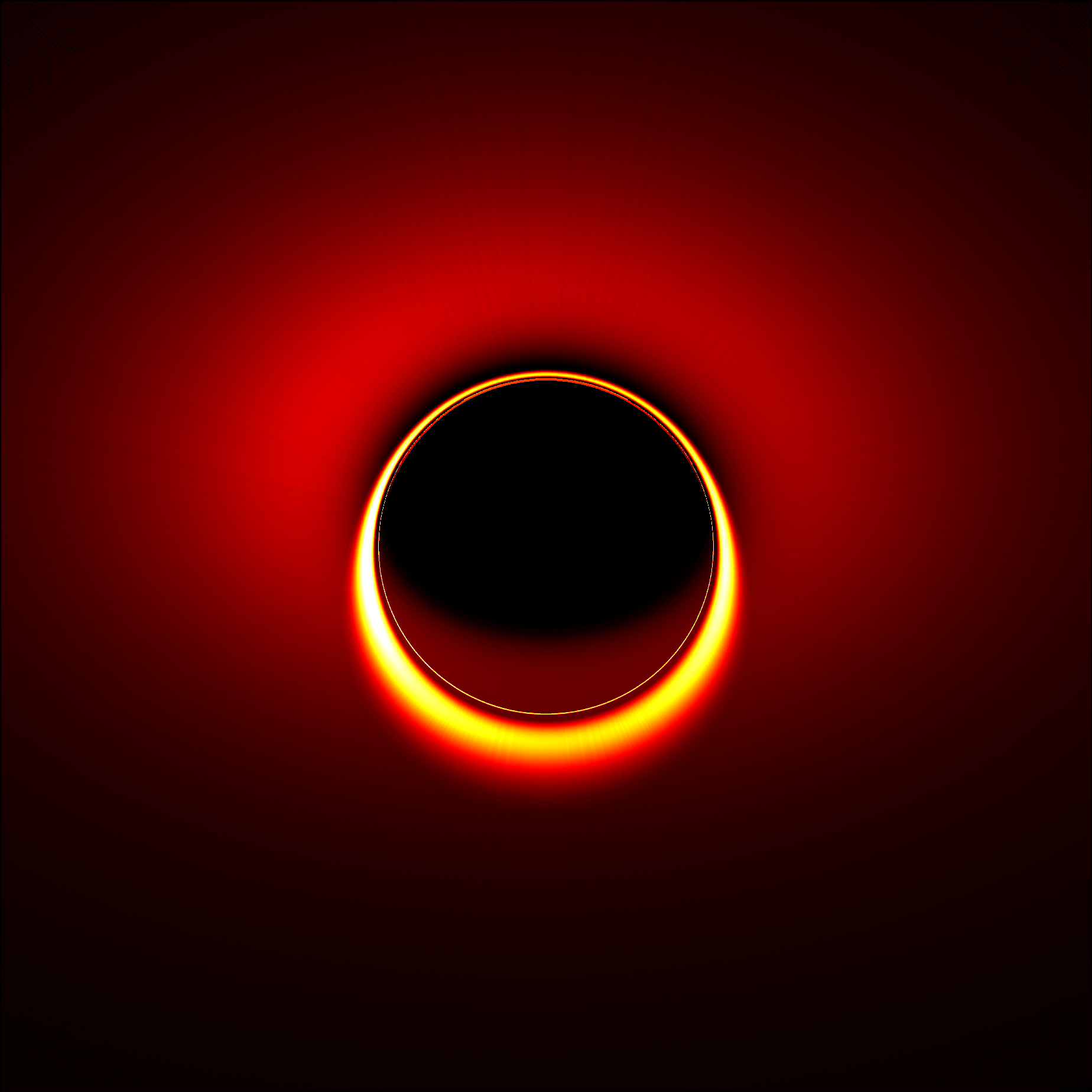} &
        \includegraphics[width=\linewidth]{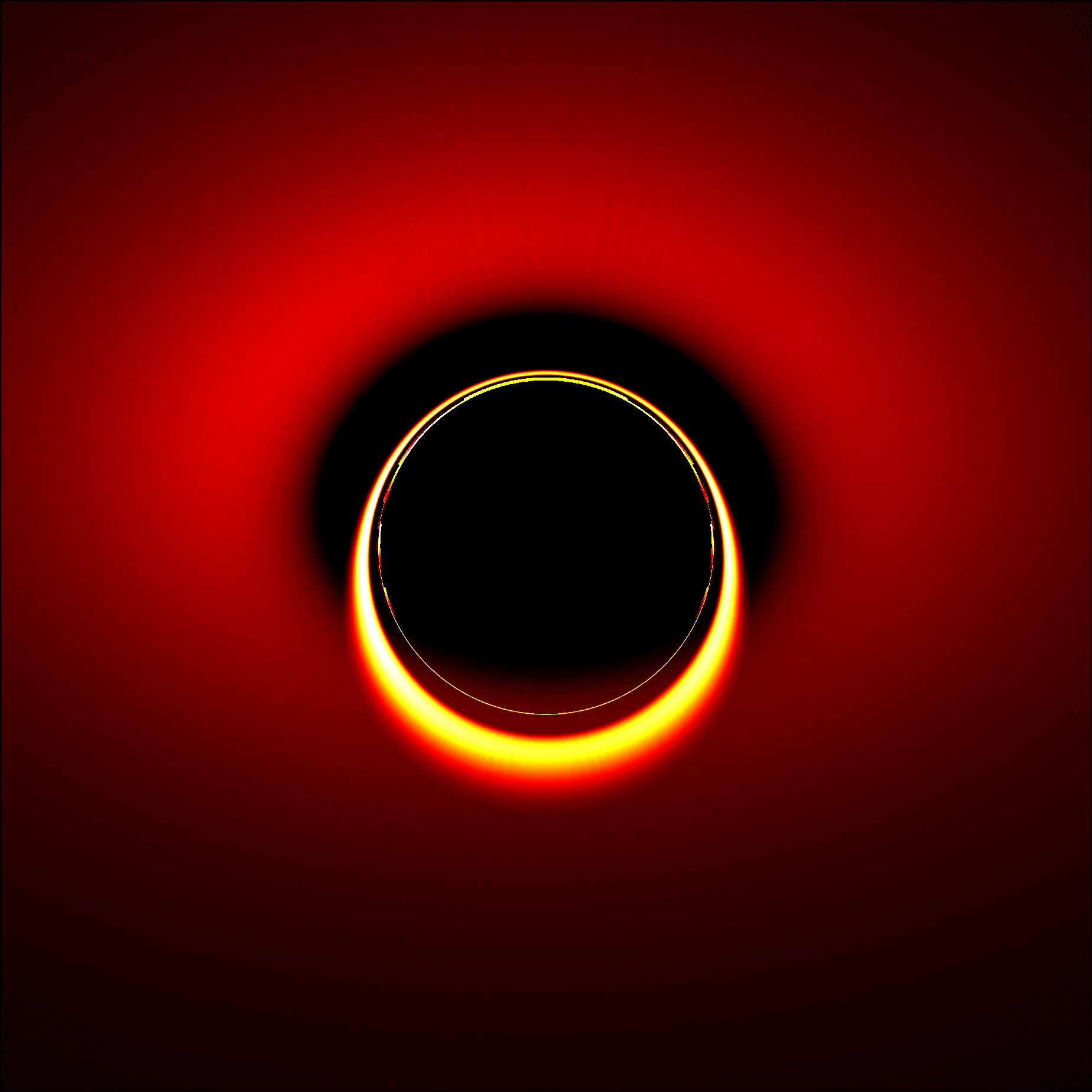} \\

        % psi = 30
        \raisebox{-0.3\height}{\rotatebox{90}{$\psi_0=30^\circ$}} &
        \includegraphics[width=\linewidth]{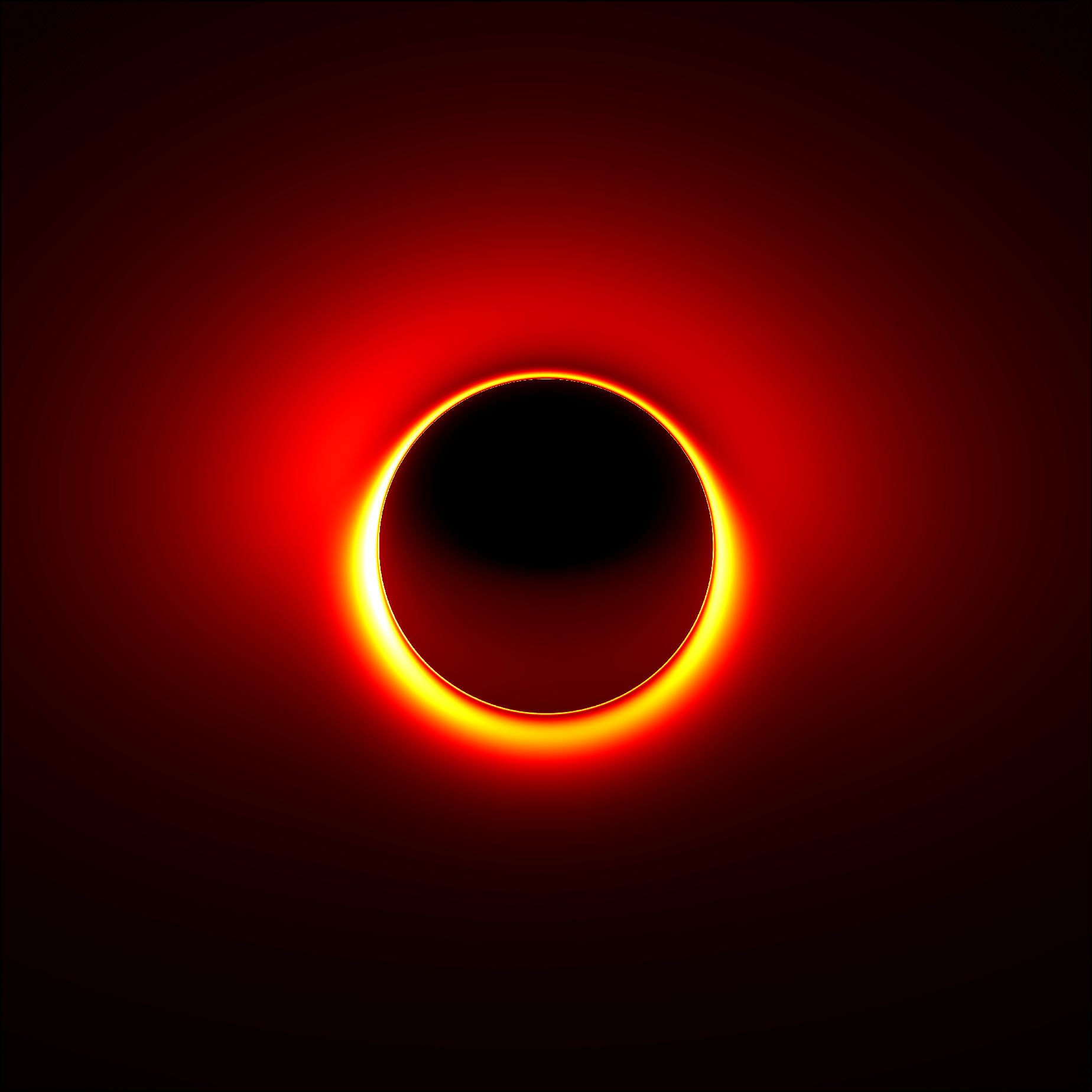} &
        \includegraphics[width=\linewidth]{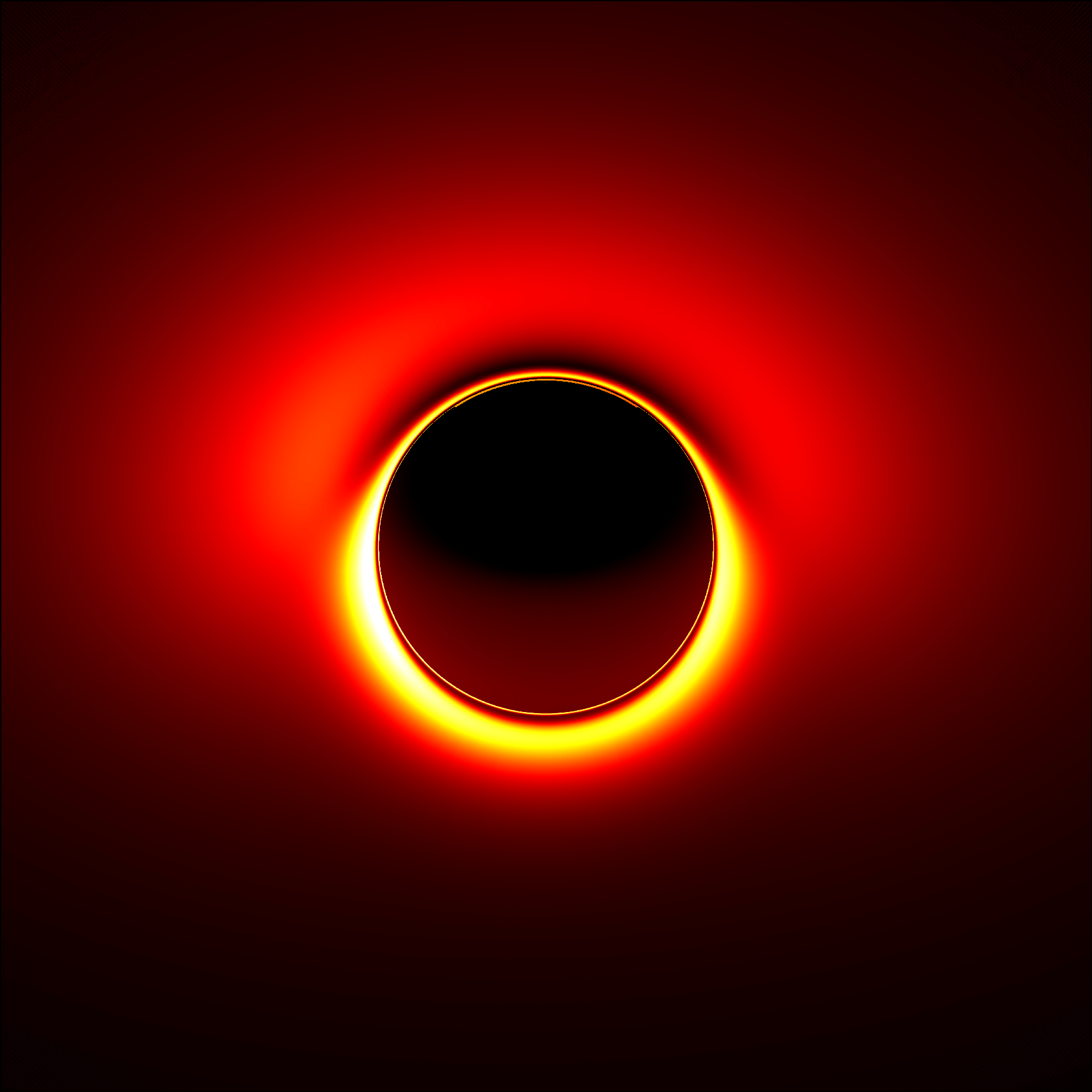} &
        \includegraphics[width=\linewidth]{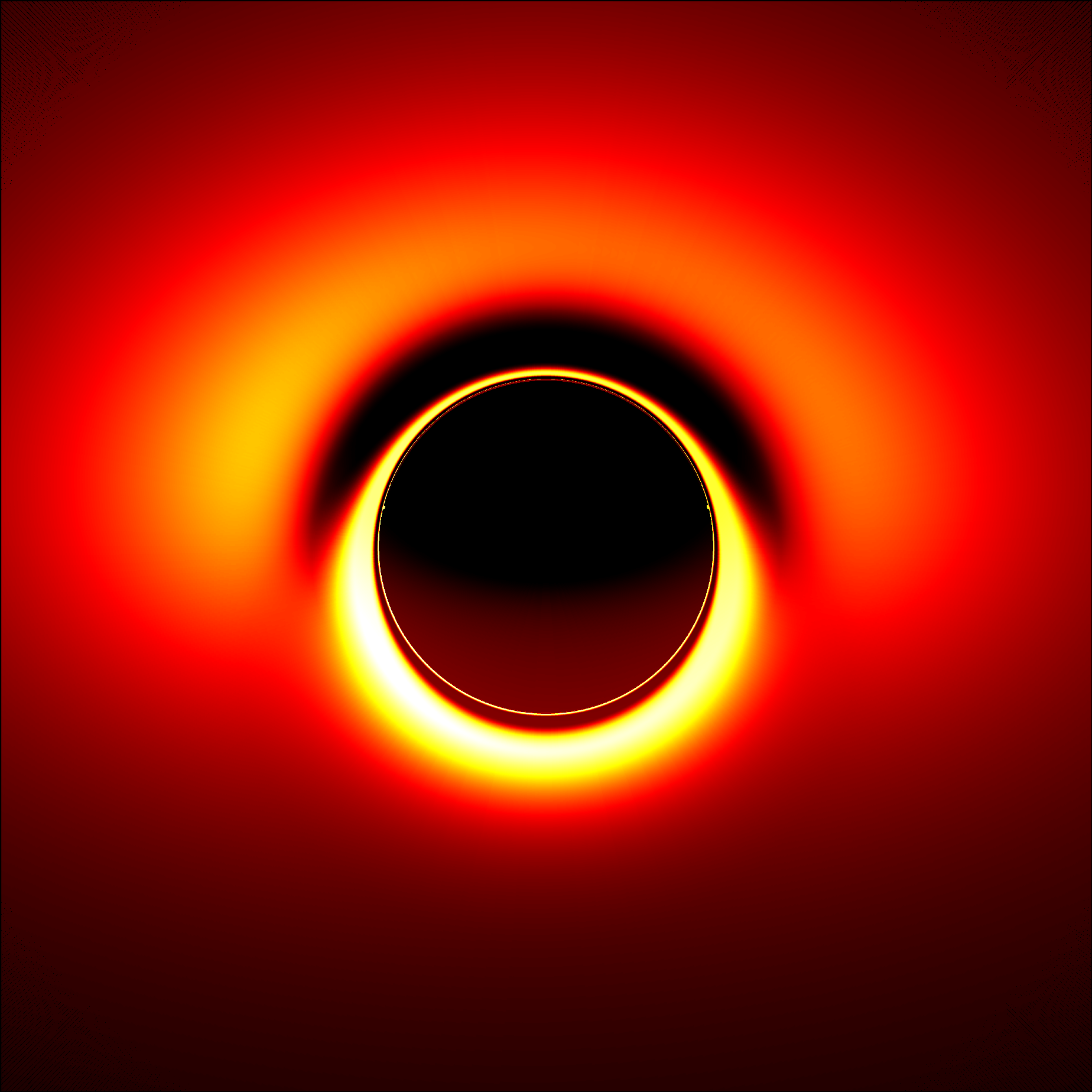} \\

        % psi = 60
        \raisebox{-0.3\height}{\rotatebox{90}{$\psi_0=60^\circ$}} &
        \includegraphics[width=\linewidth]{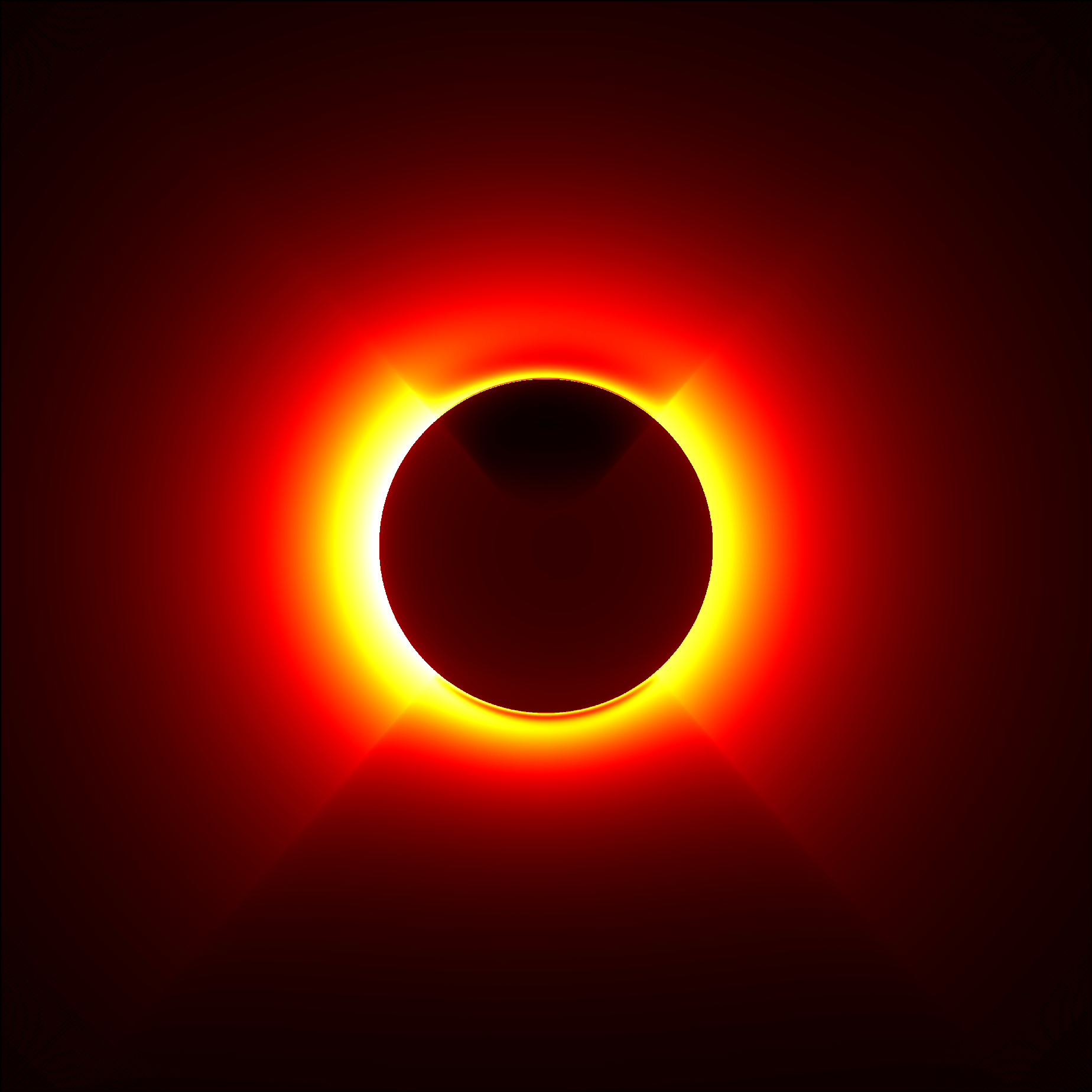} &
        \includegraphics[width=\linewidth]{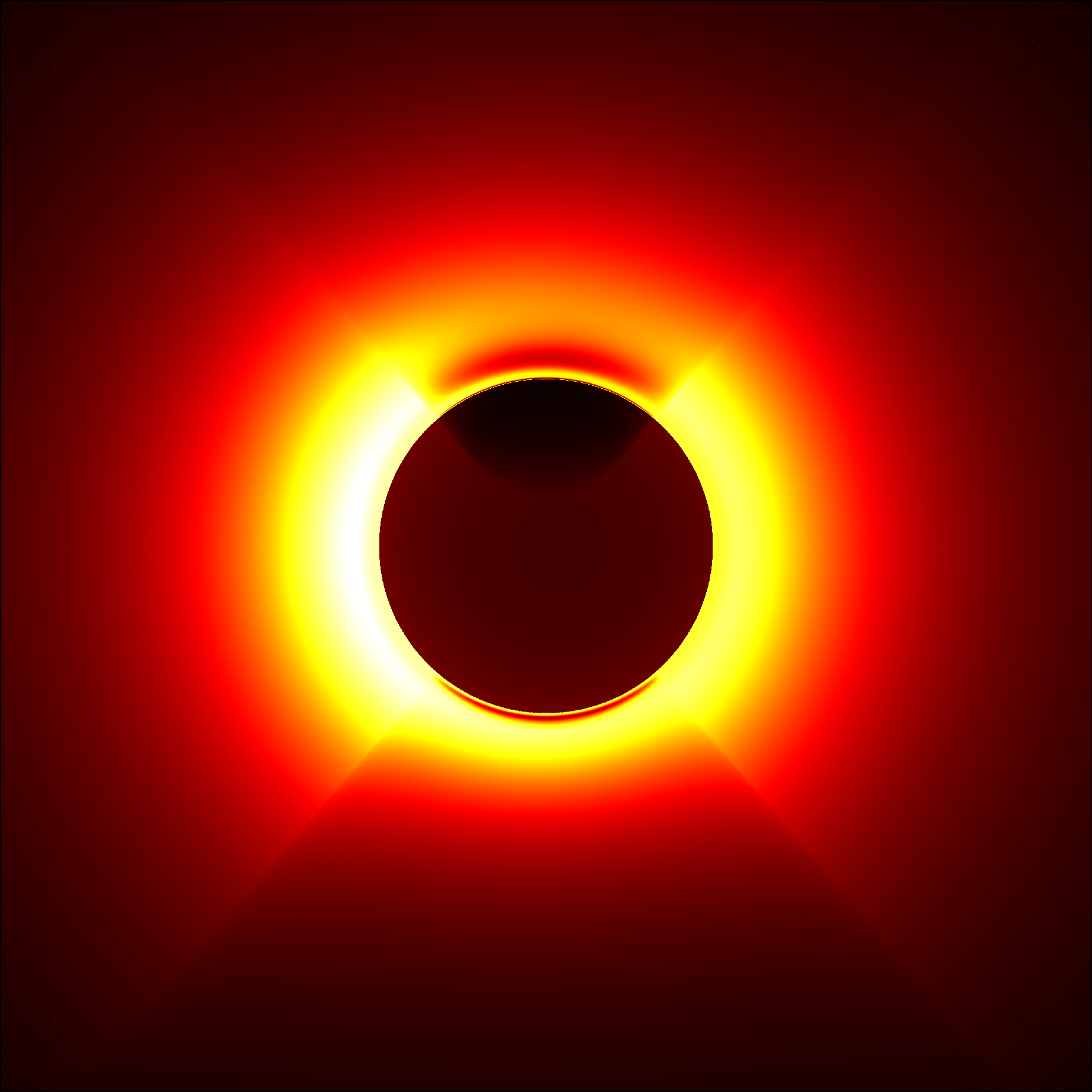} &
        \includegraphics[width=\linewidth]{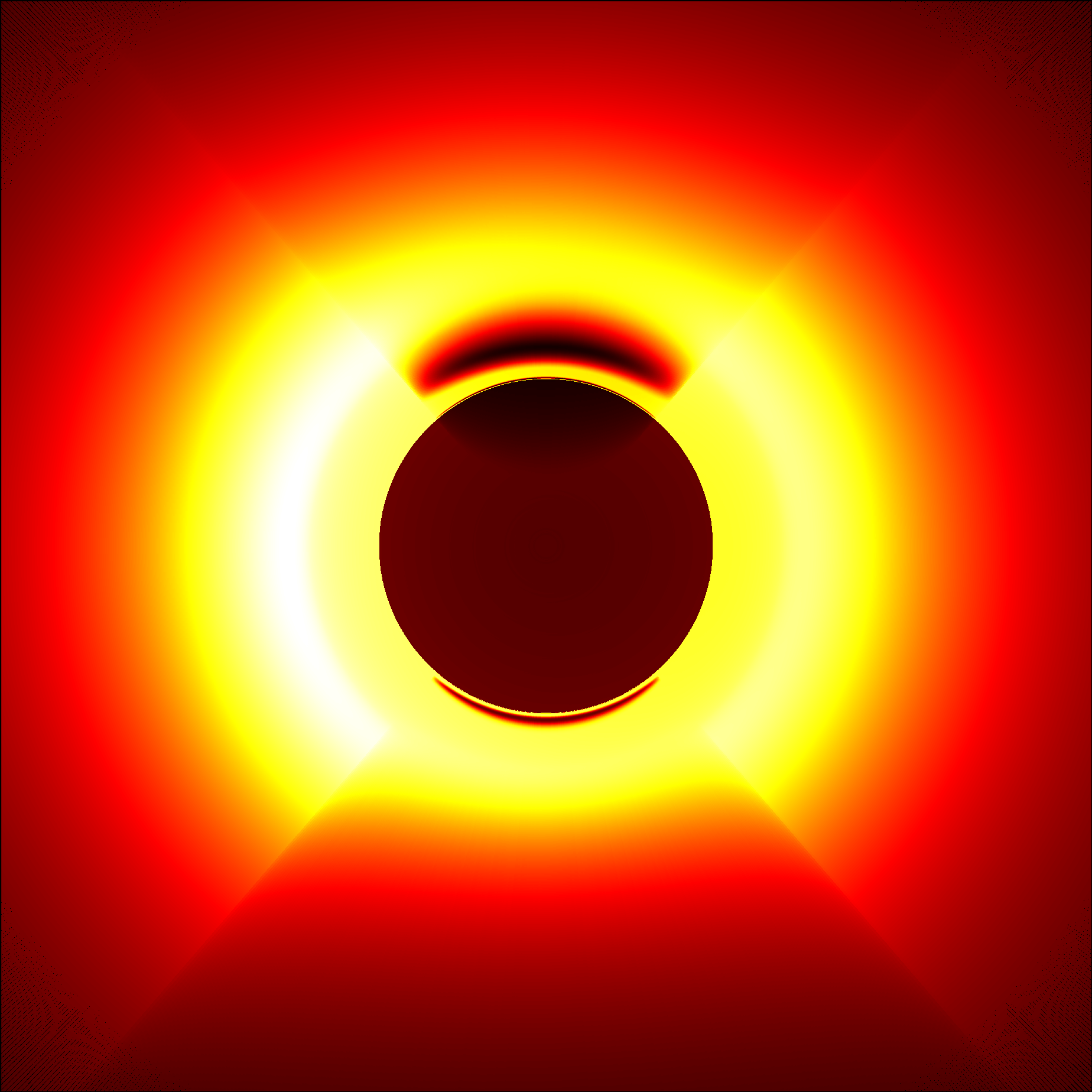} \\

        % psi = 90
        \raisebox{-0.3\height}{\rotatebox{90}{$\psi_0=90^\circ$}} &
        \includegraphics[width=\linewidth]{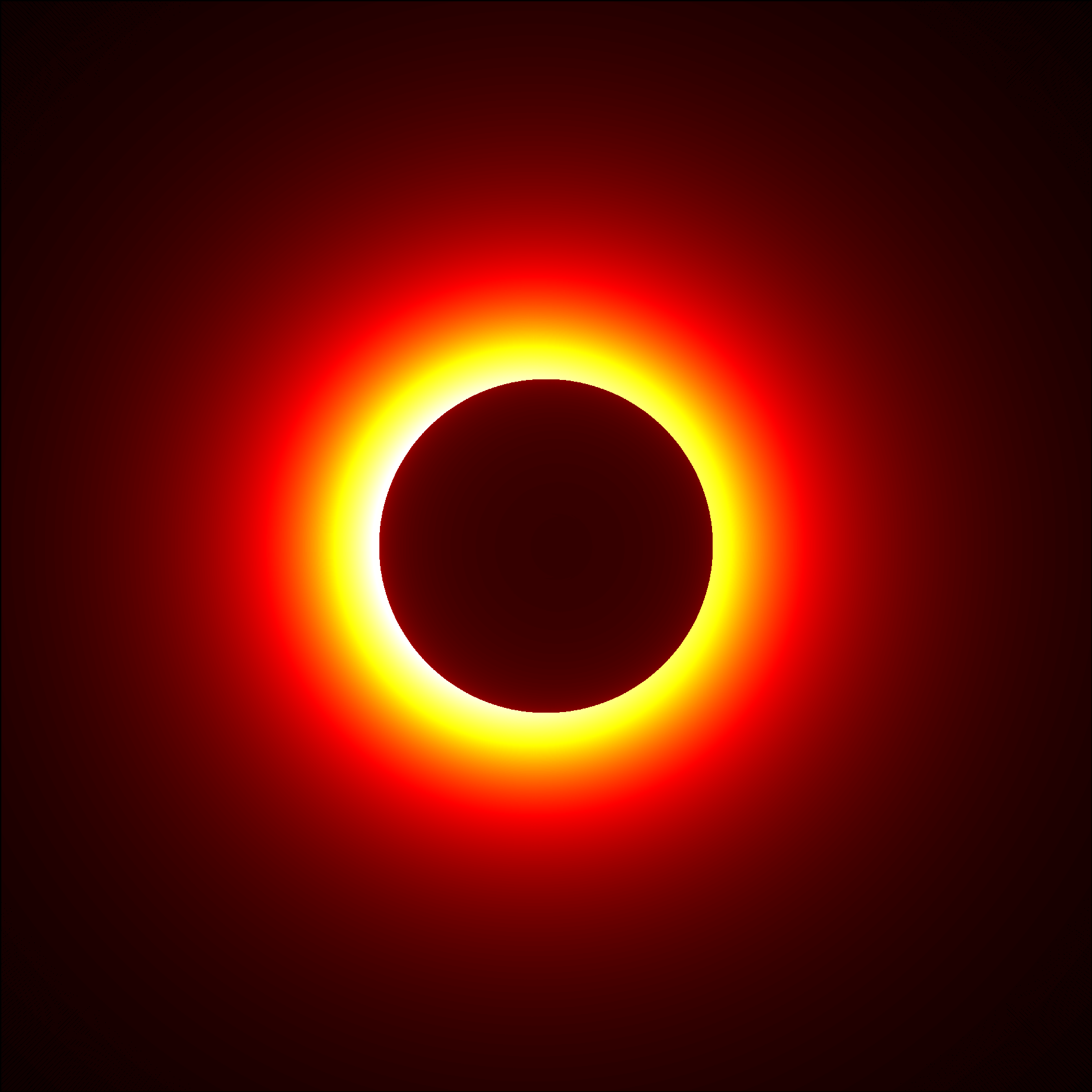} &
        \includegraphics[width=\linewidth]{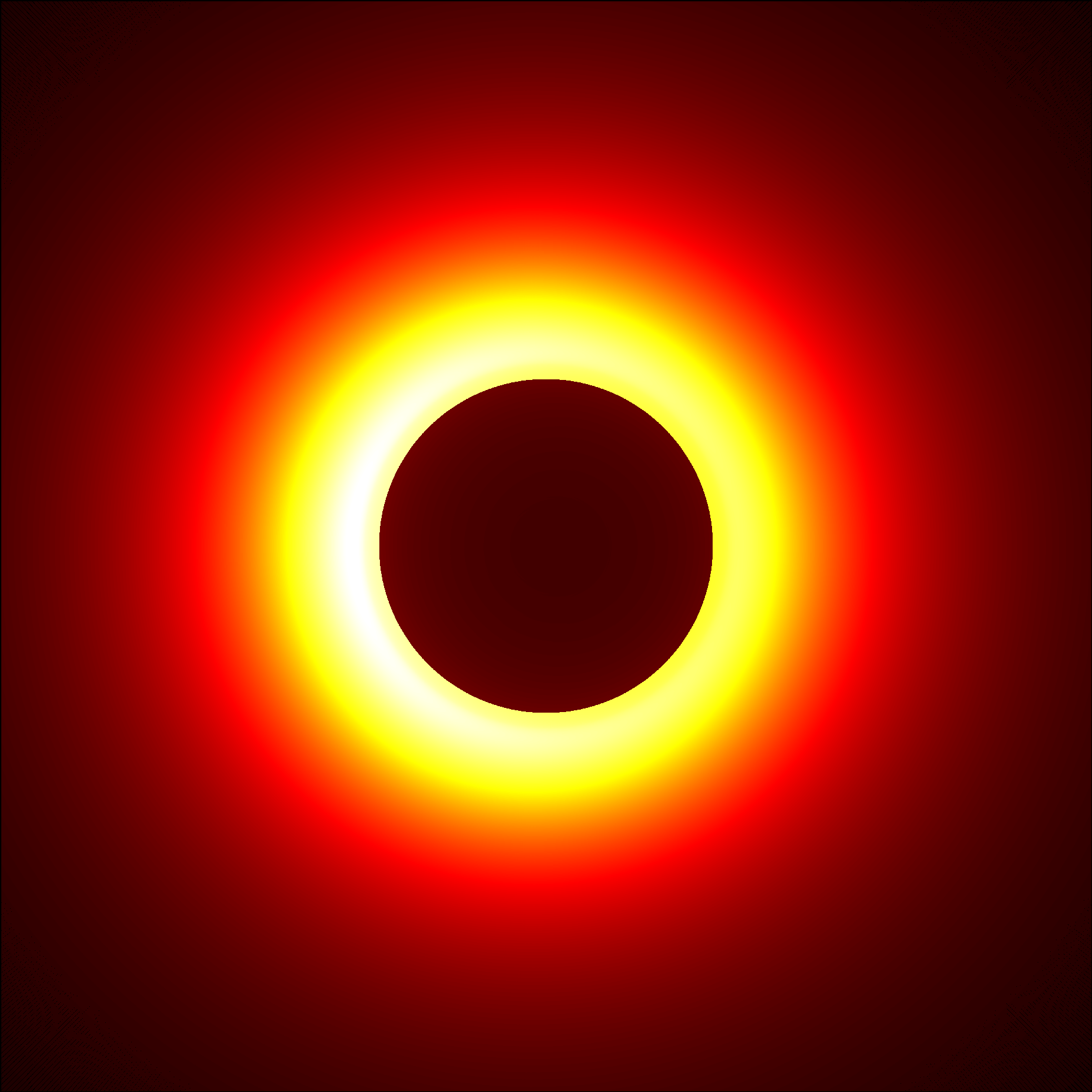} &
        \includegraphics[width=\linewidth]{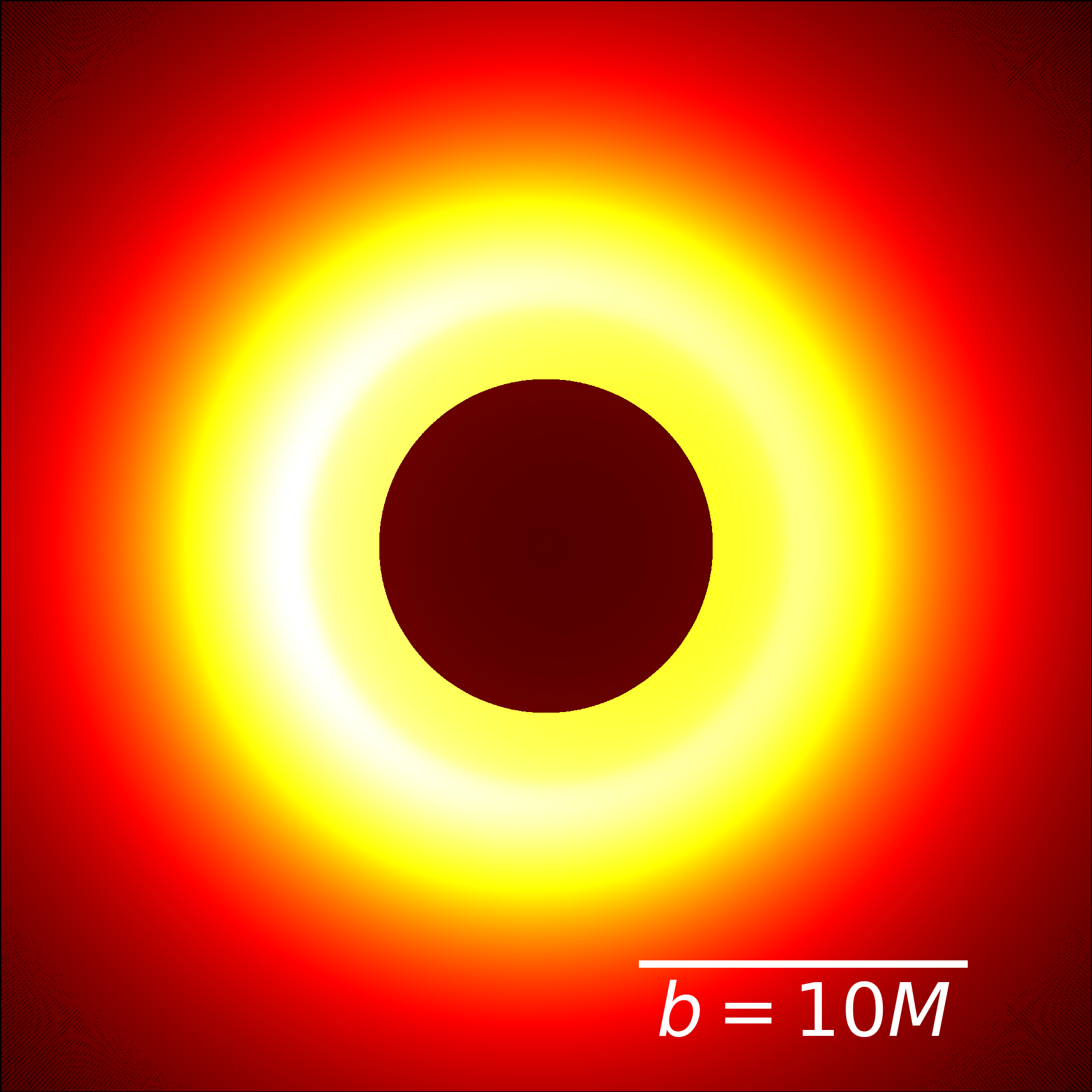} \\
    \end{tabular} 
\noindent \hspace*{0.36cm} \includegraphics[width=0.98\textwidth]{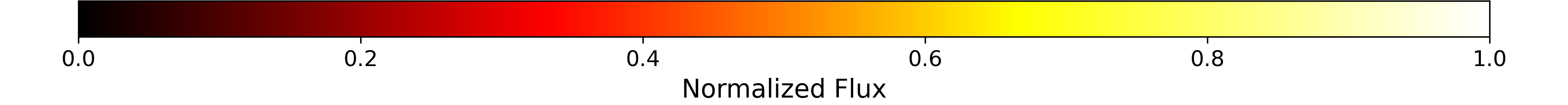}
    \caption{Observed flux for optically thin disks with an inclination angle of $\theta_0 = 50^\circ$,  with varying inner radius $r_{\rm in}$ and half-opening angle $\psi_0$. Rows correspond to different values of $\psi_0$, while columns represent different values of $r_{\rm in}$.  All disks assume $\kappa_{\rm ff} = 0.5$ and $\kappa_{\rm K} = 0.1$. The flux distributions are normalized by their respective maximum values. }
    \label{fig:fig11}
\end{figure}

The observed flux along other directions, i.e., at different values of $\alpha$, exhibits behavior intermediate between those discussed along the $z'$ and $y'$ axes, and the underlying analysis remains qualitatively similar. In Fig.~\ref{fig:fig11}, we present the observed images for optically thin disks with an inclination angle of $\theta_0 = 50^\circ$, illustrating the dependence on both the inner radius $r_{\rm in}$ and the half-opening angle $\psi_0$. {In these images, the clearly visible bright ring corresponds to the lensing ring. Whether an individual photon ring can be distinguished depends on the specific configuration. Several qualitative trends can be identified: First, for a fixed inner radius, increasing the half-opening angle causes the inner edge of the lensing ring to gradually approach the photon ring, while its outer edge progressively merges with the region of direct emission.}

{Second, for a fixed and relatively small half-opening angle, increasing the inner radius makes the photon ring more prominent. This occurs because the inner boundary of the lensing ring shifts outward, moving further away from the photon ring. Third, for larger half-opening angles, a relatively dim region consistently appears, enclosed by the photon ring. These features can be well understood in terms of the general behavior of the transfer functions discussed earlier. }

\begin{figure}[htbp]
    \centering
    \renewcommand{\thesubfigure}{} % 取消 subcaption 编号
    \setlength{\tabcolsep}{4pt} % 图间距微调
    \begin{tabular}{c *{3}{>{\centering\arraybackslash}m{0.28\textwidth}}}
        % Top header row
        & $r_{\mathrm{in}}=3$ & $r_{\mathrm{in}}=4$ & $r_{\mathrm{in}}=6$ \\

        % psi = 5
        \raisebox{-0.3\height}{\rotatebox{90}{$\psi_0=5^\circ$}} &
        \includegraphics[width=\linewidth]{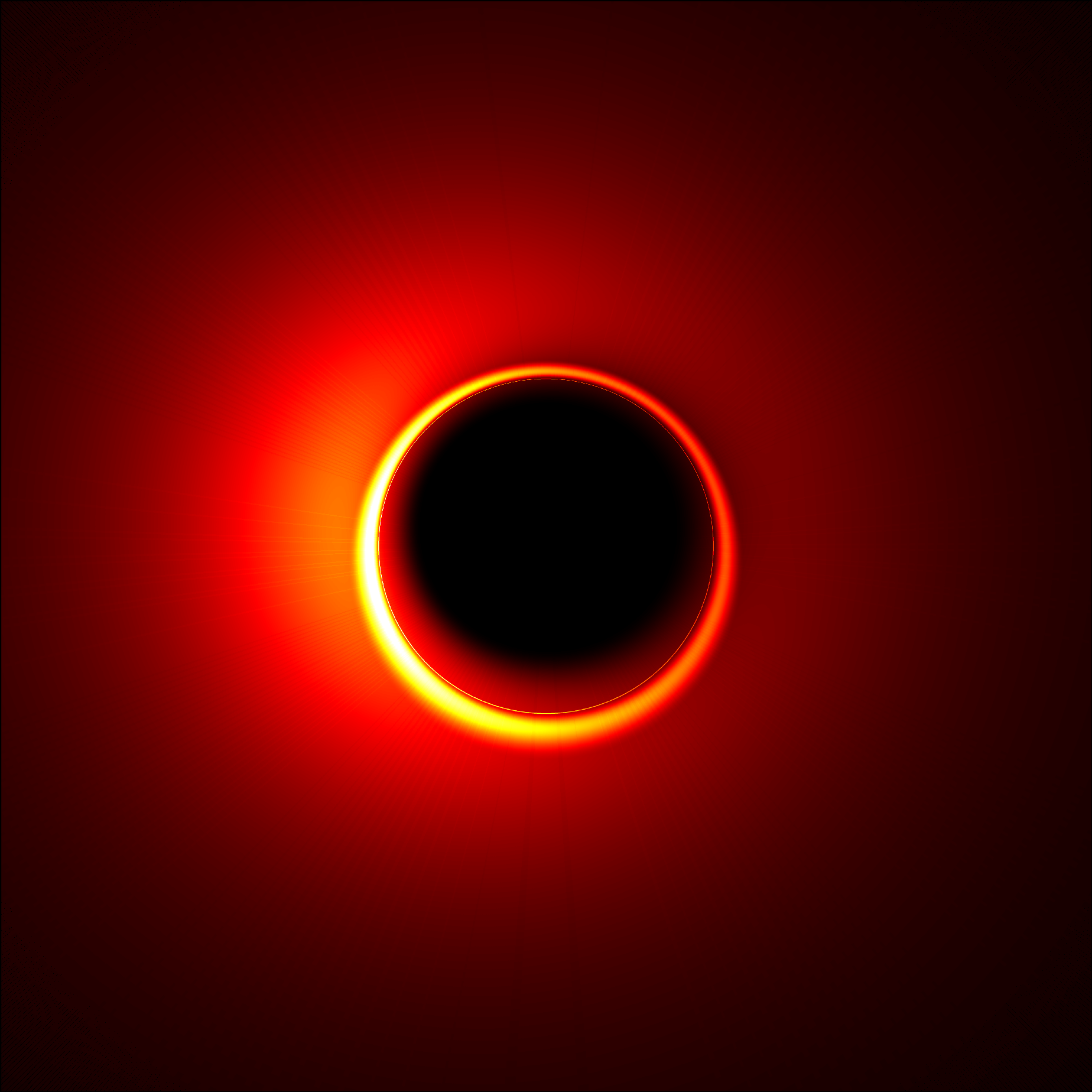} &
        \includegraphics[width=\linewidth]{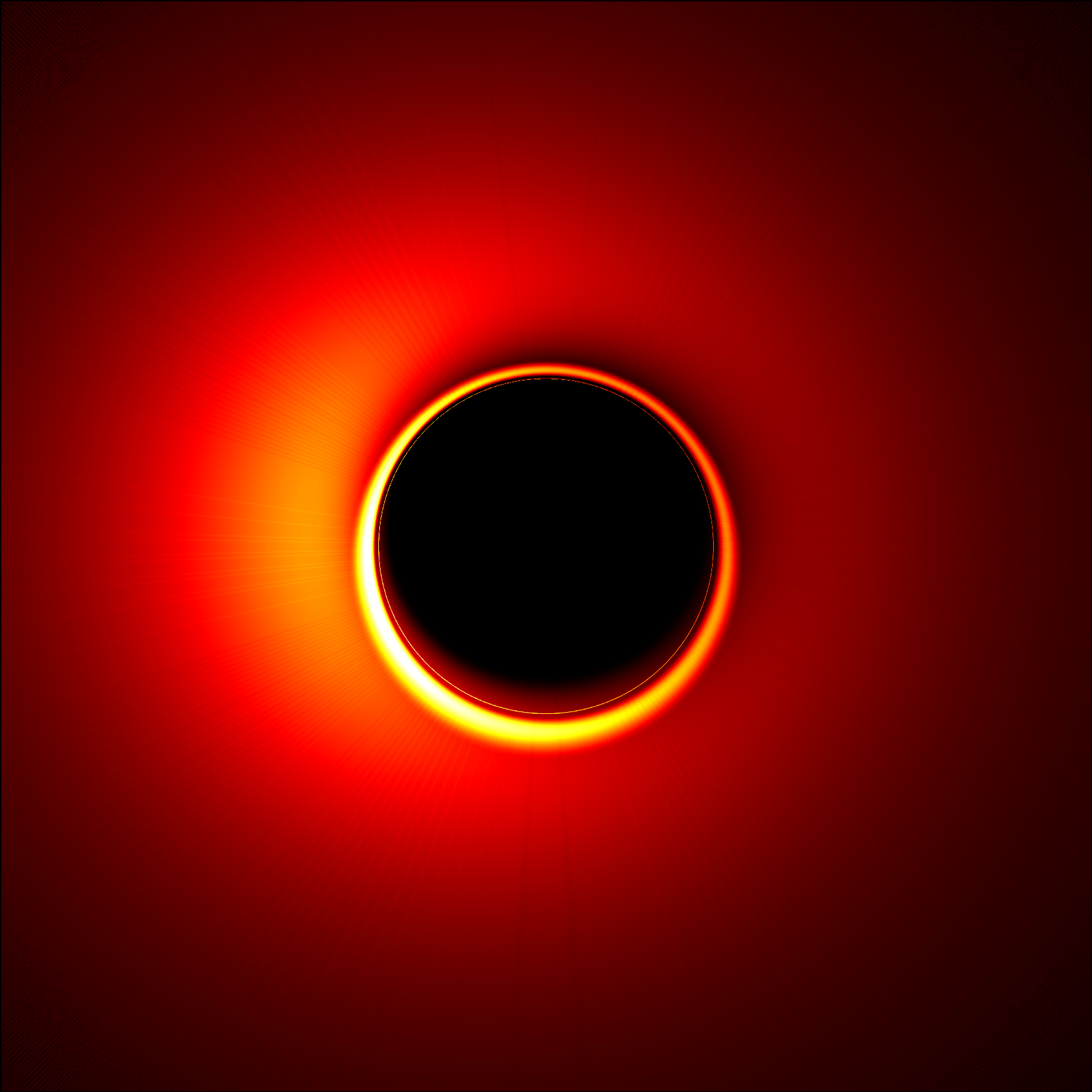} &
        \includegraphics[width=\linewidth]{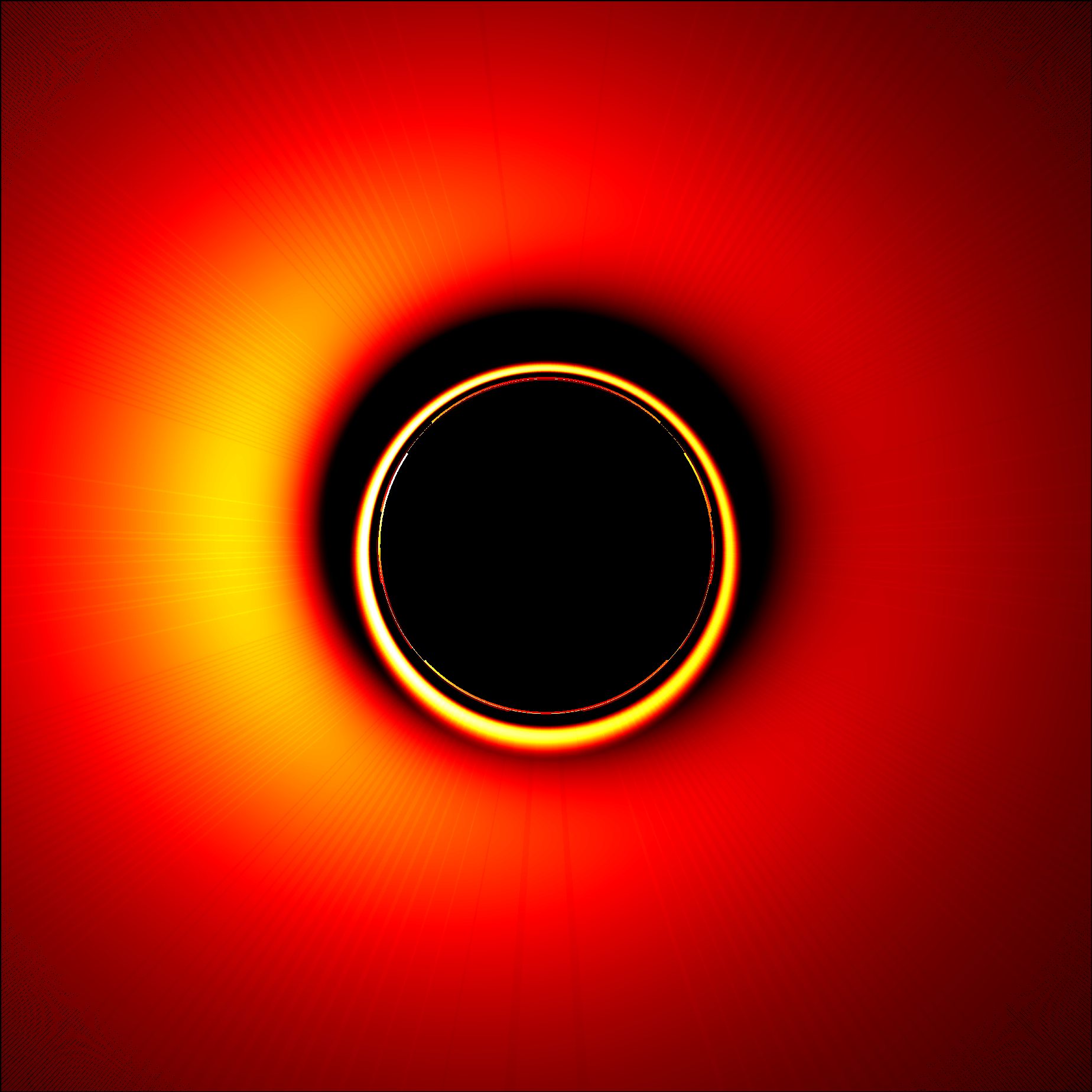} \\

        % psi = 30
        \raisebox{-0.3\height}{\rotatebox{90}{$\psi_0=30^\circ$}} &
        \includegraphics[width=\linewidth]{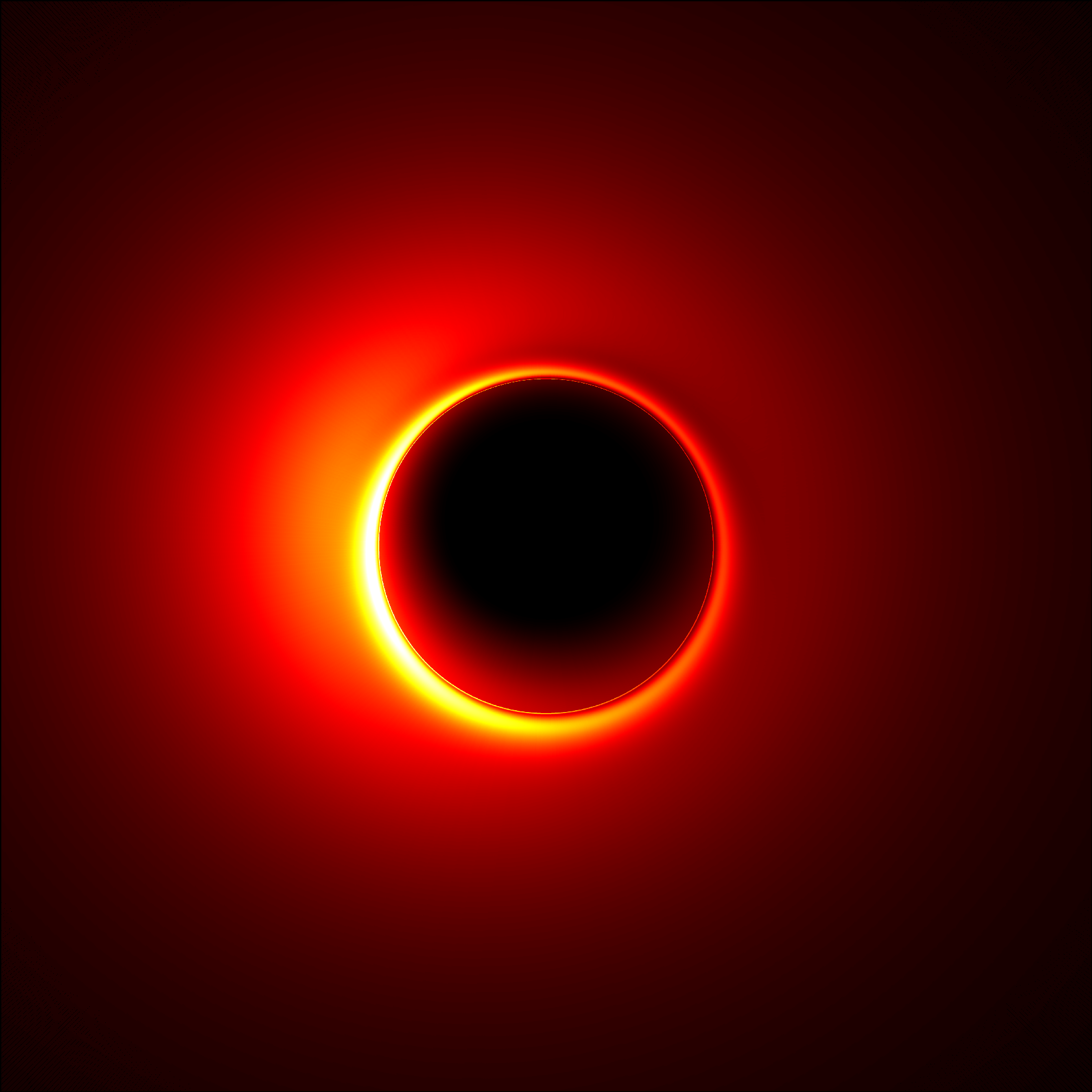} &
        \includegraphics[width=\linewidth]{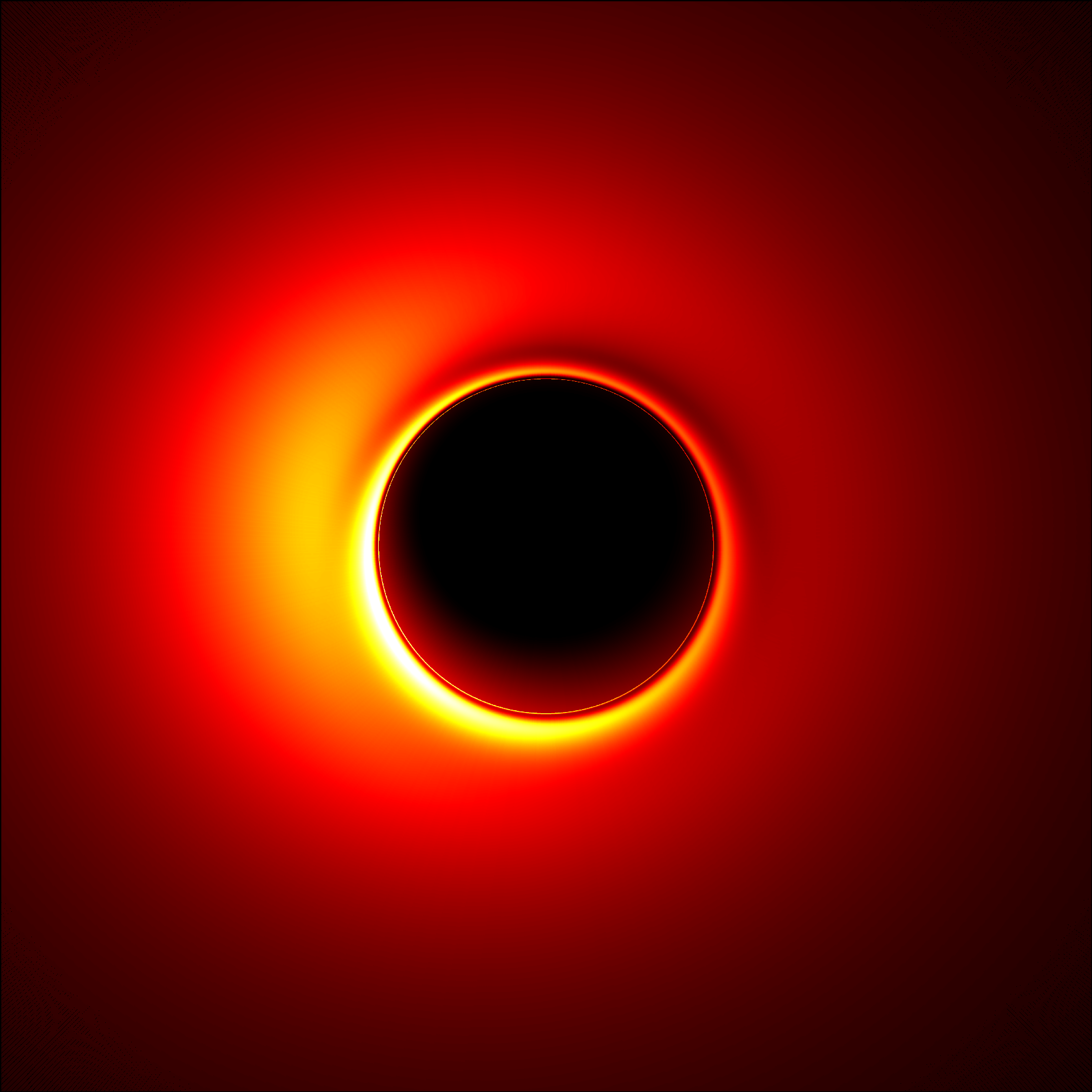} &
        \includegraphics[width=\linewidth]{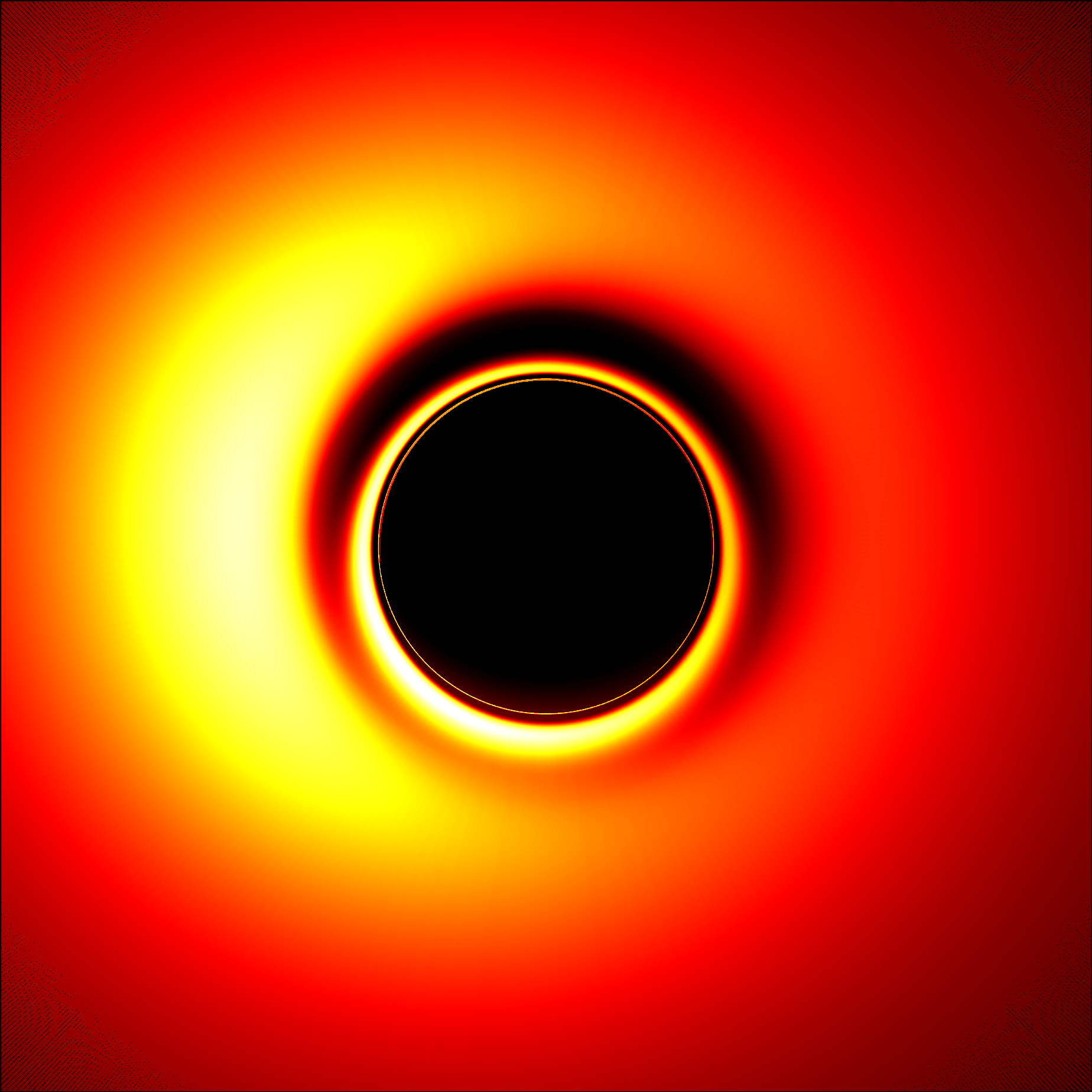} \\

        % psi = 60
        \raisebox{-0.3\height}{\rotatebox{90}{$\psi_0=60^\circ$}} &
        \includegraphics[width=\linewidth]{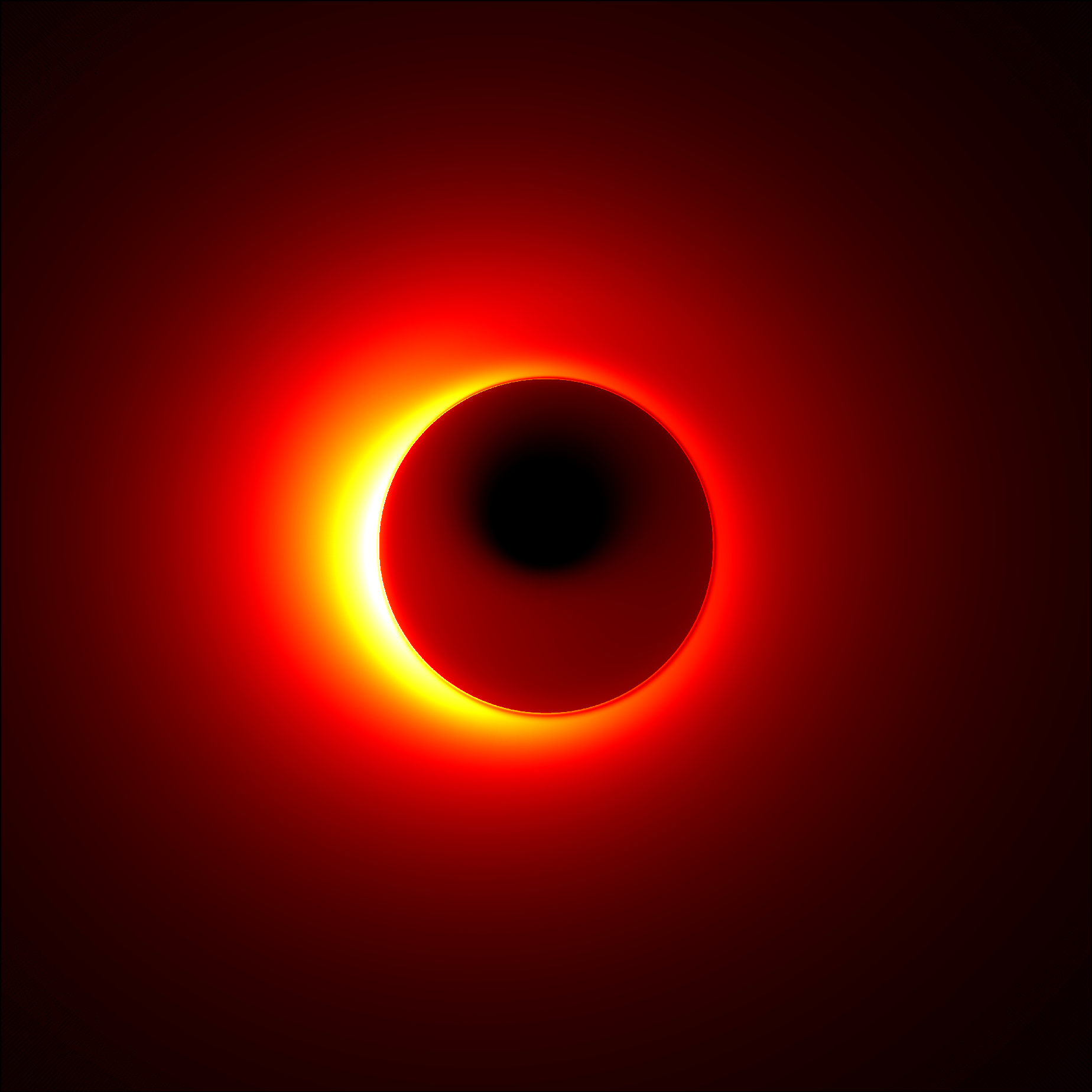} &
        \includegraphics[width=\linewidth]{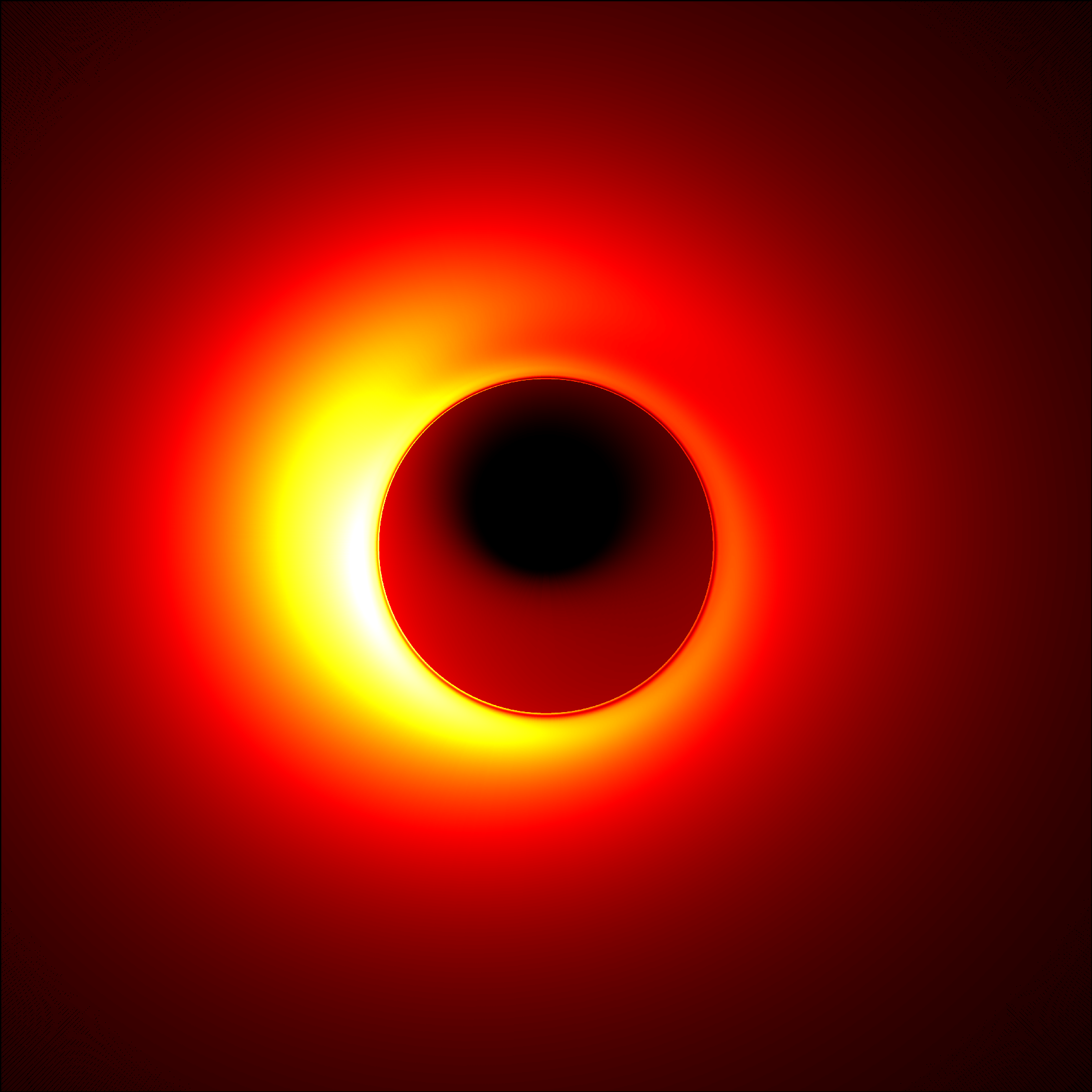} &
        \includegraphics[width=\linewidth]{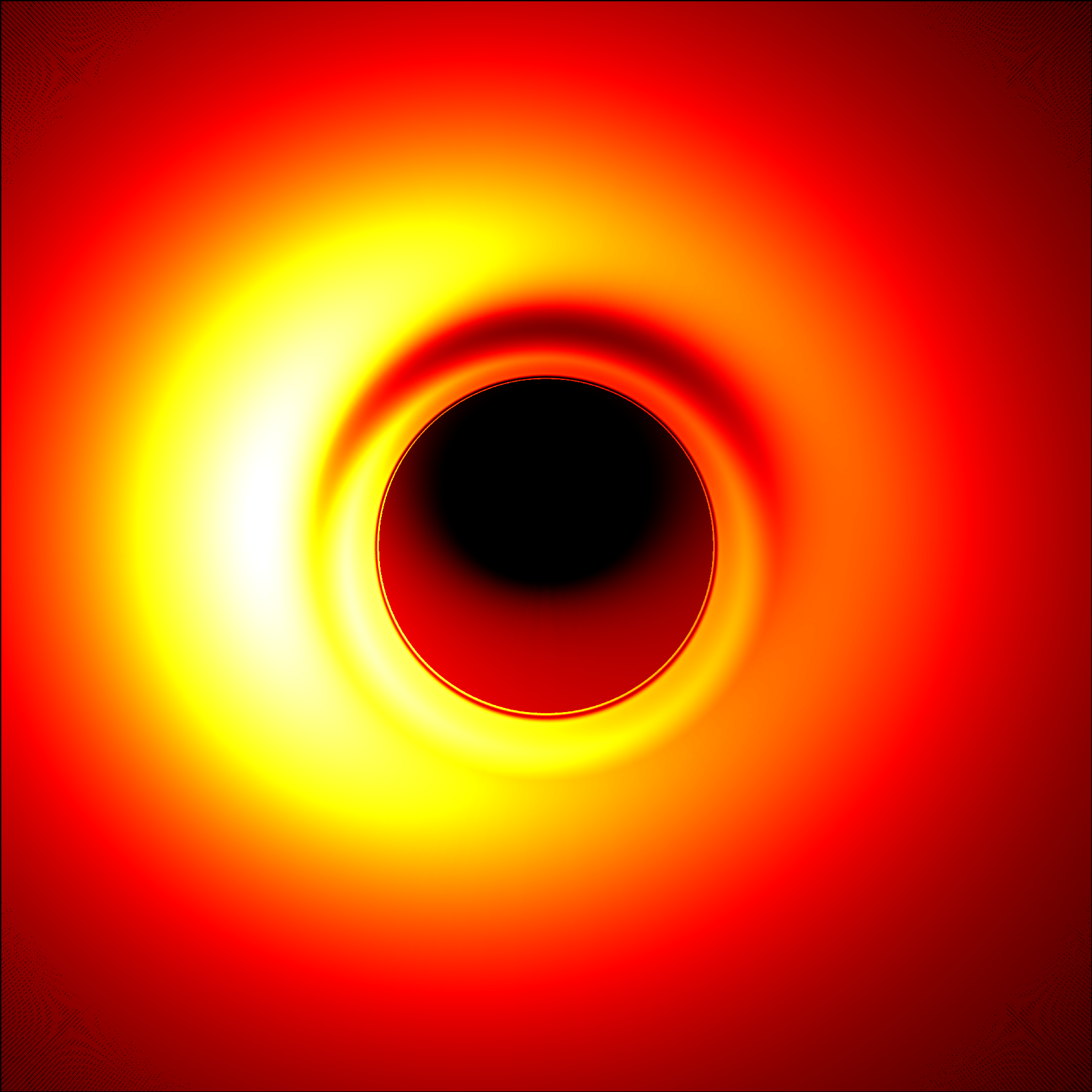} \\
        % psi = 90
        \raisebox{-0.3\height}{\rotatebox{90}{$\psi_0=90^\circ$}} &
        \includegraphics[width=\linewidth]{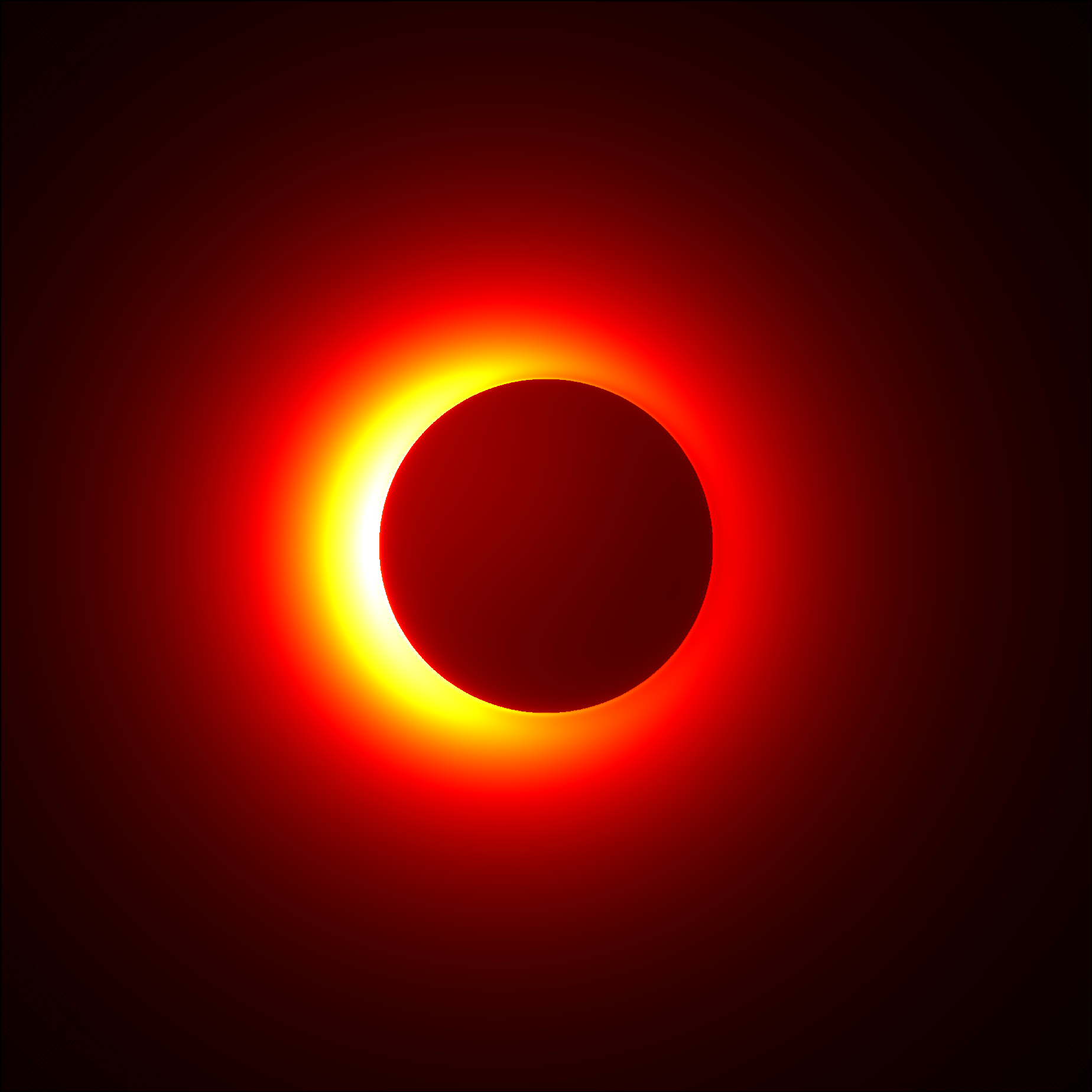} &
        \includegraphics[width=\linewidth]{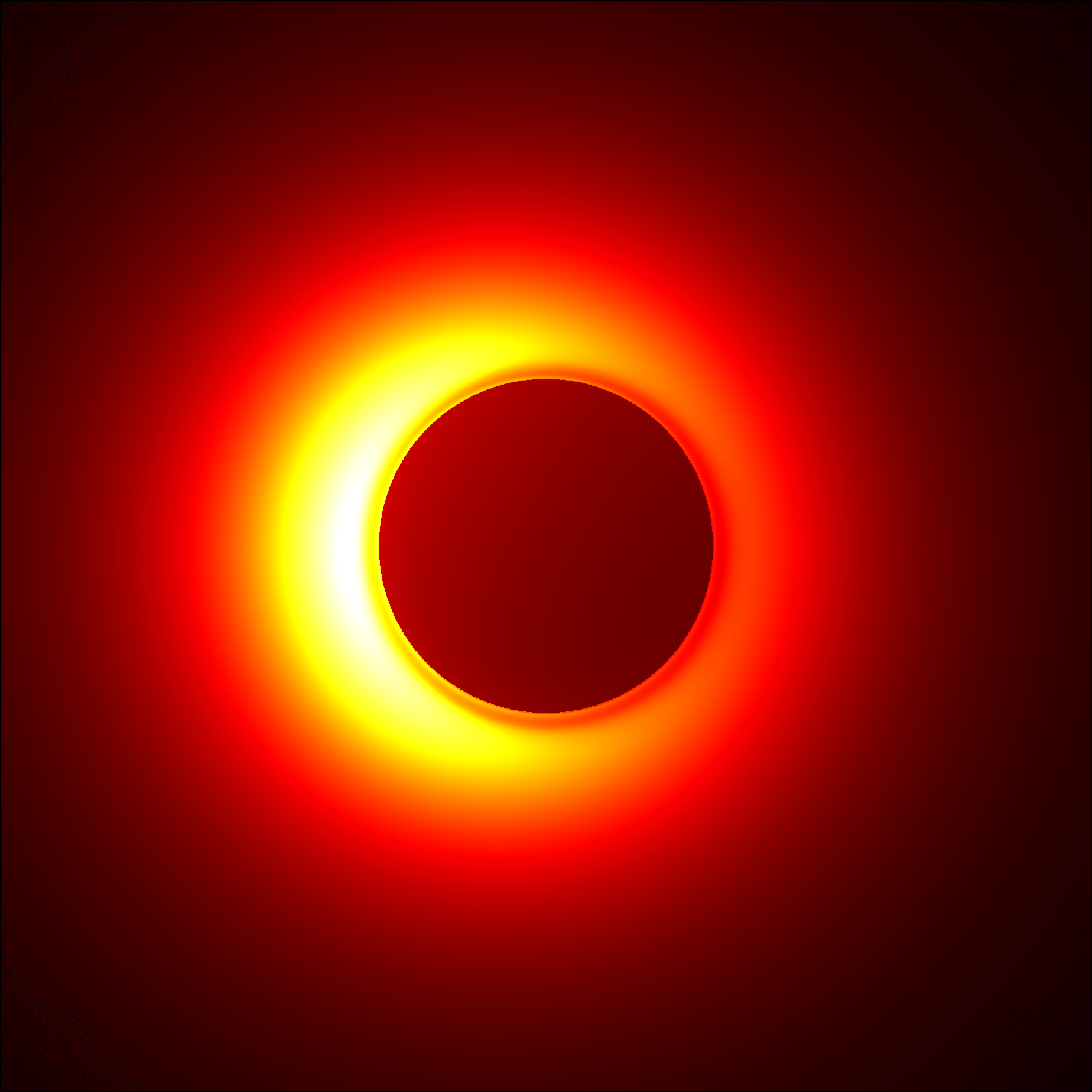} &
        \includegraphics[width=\linewidth]{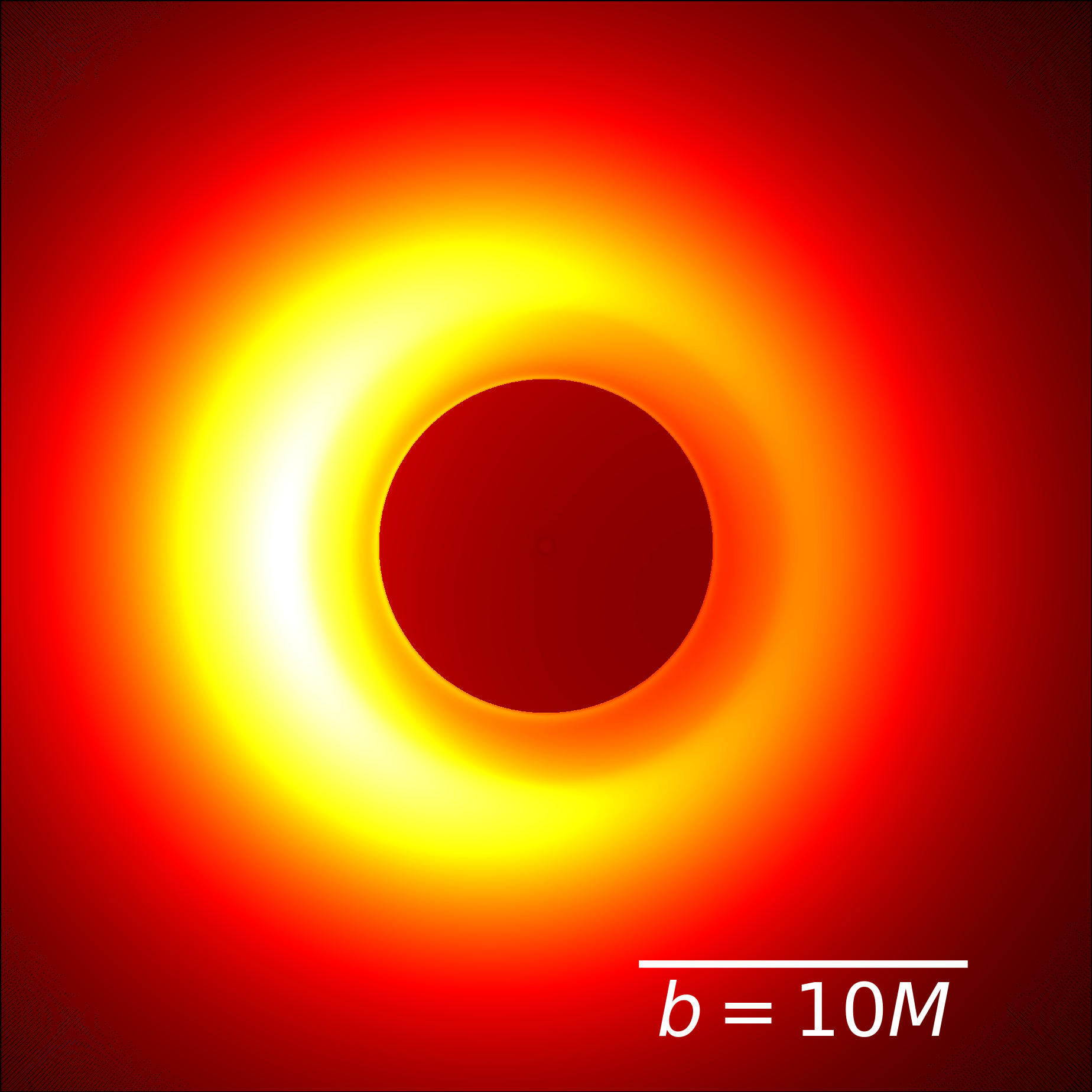} \\
    \end{tabular}
    \noindent \hspace*{0.36cm} \includegraphics[width=0.98\textwidth]{fig11m.png}
    \caption{Observed images for optically thin disks with an inclination angle of $\theta_0 = 20^\circ$,  with varying inner radius $r_{\rm in}$ and half-opening angle $\psi_0$. Rows correspond to different values of $\psi_0$, while columns represent different values of $r_{\rm in}$. All disks assume $\kappa_{\rm ff} = 0.1$ and $\kappa_{\rm K} = 0.9$. The flux distributions are normalized by their respective maximum values.}
    \label{fig:fig12}
\end{figure}

{Similar features can also be observed in Fig.~\ref{fig:fig12}, where we present the observed images for optically thin disks with an inclination angle of $\theta_0 = 20^\circ$.} Note that the disk flows in the two cases (cf. Fig.~\ref{fig:fig11} and~\ref{fig:fig12}) exhibit different dynamics: in the former, the motion is characterized by moderate infall with $\kappa_{\rm ff} = 0.5$ and low angular velocity $\kappa_{\rm K} = 0.1$, while in the latter, the flow moves nearly with Keplerian angular velocity, corresponding to $\kappa_{\rm ff} = 0.1$ and $\kappa_{\rm K} = 0.9$. {Therefore, a pronounced asymmetry in the observed flux can be clearly seen in Fig.~\ref{fig:fig12}.} Note that for $\psi_0 = 60^\circ$ in Fig.~\ref{fig:fig11}, one can observe cross-shaped lines separating regions of differing brightness. These lines correspond to the angular boundaries of the disk---the edges of the emitting region in polar angle---which become visible when the observer-black hole axis is covered by the disk.

\subsection{Optically thick disks}
\label{sec:Optically thick disks}

Now, let us turn to the imaging of optically thick disks. In this case, the observed flux along a given light trajectory is determined by the first intersection with the disk surface. As a result, the direct image becomes the dominant contribution to the final observed flux.

For a line of sight along a fixed $\alpha$ direction, whether the photon ring (peak) appears can be decided on on the condition derived from the analytical expression~\eqref{eq:critical photon orbit2}. Specifically, if the segment with the minimum inclination angle along this line of sight exceeds the right-hand side of Eq.~\eqref{eq:critical photon orbit2}, the photon ring becomes visible. Conversely, if the photon ring is absent, the lensing ring typically does not appear either, since it occurs at larger apparent radii. Another convenient criterion for identifying the presence of the photon ring (and the lensing ring) is to examine the inner edge of the direct image. If the inner edge lies at an impact parameter smaller than the critical value $b_c=3\sqrt{3}M$, then neither the photon ring nor the lensing ring will appear in the observed image. 

Using this criterion, one can readily infer the expected appearance of the corresponding optically thick image by referring to the optically thin cases presented in the previous section. Specifically, the region inside the inner edge of the direct image, together with the direct image itself in the optically thin case, can be approximately taken as the observed image in the optically thick case. While the overall flux magnitudes generally differ between the two scenarios, the qualitative shape of the flux profile remains qualitatively similar.

\begin{figure}[htbp]
    \centering
    \renewcommand{\thesubfigure}{} % 取消 subcaption 编号
    \setlength{\tabcolsep}{3pt} % 图间距微调
    \begin{tabular}{*{4}{>{\centering\arraybackslash}m{0.3\textwidth}}}
            \includegraphics[width=\linewidth]{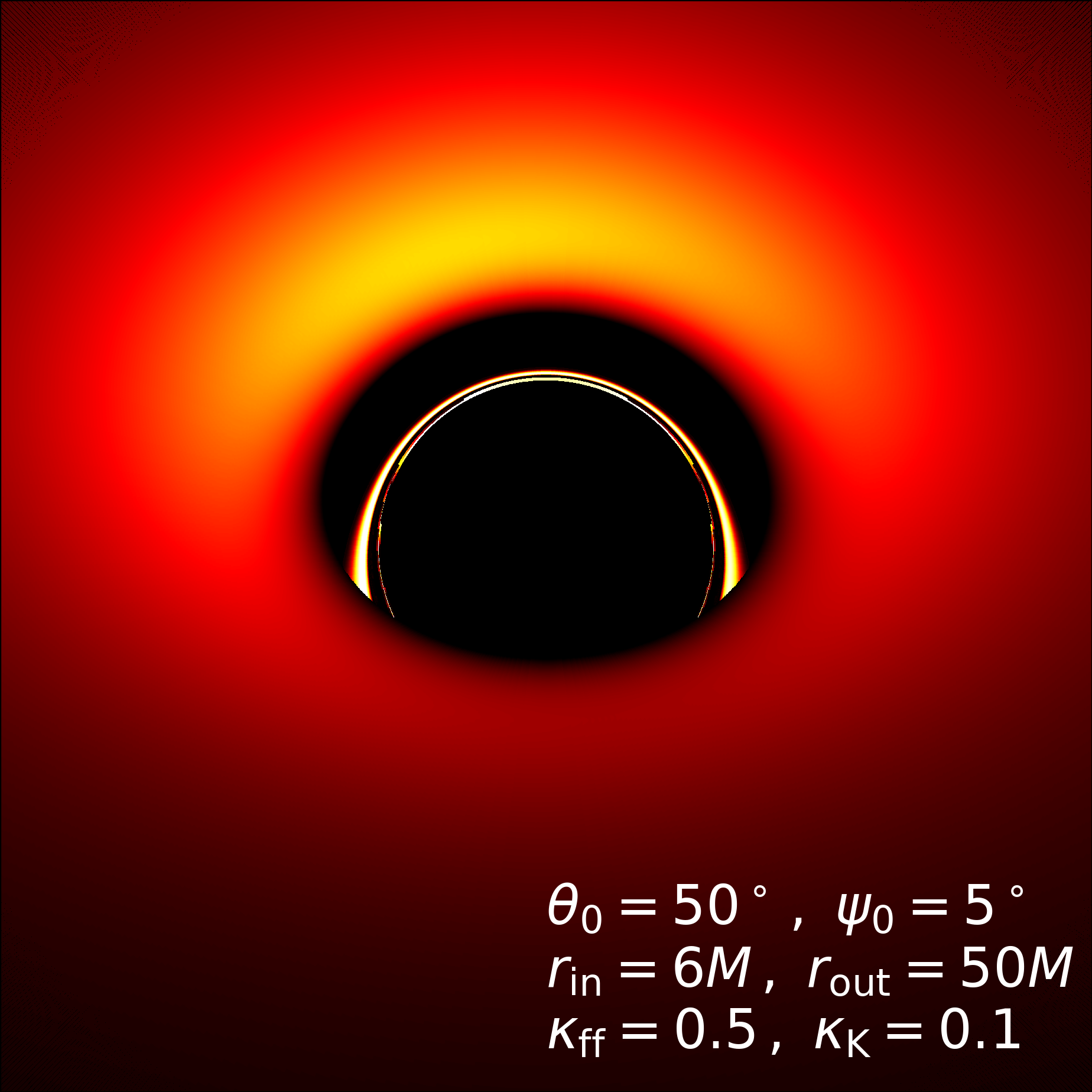} &
        \includegraphics[width=\linewidth] 
{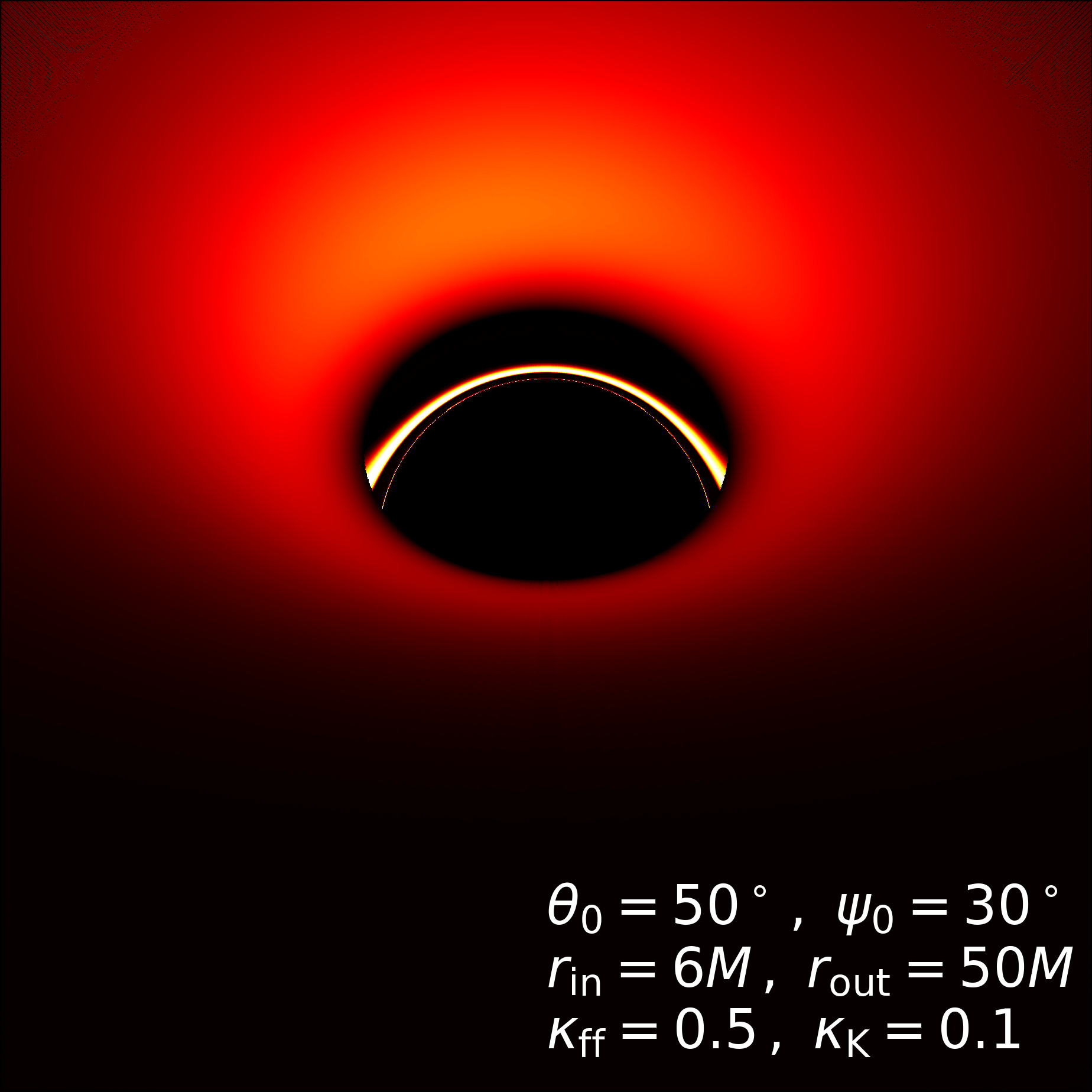} &
        \includegraphics[width=\linewidth] {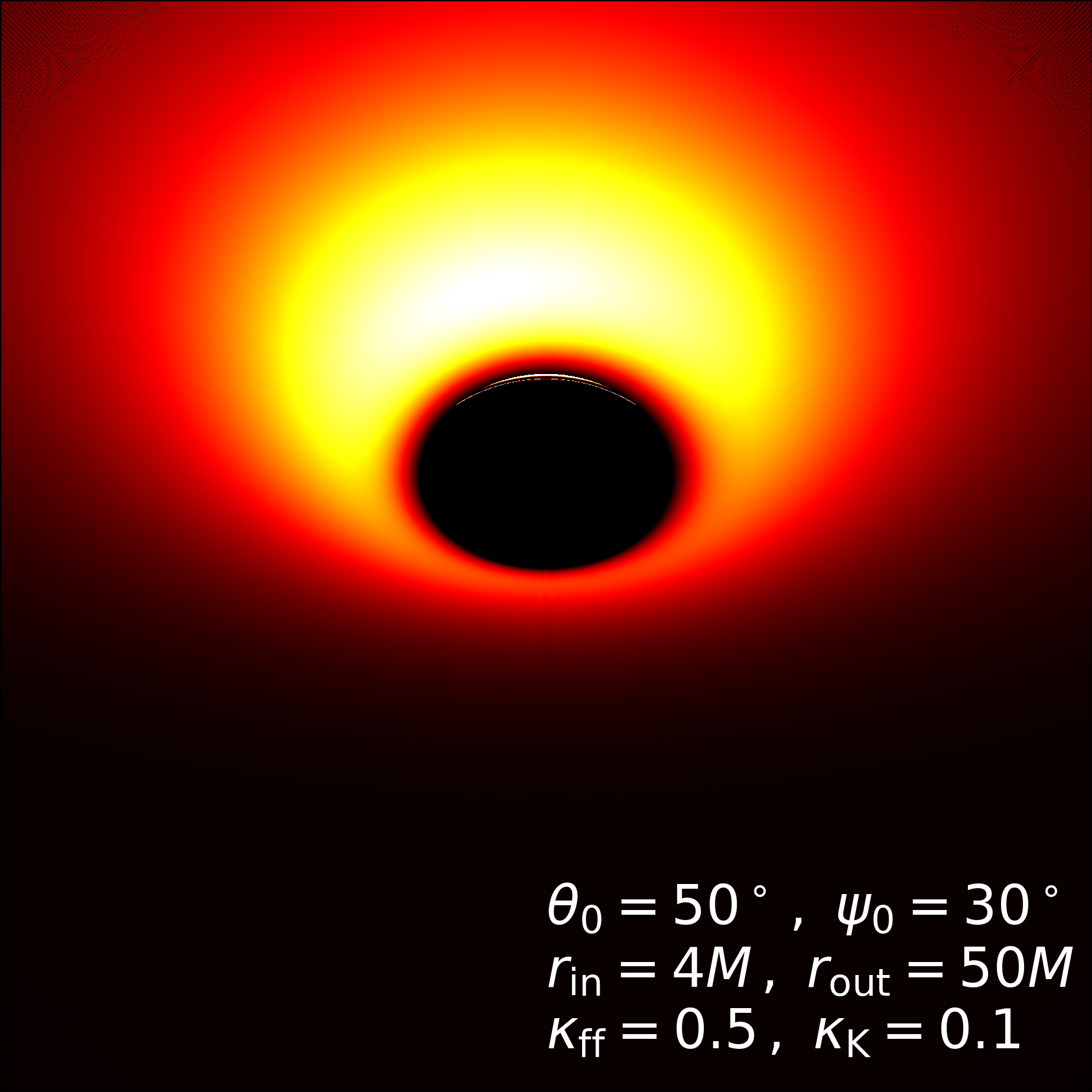}\\

        \includegraphics[width=\linewidth]
{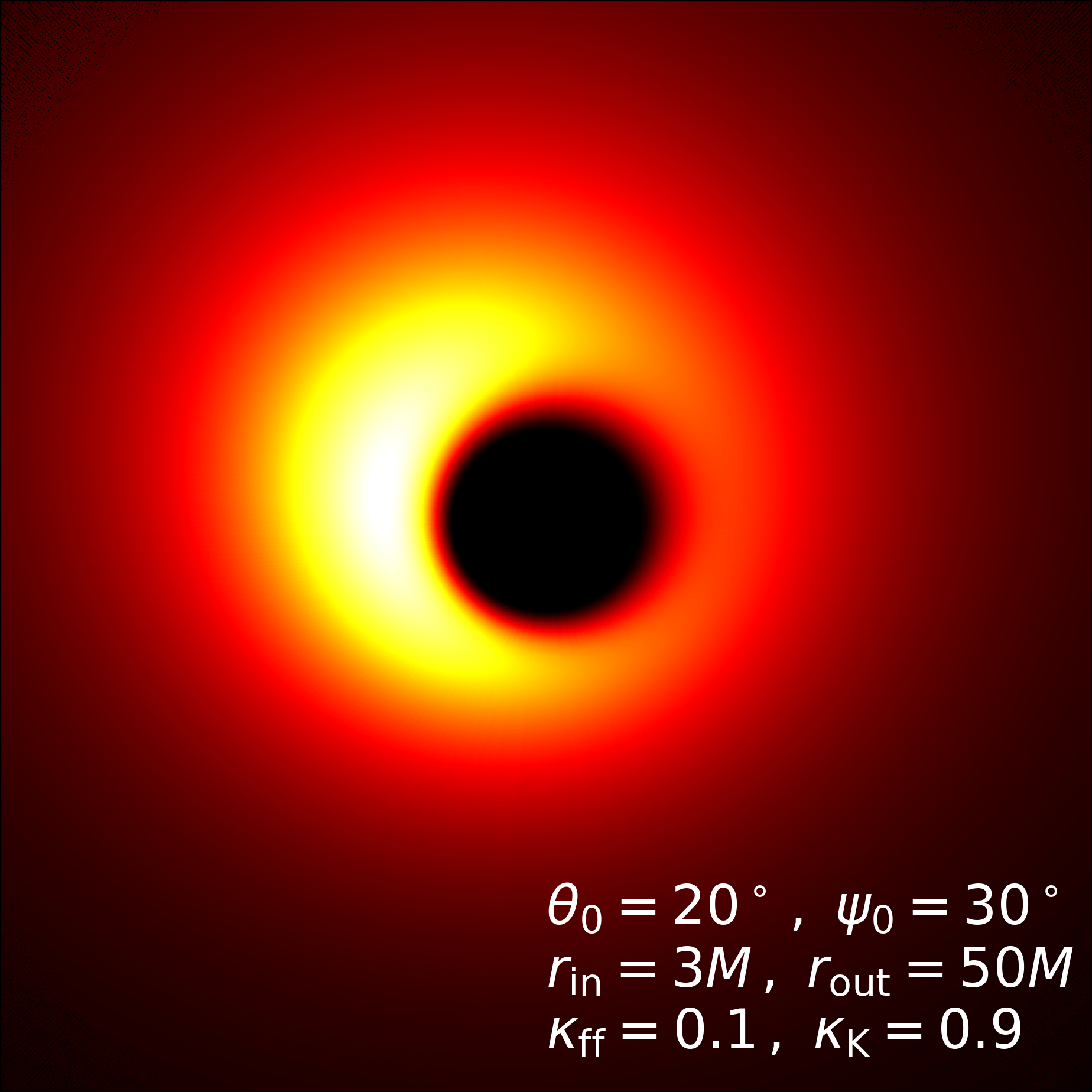} &
        \includegraphics[width=\linewidth] 
{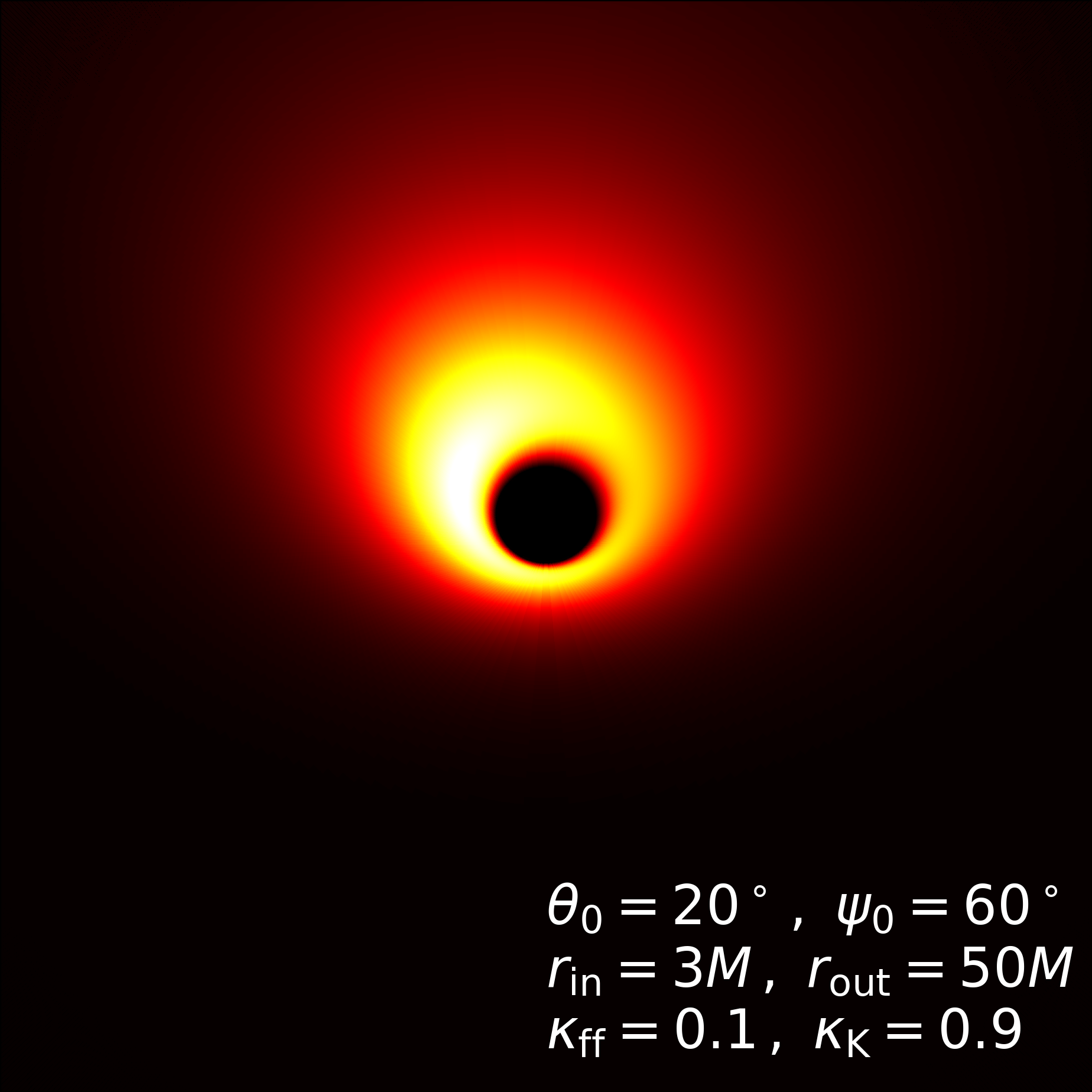} &
        \includegraphics[width=\linewidth] 
{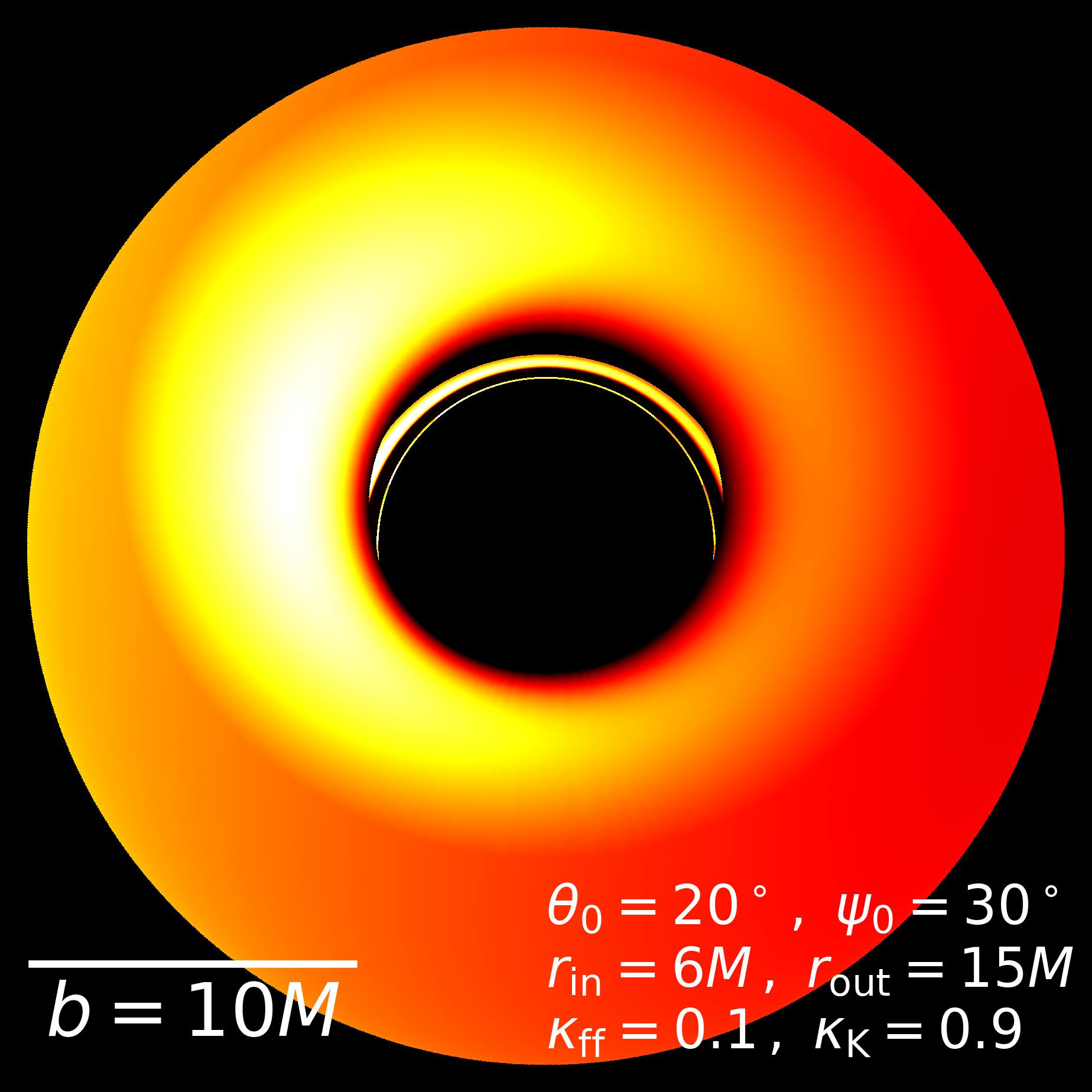} \\
    \end{tabular}
\hspace*{-0.3cm} \includegraphics[width=1.03\textwidth]{fig11m.png}
    \caption{Observed images for optically thick disks. The flux distributions are normalized by their respective maximum values. The case with $\psi_0=90^\circ$ is omitted, as it corresponds to a fully spherical emitting surface, producing a star-like image that lacks prominent relativistic lensing features and a clearly defined black hole shadow.}
    \label{fig:flux_matrix_optical_thick}
\end{figure}

In Fig.~\ref{fig:flux_matrix_optical_thick}, we present several representative images of optically thick accretion disks with varying parameters. All these disks have an outer edge at $r_{\rm out}=50M$, except for the bottom-right panel, where the disk is truncated at $r_{\rm out}=15M$. In this case, the outer boundary is clearly visible in the observed image.
For a disk with half-opening angle $\psi_0=90^\circ$, the geometry becomes spherically symmetric, and the image resembles that of a luminous sphere, much like the Sun, in the sense that all visible radiation originates from a single, optically thick spherical surface. As a result, it offers limited insight into the structure of the black hole shadow. For this reason, we do not include this configuration in our image set.

\subsection{Partially optically thick disks}

In realistic astrophysical environment, accretion flows around black holes are neither perfectly optically thin nor completely optically thick. Instead, the optical depth varies depending on local conditions such as density, temperature, magnetic field strength, and ionization state. Consequently, the observed black hole image is often shaped by a complex interplay of absorption, emission, and light bending, which cannot be fully captured by the limiting cases of optically thin or thick disks.

As outlined in the Sec.~\ref{sec:Coordinate systems and accretion disk models}, we introduce an approximate radiative transfer model that allows for partial absorption along photon geodesics. Specifically, we use the recurrence relation in Eq.~\eqref{eq:F_lambda}, repeated here for clarity:
\begin{align}\notag
F(r_2) = F(r_1) e^{-\chi  \delta l} \left[ \frac{g(r_1)}{g(r_2)} \right]^4 + \delta F_{\rm em}(r_2)\,.
\end{align}
Here, $\chi$ represents a constant effective absorption coefficient, with the limiting cases $\chi \to 0$ and $\chi \to \infty$ reproducing the optically thin and thick scenarios, respectively. 

It should be noted that a key limitation of this model lies in the assumption of a constant, frequency-independent absorption coefficient $\chi$. In realistic accretion flows~\cite{frank2002accretion}, the effective absorption is governed by microphysical processes such as free-free absorption and synchrotron self-absorption, which depend sensitively on the photon frequency and energy density of the accretion flows. As a result, the absorption coefficient $\chi$ may vary by orders of magnitude across the disk. This simplification means that our model cannot capture frequency-dependent spectral features or spatially resolved variations in opacity.

Nevertheless, this simplified approach allows us to interpolate between the two extreme cases previously studied in Secs.~\ref{sec:Optically thin disks} and \ref{sec:Optically thick disks}, provides useful qualitative insight into how partial absorption modifies the observed flux and image.  

\begin{figure}[htbp]
\centering
\includegraphics[width=.85\textwidth]{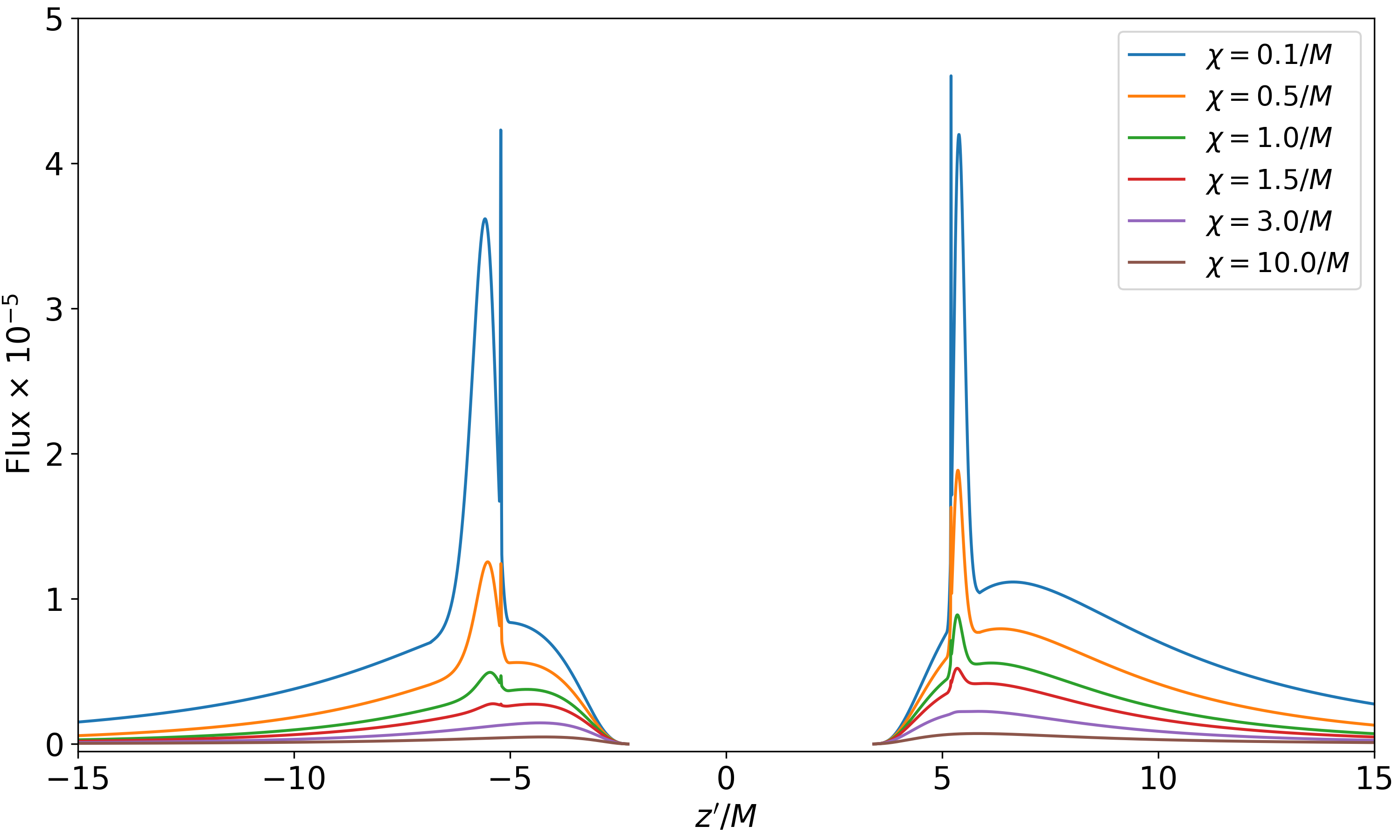}
\caption{Observed flux along the $z'$ axis of the image plane for partially optically thick disks with an inclination angle $\theta_0 = 20^\circ$, an inner radius of $r_{\rm in} = 2.3 M$, a half-opening angle $\psi_0 = 10^\circ$ and varying effective absorption coefficients. All disks assume $\kappa_{\rm ff} = 0.5$ and $\kappa_{\rm K} = 0.3$. \label{fig:fig14}}
\end{figure}

\begin{figure}[htbp]
\centering
\includegraphics[width=.85\textwidth]{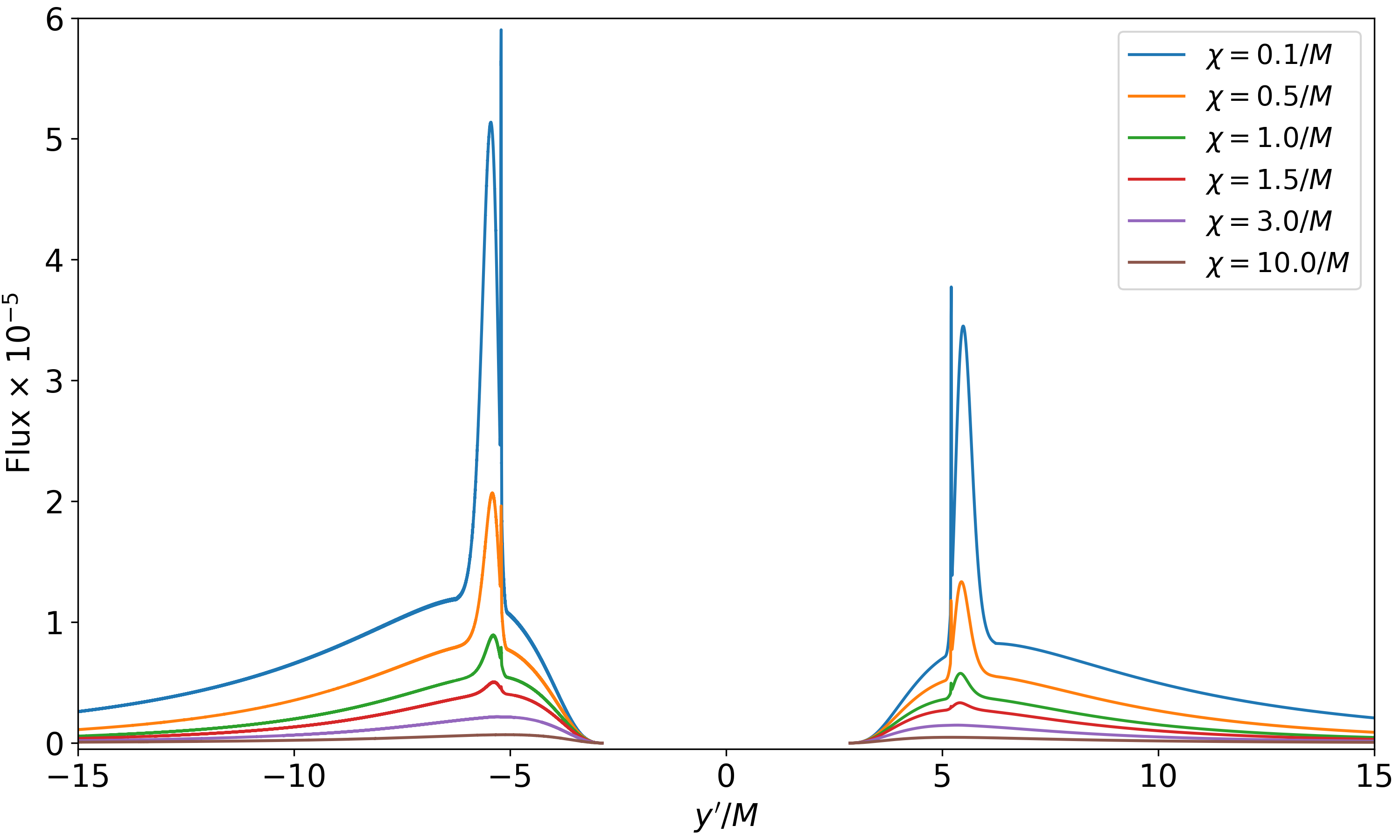}
\caption{Same as Fig.~\ref{fig:fig14}, but with observed flux along the $y'$ axis of the image plane. \label{fig:fig15}}
\end{figure}

In Figs.~\ref{fig:fig14} and \ref{fig:fig15}, we present the observed flux profiles along the $z'$ and $y'$ axes of the image plane for partially optically thick disks with a half-opening angle of $\psi_0 = 10^\circ$ and $r_{\rm in}=2.3M$. We choose a small inner radius of $r_{\rm in}=2.3M$, as this configuration maximizes the contrast between the optically thin and optically thick cases. Specifically, when  $r_{\rm in} < r_{\rm sp}$, the inner edge of the direct image always lies within the critical impact parameter $b = b_c$, as shown by the transfer functions.  As a result, both the photon ring and the lensing ring are entirely absent in the optically thick case.  

\begin{figure}[htbp]
    \centering
    \renewcommand{\thesubfigure}{} % 取消 subcaption 编号
    \setlength{\tabcolsep}{1pt} % 图间距微调
    \begin{tabular}{*{4}{>{\centering\arraybackslash}m{0.23\textwidth}}}
        \includegraphics[width=\linewidth]
{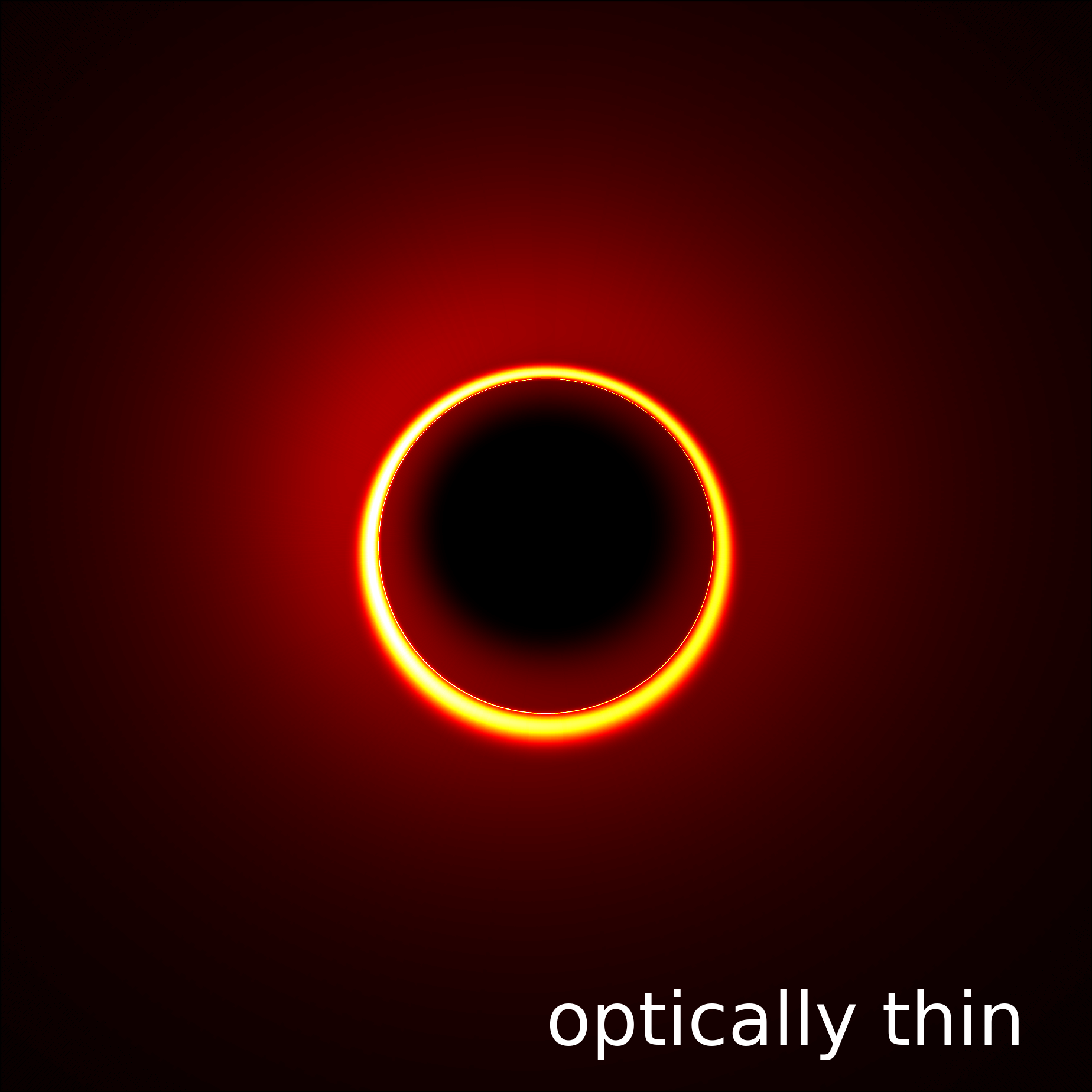} &
        \includegraphics[width=\linewidth] 
{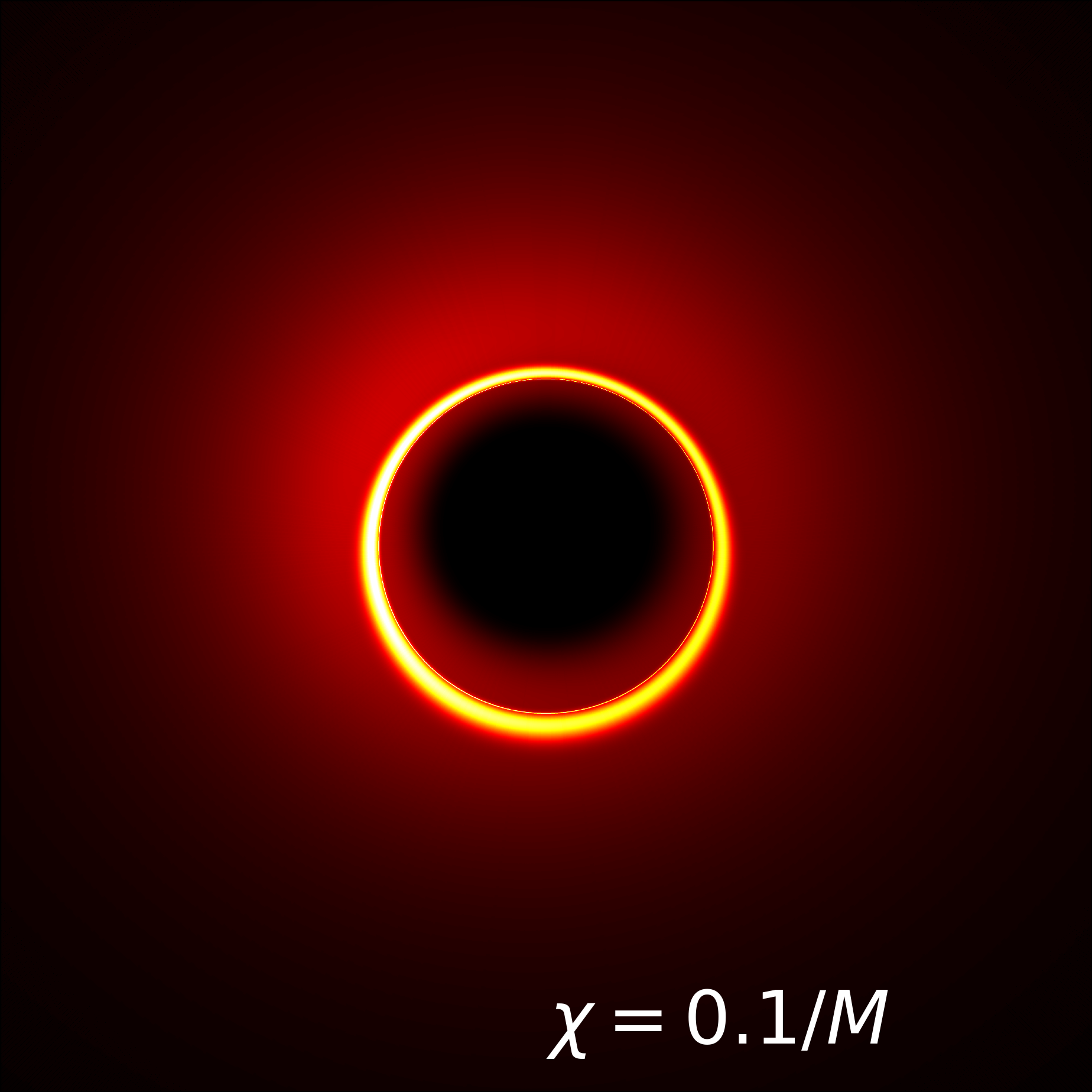} &
       \includegraphics[width=\linewidth] 
{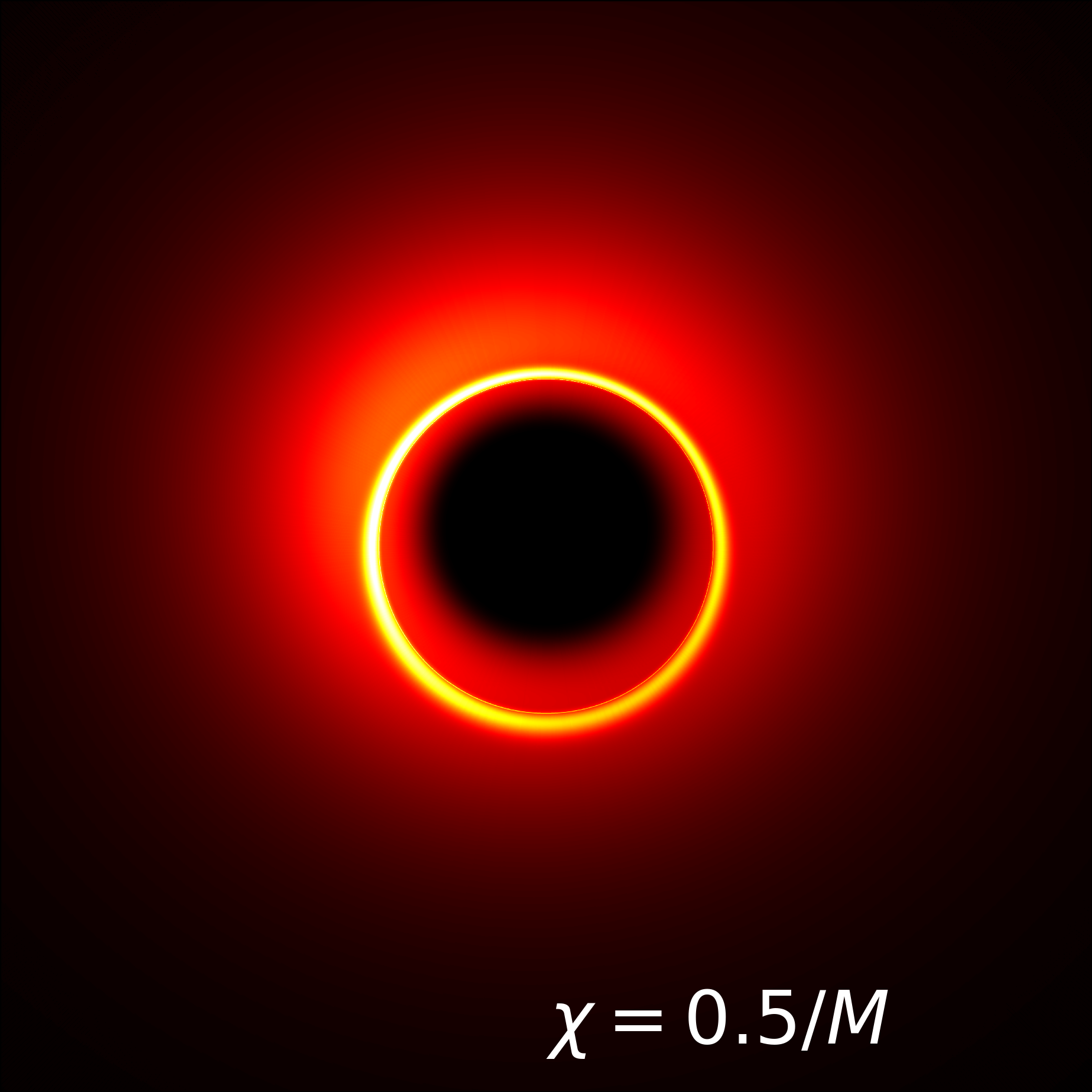} &       
       \includegraphics[width=\linewidth] {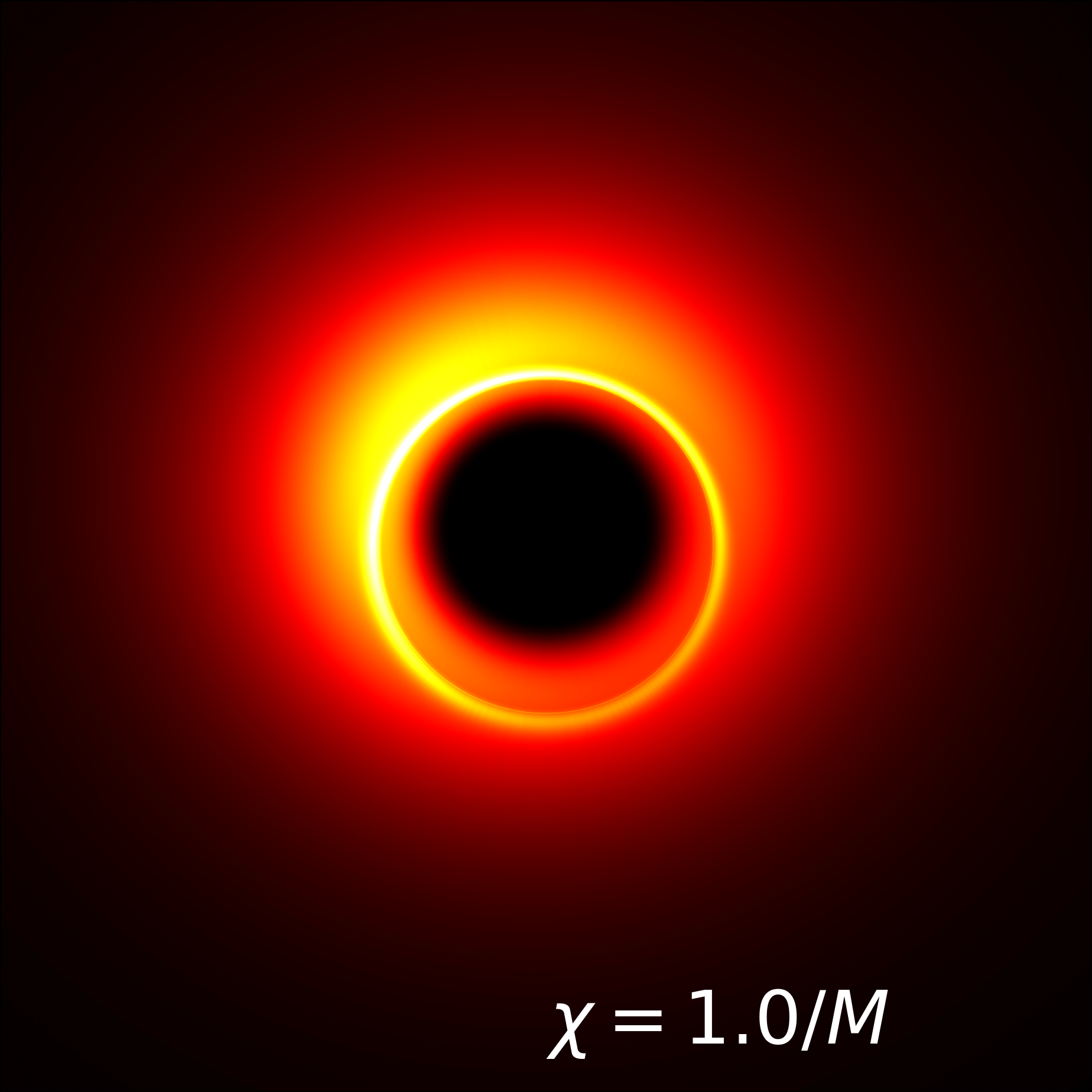}\\

        \includegraphics[width=\linewidth] 
{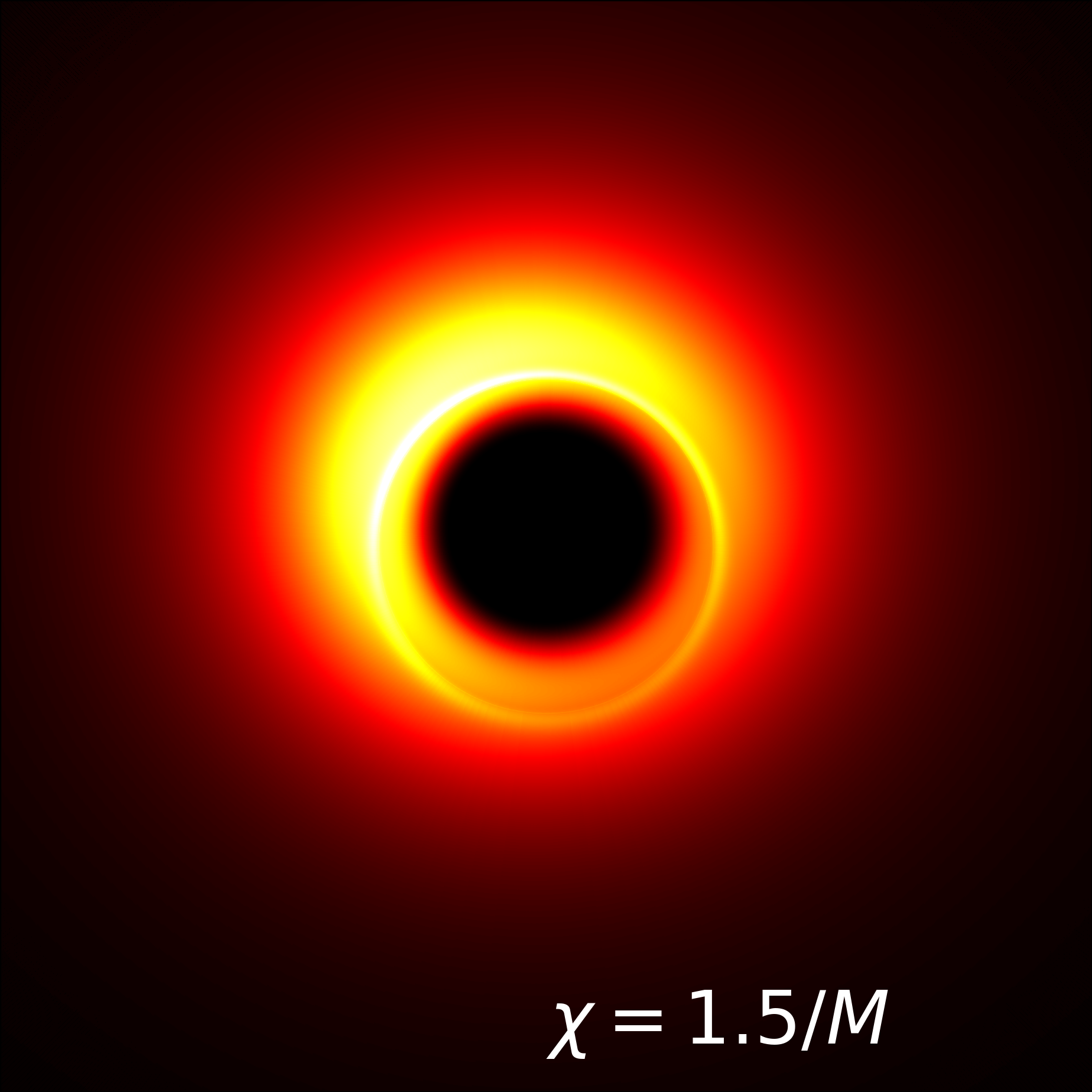} &
        \includegraphics[width=\linewidth] 
{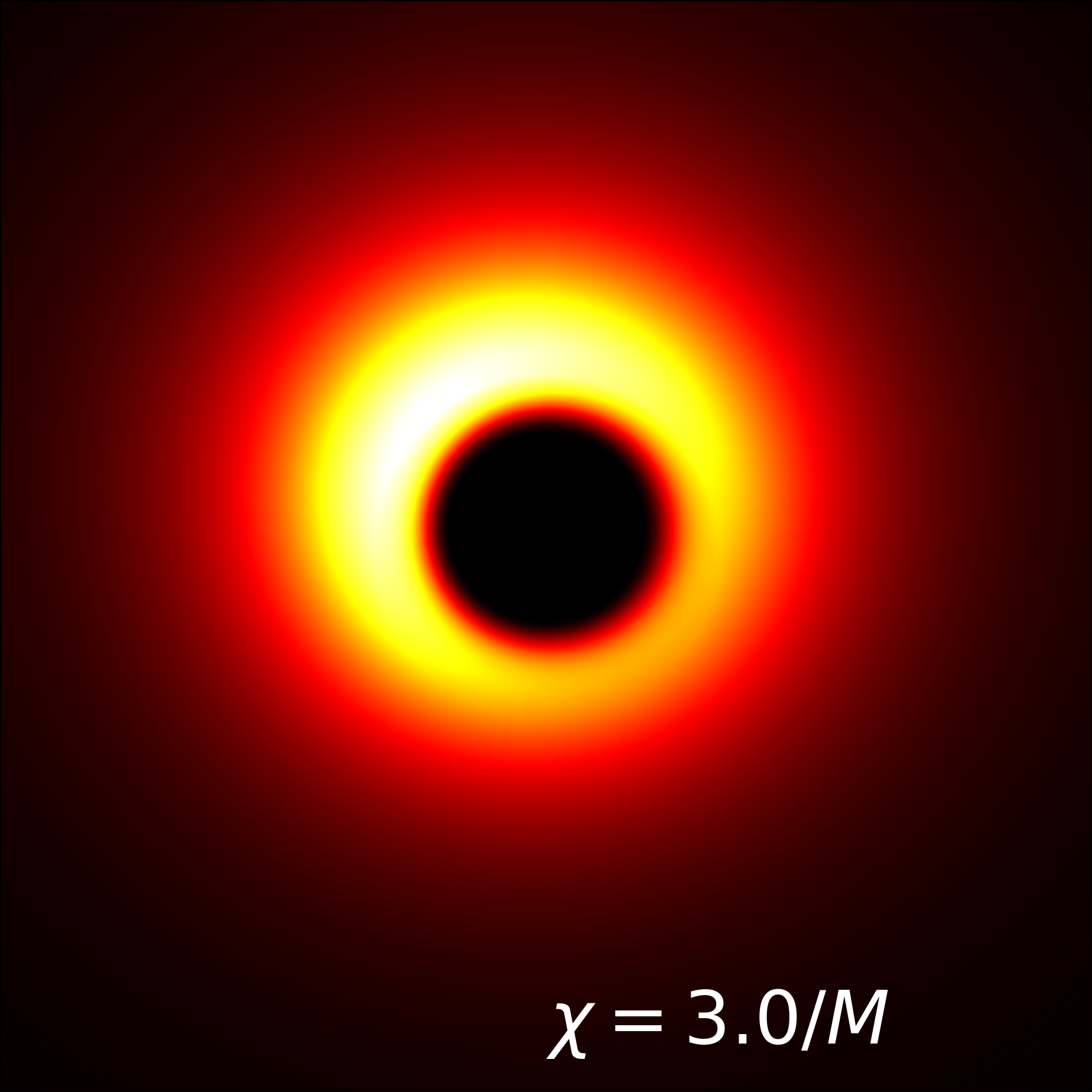} &
        \includegraphics[width=\linewidth] 
{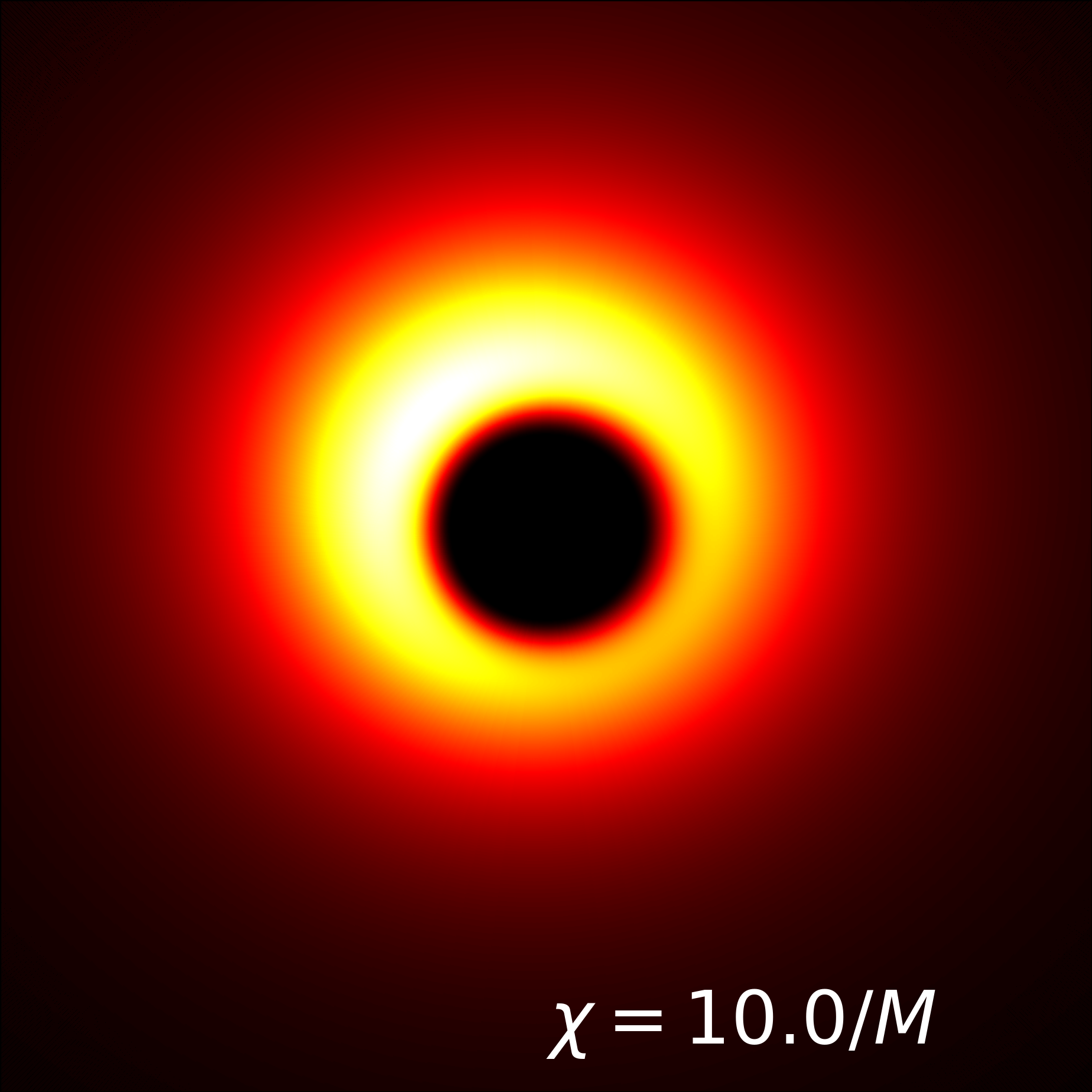} &
        \includegraphics[width=\linewidth] 
{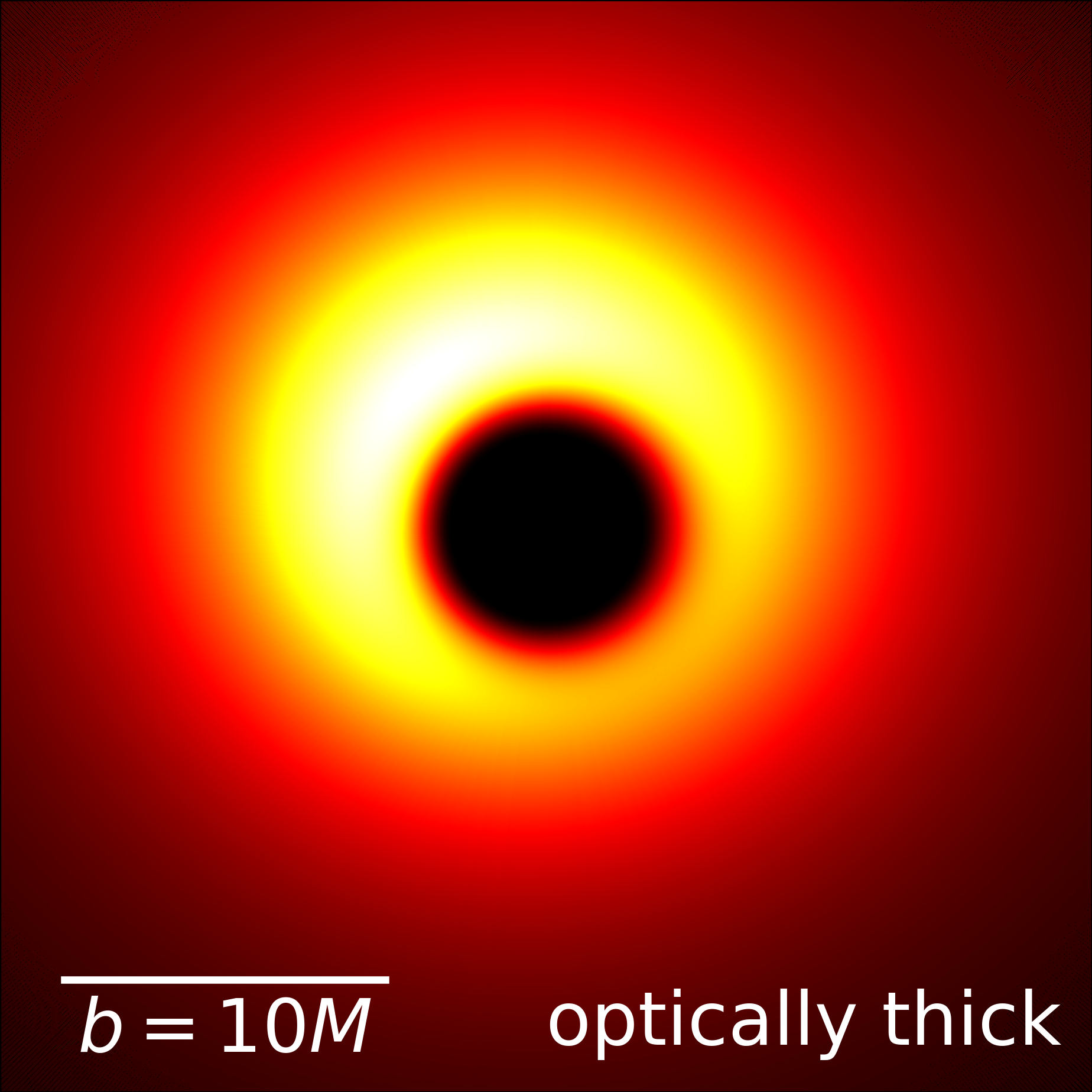} \\

    \end{tabular}
\hspace*{-0.4cm} \includegraphics[width=1.04\textwidth]{fig11m.png}
    \caption{Observed images for partially optically thick disks with an inclination angle $\theta_0 = 20^\circ$, an inner radius of $r_{\rm in} = 2.3 M$, a half-opening angle $\psi_0 = 10^\circ$ and varying effective absorption coefficients. All disks assume $\kappa_{\rm ff} = 0.5$ and $\kappa_{\rm K} = 0.3$. Two limiting cases---the optically thin and optically thick regimes---are also presented for comparison.  The flux distributions are normalized by their respective maximum values.}
    \label{fig:flux_matrix_optical_intermediate}
\end{figure}

In Figs.~\ref{fig:fig14} and \ref{fig:fig15}, the effective absorption coefficient $\chi$ is varied from $\chi = 0.1/M$ to $\chi = 10/M$. As $\chi$ increases, the overall observed flux decreases, and both the photon ring and lensing ring become  less pronounced. Notably, for $\chi \gtrsim 1/M$, these rings nearly vanish. This behavior is more evident in the full images shown in Fig.~\ref{fig:flux_matrix_optical_intermediate}, where the limiting cases of optically thin and optically thick disks are also included for comparison.  A noticeable transition in the image occurs around $\chi \sim 1/M$.

This transition can be understood as follows. For geometrically thick disks with a  half-opening angle of $\psi_0 = 10^\circ$, the typical path length of a photon with impact parameter $ b_c$ traversing the disk once is approximately $\Delta s \sim 2r_{\rm sp} \psi_0 \sim M$.  When the absorption length $1/\chi$ is comparable to this typical path length, the emitted flux is attenuated by a factor of  $1-\ee ^{-1}\simeq 0.63$.  When  $1/\chi$ is much larger than  this scale, the disk behaves as optically thin, and photons can accumulate flux through multiple disk crossings. Conversely, when $1/\chi \ll \Delta s$, most emission is absorbed before escaping, and the disk behaves as optically thick. Therefore, $\chi \sim 1/M$ marks the transition between these two regimes, and the image structure changes significantly. 

\begin{figure}[htbp]
\centering
\includegraphics[width=.85\textwidth]{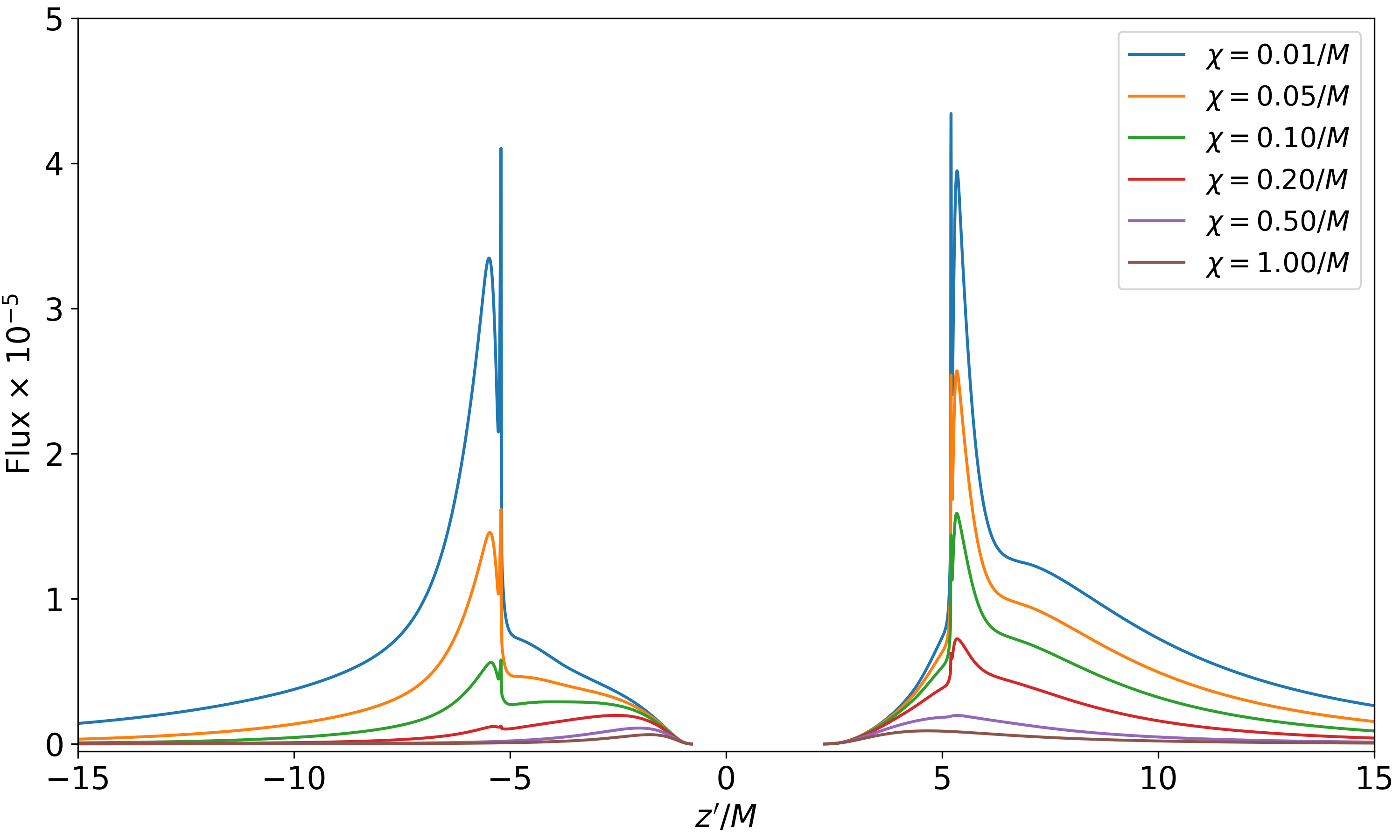}
\caption{Same as Fig.~\ref{fig:fig14}, but for a half-opening angle of $\psi_0 = 50^\circ$. A different set of effective absorption coefficients $\chi$ is used to illustrate the effect of optical depth under broader disk geometry. \label{fig:fig17}}
\end{figure}

Note that the typical path length through the disk increases with the half-opening angle $\psi_0$, leading to a corresponding shift in the transition between optical regimes. In Fig.~\ref{fig:fig17}, we present the observed flux profiles along the $z'$ axis of the image plane for partially optically thick disks with a half-opening angle of $\psi_0 = 50^\circ$. In this case, the transition from optically thin to optically thick behavior occurs at approximately $\chi \sim 0.2/M$. This shift arises because photons traverse longer paths through the emitting material at larger $\psi_0$, increasing the likelihood of absorption and thus requiring a smaller absorption coefficient to achieve significant attenuation.

\section{Conclusions and Discussion}
\label{sec:Conclusions and Discussion}

In this work, we focused on geometrically thick accretion disks, whose structure is characterized by an inner radius $r_{\rm in}$ and a half-opening angle $\psi_0$, as illustrated in Fig.~\ref{fig:1}. To study their imaging properties, we have developed a systematic framework based on decomposing the emitting region into one-dimensional luminous segments. Each segment is defined by its inner radius $r_{\rm in}$ and an inclination angle $\varphi_0$.

A central component of our approach was the use of transfer functions associated with these segments. These transfer functions play two critical roles: they determine the inner boundaries of the direct image, lensing ring, and photon ring; and they govern the width of the lensing ring in the case of geometrically thick disks. Both aspects are essential for understanding the structure of black hole shadows. One crucial outcome of this analysis is that, as the half-opening angle $\psi_0$ increases, the lensing ring systematically broadens and gradually approaches the photon ring. This behavior is a generic feature of optically thin disks. However, the initial width of the lensing ring primarily depends on the disk inclination angle $\theta_0$ and the inner radius $r_{\rm in}$, {while the rate at which it broadens with increasing $\psi_0$ varies with the image-plane polar angle $\alpha$.} This dependence arises because a geometrically thicker disk includes emitting segments spanning a broader range of inclination angles $\varphi_0$. Specifically, segments with smaller $\varphi_0$ tend to produce lensing rings with more extended outer edges, as revealed by the transfer function in Fig.~\ref{fig:transfer function}, whereas segments with larger $\varphi_0$ generate lensing rings located closer to the photon ring. Moreover, for a given disk configuration, the range of contributing $\varphi_0$ segments {varies with the image-plane polar angle $\alpha$}, resulting in variations in the lensing ring width across different regions of the image plane. 

The above results provide a concrete resolution to the differing conclusions presented in Refs.~\cite{Gralla:2019xty} and~\cite{Narayan:2019imo} regarding the observability and significance of the photon ring in the images of optically thin accretion flows around a Schwarzschild black hole. Both conclusions are valid within their respective contexts: Gralla et al.~\cite{Gralla:2019xty} focus on geometrically thin accretion disks, {whereas Narayan et al.~\cite{Narayan:2019imo} examine spherical accretion flows. In fact, our study bridges the gap between these two scenarios. Our analysis demonstrates that as the disk transitions from a geometrically thin configuration to a spherical flow,} the width of the photon ring remains nearly constant, typically $\Delta  \lesssim 0.1M$. In contrast, the width of the lensing ring grows significantly---with variations typically on the order of $M$---and gradually bridges the photon ring and the direct image. From an observational perspective, it is therefore the lensing ring that provides a prominent and evolving signature as the disk geometry varies. This finding supports the viewpoint presented in Ref.~\cite{Gralla:2019xty}, namely that the lensing ring can provide a more prominent observational feature than the photon ring.   Nevertheless, this does not diminish the physical importance of the photon ring. On the contrary, the photon ring precisely determines the innermost edge of the lensing ring and effectively defines the outer boundary of the black hole shadow, particularly in the case of geometrically thick accretion disks. Its presence is foundational to the overall image structure, even though its flux contribution may be subdominant or observationally blended with that of the lensing ring.
      
In the optically thick regime, the observed flux is predominantly determined by the direct image, which corresponds to the first intersection between the light ray and the emitting surface when tracing backward from the observer. Whether the lensing ring and photon ring can appear is {determined by the image-plane polar angle $\alpha$ }and the inclination angle of the emitting segment, as quantified by the analytical criterion given in Eq.~\eqref{eq:critical photon orbit2}. A practical criterion is that if the inner edge of the direct image---obtainable from the transfer function---lies beyond the critical impact parameter  $b_c=3\sqrt{3}$, then neither the photon ring nor the lensing ring will appear in the observed image. Using this criterion, one can infer the qualitative structure of optically thick images from their optically thin counterparts.

For partially optically thick disks, we investigated how the observed image responds to changes in the effective absorption coefficient $\chi$. We focused on configurations with $r_{\rm in} \leq r_{\rm sp}$, as this choice maximizes the difference between the optically thin and optically thick cases.  As $\chi$ increases, the average flux diminishes and the characteristic features associated with the lensing and photon rings begin to fade. A pronounced transition in the observed image occurs around $\chi 
\sim (6M\psi_0)^{-1}$, which corresponds to the regime where the absorption length becomes comparable to the typical path length of photons traversing the disk. At this threshold, most photons are absorbed before accumulating significant flux through multiple crossings, effectively suppressing the contributions from high-deflection orbits. As a result, the image becomes increasingly dominated by direct emission, approaching the appearance of an optically thick disk. 

Although our study incorporated a broad range of parameters in modeling the images of accretion flows, it remained a simplified approach that did not fully capture the complexity of realistic astrophysical environments. The accretion disk structure was treated as a prescribed input rather than derived self-consistently from first-principles physical models. The radiative transfer was approximated by a simplified, frequency-independent absorption scheme, and potentially important components such as jets, winds, and magnetic fields were not taken into account. Despite these limitations, the segment-based transfer function framework provided a useful and flexible tool for exploring key imaging features. 

At last, we note that although this work focused on the Schwarzschild black hole, the method can be readily generalized to other spherically symmetric spacetimes, such as those arising in alternative theories of gravity or semiclassical quantum gravity models~\cite{Capozziello:2011et,Battista2023,Casadio:2025lkn,Wang:2025fmz,Li:2025zdt,Tan:2024jkj}. This generalization is particularly relevant given that many modified gravity scenarios predict deviations from the Schwarzschild geometry while maintaining spherical symmetry~\cite{Vagnozzi:2022moj,Wang:2022ews,Cao:2024kht,Wang:2023rjl,Li:2024abk,Gogoi:2025rcn,Lim:2025cne}. Such extensions may provide a useful framework for exploring potential observational signatures that distinguish general relativity from its alternatives.

\appendix
\section{Local Proper Reference Frame} 
\label{appA}

The local proper reference frame $(t', x', y', z')$ forms an orthonormal (dual) basis, which can be expressed in terms of the coordinate (dual) basis as follows~\cite{Nakahara2003}: 
\begin{subequations} \label{eq:tetrads} 
\begin{align} \frac{\partial}{\partial {x'}^{a}} &= e_a{}^{\mu} \frac{\partial}{\partial x^{\mu}}, \\ 
\dd {x'}^{a} &= e^a{}_{\mu} \dd x^{\mu}, 
\end{align} 
\end{subequations} 
where Latin indices ($a=0,1,2,3$) refer to the local frame components with $x'^0 = t'$, $x'^1 = x'$, $x'^2 = -z'$, and $x'^3 = y'$, while Greek indices ($\mu = 0,1,2,3$) correspond to the spacetime coordinates $(t, r, \theta, \phi)$. Here, $e_a{}^{\mu}$ denotes the inverse of $e^{a}{}_{\mu}$, and both are commonly referred to as tetrads (or vielbeins). The tetrads form a $4\times 4$ matrix satisfying 
\begin{align}
  g_{\mu \nu} =e^{a}{}_{\mu}e^{b}{}_{\nu}\,\eta_{ab}\,,
\end{align}
with $\eta_{ab}= {\rm diag}(-1,1,1,1)$ being the Lorentz metric.  For the Schwarzschild metric~\eqref{eq:metric}, the nonzero components of the orthonormal tetrads outside the black hole are given by
\begin{align} e^{a}{}_{\mu} = \operatorname{diag}\left( \sqrt{1-\frac{2M}{r}}, \frac{1}{\sqrt{1-\frac{2M}{r}}}, r, r \sin \theta \right)\,. 
\end{align}

In the local reference frame, we have form Fig.~\ref{fig:2} that
\begin{subequations}\label{eq:alpha&beta_0}
\begin{align}
\tan ^2 \beta & =\frac{\dd {y'}^2 +\dd {z'}^2}{ \dd {x'}^2 } \notag \\
             &= \left(\frac{e^{3}{}_{3}}{e^{1}{}_{1}}\right)^2 \left(\frac{\dd \phi}{\dd r}\right)^2 + \left(\frac{e^{2}{}_{2}}{e^{1}{}_{1}}\right)^2 \left(\frac{\dd \theta}{\dd r}\right)^2 \,, \\
\tan ^2 \alpha &= \frac{\dd {z'}^2}{\dd {y'}^2} \notag \\
               &= \left(\frac{e^{2}{}_{2}}{e^{3}{}_{3}}\right)^2 \left(\frac{\dd \theta }{\dd \phi}\right)^2 \,.
\end{align}
\end{subequations} 
Note that the Schwarzschild spacetime possesses four Killing vectors associated with symmetries under time translations and rotations about the $x$-, $y$-, and $z$-axes~\cite{Carroll2004}. Each Killing vector corresponds to a conserved quantity along geodesics, namely, the energy $E$ and the components of angular momentum $L_x$, $L_y$, and $L_z$. For our analysis, the relevant conserved quantities are
\begin{subequations}\label{eq:conserved_quantities}
\begin{align}
 E &= \left(1-\frac{2M}{r}\right) \frac{\dd t}{\dd \lambda}\,,  \\ 
 L_z &= r^2 \sin^2 \theta \frac{\dd \phi}{\dd \lambda}\,,  \\
 L^2 &= L_x^2 + L_y^2 + L_z^2 = r^4 \left[ \left( \frac{\dd \theta}{\dd \lambda} \right)^2 + \sin^2 \theta \left( \frac{\dd \phi}{\dd \lambda} \right)^2 \right]\,, 
 \end{align} 
\end{subequations} 
where $\lambda$ is the affine parameter along the null geodesic. Denote the four-momentum of the photon by $k^{\mu} =\dd x^{\mu} /\dd \lambda$, satisfying the null condition $g_{\mu \nu}k^{\mu} k^{\nu}=0$. Using the conserved quantities given in Eq.~\eqref{eq:conserved_quantities}, we can express Eq.~\eqref{eq:alpha&beta_0} as 
\begin{subequations}\label{eq:alpha&beta_1} 
\begin{align}
\tan ^2 \beta & =\frac{1}{r^2} \left(\frac{E^2}{L^2(1-2M/r)}- \frac{1}{r^2}\right)^{-1}\,, \\
\tan ^2 \alpha &= \frac{L^2 \sin ^2 \theta}{L_{z}^2}-1 \,.
\end{align}
\end{subequations}
For a distant observer located at $r = r_{\rm obs} \gg M \sim b$ and $\theta = \theta_0$, we introduce the impact parameter
\begin{align}
b=L/E\,.
\end{align}
Under this situation, we obtain 
\begin{subequations}\label{eq:alpha&beta_2}
\begin{align}
\tan ^2 \beta & \simeq \frac{b^2}{r_{\rm obs}^2}\,, \\
{L_z}^2 &=L^2 \sin ^2 \theta_0 \cos ^2 \alpha\,, \label{eq:Lz/L}
\end{align}
\end{subequations}
from which one can obtain Eq.~\eqref{eq:apparent_size} and \eqref{eq:kphi/kt}. One can also derive Eq.~\eqref{eq:Lz/L} from spatial geometric by noting that $L_z$ is the projection of the angular momentum vector onto the $z$-axis.

\section{Radiation Flux Formulas}
\label{appB}

To compute the radiation flux emitted by the accretion disk, we begin by reviewing the classical Shakura-Sunyaev model-a geometrically thin, optically thick accretion disk. In this framework, the energy dissipated per unit surface area per unit time, namely the radiation flux $ F(r) $, can be derived from the conservation laws of angular momentum and energy. Our analysis mainly follows the treatment presented in Ref.~\cite{frank2002accretion}.

Due to the radial variation in angular velocity, a viscous shear force arises in the $\phi-$direction. The corresponding viscous stress component is given by
\begin{align}
\sigma_{r\phi} = \eta \left( \frac{\partial v_{\phi}}{\partial r} - \frac{v_{\phi}}{r} \right) = \eta r \frac{\dd \Omega}{\dd r}\,,
\end{align}
where $v_{\phi} = r \Omega$ is the rotational velocity, $\Omega$ is the angular velocity, and $\eta$ is the dynamic viscosity.

The total viscous torque exerted by material at radii $> r$ on the material at radius $r$ is obtained by integrating the stress over the vertical extent of the disk:
\begin{align}\label{eq:AppBG}
G(r) = 2\pi r \int r \sigma_{r\phi} \, \dd z = 2\pi r^3 \nu \Sigma \frac{\dd \Omega}{\dd r}\,,
\end{align}
where $\nu = \eta / \rho$ is the kinematic viscosity and $\Sigma = 2\rho h$ is the surface density, with $h$ being the disk half-thickness.

Consider an annulus of disk material between $r$ and $r + \dd r$. The net torque acting on this annulus is
\begin{align}
G(r + \dd r) - G(r) = \frac{\partial G}{\partial r} \, \dd r\,.
\end{align}
The corresponding rate of work done by this torque is
\begin{align}
\Omega \frac{\partial G}{\partial r} \, \dd r = \left[ \frac{\partial (G \Omega)}{\partial r} - G \frac{\dd \Omega}{\dd r} \right] \dd r\,.
\end{align}
The first term in the square brackets represents the radial transport of rotational energy and is determined by boundary conditions upon integration. The second term, $-G \frac{\dd \Omega}{\dd r} \, \dd r$, represents the local rate of mechanical energy dissipation. Assuming the dissipated energy $Q^{+}$ is radiated from both the upper and lower surfaces of the disk, the radiative flux per unit surface area is given by
\begin{align}
F(r) \approx \frac{Q^{+}}{4\pi r \dd r} = \frac{1}{4\pi r} G(r) \frac{\dd \Omega}{\dd r} = \frac{\nu \Sigma}{2} \left( r \frac{\dd \Omega}{\dd r} \right)^2\,,
\end{align}
where we have used Eq.~\eqref{eq:AppBG} in the final expression.

In a steady-state, geometrically thin accretion disc, mass conservation requires a constant mass accretion rate $\dot{M}$ throughout the disc~\cite{Shakura:1972te}:
\begin{align}\label{eq:AppB_mass_rate}
\dot{M} = -2\pi r \Sigma v_r\,.
\end{align}
On the other hand, conservation of angular momentum gives~\cite{frank2002accretion}
\begin{align}
\frac{\partial (r^3 \Sigma v_r \Omega)}{\partial r} = \frac{1}{2\pi} \frac{\partial G}{\partial r}\,,
\end{align}
integrating which yields
\begin{align}\label{eq:AppB_GC}
r^3 \Sigma v_r \Omega = \frac{G}{2\pi} + \frac{C}{2\pi}\,,
\end{align}
where $C$ is a constant related to the angular momentum flux at the inner edge of the disk, $r_{\rm in}$. If one assumes $G(r) = 0$ at $r_{\rm in}$, then
\begin{align}
C = 2\pi r_{\rm in}^3 \Sigma v_r \Omega \big|_{r_{\rm in}}\,.
\end{align}

Using Eqs.~\eqref{eq:AppBG} and \eqref{eq:AppB_mass_rate}, we obtain from Eq.~\eqref{eq:AppB_GC}:
\begin{align}\label{eq:AppB_nuSigma}
\nu \Sigma = \frac{\dot{M}}{2\pi r} \left( -\Omega + \frac{r_{\rm in}^2 \Omega|_{r_{\rm in}}}{r^2} \right) \left( \frac{\dd \Omega}{\dd r} \right)^{-1}\,.
\end{align}

For the classical thin disk with Keplerian angular velocity $\Omega = \Omega_{\rm K}$, this reduces to
\begin{align}
\nu \Sigma = \frac{\dot{M}}{3\pi} \left[ 1 - \left( \frac{r_{\rm in}}{r} \right)^{1/2} \right]\,,
\end{align}
which leads to the radiative flux per unit surface area:
\begin{align}\label{eq:AppB_F_thin} 
F(r) = \frac{3M \dot{M}}{8\pi r^3} \left[ 1 - \left( \frac{r_{\rm in}}{r} \right)^{1/2} \right]\,,
\end{align}
which is the standard result derived in Ref.~\cite{Shakura:1972te}.

For slim disks or ADAF, Eq.~\eqref{eq:AppB_F_thin} is no longer valid, since a significant fraction---or even the majority---of the dissipated energy $Q^{+}$ is transported inward by advective processes. Moreover, the orbital motion is no longer strictly Keplerian. To simplify our analysis, we adopt the following approximation for optically thick disks:
\begin{align}\label{eq:App_Fthick}
F(r) = \epsilon Q^{+} = \frac{\epsilon \nu \Sigma}{2} \left( r \frac{\dd \Omega}{\dd r} \right)^2\,,
\end{align}
and we adopt Eq.~\eqref{eq:AppB_nuSigma} as an approximation for $\nu \Sigma$ in the main text. Here, we assume that viscous dissipation is the sole source of energy in the flow. 

For optically thin disks, the relevant quantity is the dissipated energy per unit volume:
\begin{align}
q^{+} = \nu \rho \left( r \frac{\dd \Omega}{\dd r} \right)^2\,,
\end{align}
and the corresponding radiative energy loss per unit volume is approximated as
\begin{align}\label{eq:App_f}
q^- = \epsilon q^{+}\,.
\end{align}
The parameter $\epsilon$ in Eqs.~\eqref{eq:App_Fthick} and \eqref{eq:App_f} characterizes the efficiency of radiative cooling---that is, the fraction of dissipated energy that is locally radiated. For a classical thin disk, we have $\epsilon = 1$. In contrast, for ADAF, radiative cooling is highly inefficient, and $\epsilon \ll 1$. Since we are primarily interested in the envelope of the flux profile, and $\epsilon$ contributes only as an overall normalization factor, we set $\epsilon = 1$ in our numerical calculations. In this situation, we have 
\begin{align}
q^- = \frac{\dd \Omega }{\dd r}\frac{\dot{M}}{4 \pi \psi _0 }\left(-\Omega + \frac{r^2 _{\rm in} \Omega \big|_{r_{\rm in}}}{r^2} \right)\,.
\end{align}

\acknowledgments

We thank the anonymous referee for insightful comments and valuable suggestions, which have helped improve the clarity and quality of this manuscript. This work is supported by the National Natural Science Foundation of China (Grants No. 12405063).

\bibliographystyle{JHEP}
\bibliography{references}{}

\providecommand{\href}[2]{#2}\begingroup\raggedright\begin{thebibliography}{10}

\bibitem{EventHorizonTelescope:2019dse}
{\scshape Event Horizon Telescope} collaboration, \emph{{First M87 Event
  Horizon Telescope Results. I. The Shadow of the Supermassive Black Hole}},
  \href{https://doi.org/10.3847/2041-8213/ab0ec7}{\emph{Astrophys. J. Lett.}
  {\bfseries 875} (2019) L1}
  [\href{https://arxiv.org/abs/1906.11238}{{\ttfamily 1906.11238}}].

\bibitem{EventHorizonTelescope:2019ggy}
{\scshape Event Horizon Telescope} collaboration, \emph{{First M87 Event
  Horizon Telescope Results. VI. The Shadow and Mass of the Central Black
  Hole}}, \href{https://doi.org/10.3847/2041-8213/ab1141}{\emph{Astrophys. J.
  Lett.} {\bfseries 875} (2019) L6}
  [\href{https://arxiv.org/abs/1906.11243}{{\ttfamily 1906.11243}}].

\bibitem{EventHorizonTelescope:2022wkp}
{\scshape Event Horizon Telescope} collaboration, \emph{{First Sagittarius A*
  Event Horizon Telescope Results. I. The Shadow of the Supermassive Black Hole
  in the Center of the Milky Way}},
  \href{https://doi.org/10.3847/2041-8213/ac6674}{\emph{Astrophys. J. Lett.}
  {\bfseries 930} (2022) L12}
  [\href{https://arxiv.org/abs/2311.08680}{{\ttfamily 2311.08680}}].

\bibitem{Vagnozzi:2022moj}
S.~Vagnozzi et~al., \emph{{Horizon-scale tests of gravity theories and
  fundamental physics from the Event Horizon Telescope image of Sagittarius
  A}}, \href{https://doi.org/10.1088/1361-6382/acd97b}{\emph{Class. Quant.
  Grav.} {\bfseries 40} (2023) 165007}
  [\href{https://arxiv.org/abs/2205.07787}{{\ttfamily 2205.07787}}].

\bibitem{Virbhadra:1999nm}
K.S.~Virbhadra and G.F.R.~Ellis, \emph{{Schwarzschild black hole lensing}},
  \href{https://doi.org/10.1103/PhysRevD.62.084003}{\emph{Phys. Rev. D}
  {\bfseries 62} (2000) 084003}
  [\href{https://arxiv.org/abs/astro-ph/9904193}{{\ttfamily
  astro-ph/9904193}}].

\bibitem{Adler:2022qtb}
S.L.~Adler and K.S.~Virbhadra, \emph{{Cosmological constant corrections to the
  photon sphere and black hole shadow radii}},
  \href{https://doi.org/10.1007/s10714-022-02976-7}{\emph{Gen. Rel. Grav.}
  {\bfseries 54} (2022) 93} [\href{https://arxiv.org/abs/2205.04628}{{\ttfamily
  2205.04628}}].

\bibitem{Zhang:2024lsf}
Z.~Zhang, Y.~Hou, M.~Guo and B.~Chen, \emph{{Imaging thick accretion disks and
  jets surrounding black holes}},
  \href{https://doi.org/10.1088/1475-7516/2024/05/032}{\emph{JCAP} {\bfseries
  05} (2024) 032} [\href{https://arxiv.org/abs/2401.14794}{{\ttfamily
  2401.14794}}].

\bibitem{Sokoliuk:2022owk}
O.~Sokoliuk, S.~Praharaj, A.~Baransky and P.K.~Sahoo, \emph{{Accretion flows
  around exotic tidal wormholes - I. Ray-tracing}},
  \href{https://doi.org/10.1051/0004-6361/202244358}{\emph{Astron. Astrophys.}
  {\bfseries 665} (2022) A139}
  [\href{https://arxiv.org/abs/2207.07193}{{\ttfamily 2207.07193}}].

\bibitem{Virbhadra:2022iiy}
K.S.~Virbhadra, \emph{{Distortions of images of Schwarzschild lensing}},
  \href{https://doi.org/10.1103/PhysRevD.106.064038}{\emph{Phys. Rev. D}
  {\bfseries 106} (2022) 064038}
  [\href{https://arxiv.org/abs/2204.01879}{{\ttfamily 2204.01879}}].

\bibitem{Hou:2023bep}
Y.~Hou, Z.~Zhang, M.~Guo and B.~Chen, \emph{{A new analytical model of
  magnetofluids surrounding rotating black holes}},
  \href{https://doi.org/10.1088/1475-7516/2024/02/030}{\emph{JCAP} {\bfseries
  02} (2024) 030} [\href{https://arxiv.org/abs/2309.13304}{{\ttfamily
  2309.13304}}].

\bibitem{Li:2025ixk}
Z.~Li and X.-K.~Guo, \emph{{Thin accretion disk around rotating hairy black
  hole: radiative property and optical appearance}},
  \href{https://doi.org/10.1140/epjc/s10052-025-14423-3}{\emph{Eur. Phys. J. C}
  {\bfseries 85} (2025) 679}
  [\href{https://arxiv.org/abs/2501.01018}{{\ttfamily 2501.01018}}].

\bibitem{Sui:2023yay}
T.-T.~Sui, Z.-L.~Wang and W.-D.~Guo, \emph{{The effect of scalar hair on the
  charged black hole with the images from accretions disk}},
  \href{https://doi.org/10.1140/epjc/s10052-024-12807-5}{\emph{Eur. Phys. J. C}
  {\bfseries 84} (2024) 441}
  [\href{https://arxiv.org/abs/2311.10946}{{\ttfamily 2311.10946}}].

\bibitem{Li:2024oyc}
Y.-Z.~Li and X.-M.~Kuang, \emph{{Trajectories of photons around a rotating
  black hole with unusual asymptotics}},
  \href{https://doi.org/10.1140/epjc/s10052-024-12627-7}{\emph{Eur. Phys. J. C}
  {\bfseries 84} (2024) 271}
  [\href{https://arxiv.org/abs/2401.07495}{{\ttfamily 2401.07495}}].

\bibitem{Gralla:2019xty}
S.E.~Gralla, D.E.~Holz and R.M.~Wald, \emph{{Black Hole Shadows, Photon Rings,
  and Lensing Rings}},
  \href{https://doi.org/10.1103/PhysRevD.100.024018}{\emph{Phys. Rev. D}
  {\bfseries 100} (2019) 024018}
  [\href{https://arxiv.org/abs/1906.00873}{{\ttfamily 1906.00873}}].

\bibitem{Narayan:2019imo}
R.~Narayan, M.D.~Johnson and C.F.~Gammie, \emph{{The Shadow of a Spherically
  Accreting Black Hole}},
  \href{https://doi.org/10.3847/2041-8213/ab518c}{\emph{Astrophys. J. Lett.}
  {\bfseries 885} (2019) L33}
  [\href{https://arxiv.org/abs/1910.02957}{{\ttfamily 1910.02957}}].

\bibitem{Bondi:1952}
H.~Bondi, \emph{On spherically symmetrical accretion}, {\emph{Monthly Notices
  of the Royal Astronomical Society} {\bfseries 112} (1952) 195}.

\bibitem{Abramowicz:2011xu}
M.A.~Abramowicz and P.C.~Fragile, \emph{{Foundations of Black Hole Accretion
  Disk Theory}}, \href{https://doi.org/10.12942/lrr-2013-1}{\emph{Living Rev.
  Rel.} {\bfseries 16} (2013) 1}
  [\href{https://arxiv.org/abs/1104.5499}{{\ttfamily 1104.5499}}].

\bibitem{kato2008black}
S.~Kato, J.~Fukue and S.~Mineshige, \emph{Black-Hole Accretion Disks---Towards
  a New Paradigm---}, Kyoto University Press (2008).

\bibitem{Yuan:2014gma}
F.~Yuan and R.~Narayan, \emph{{Hot Accretion Flows Around Black Holes}},
  \href{https://doi.org/10.1146/annurev-astro-082812-141003}{\emph{Ann. Rev.
  Astron. Astrophys.} {\bfseries 52} (2014) 529}
  [\href{https://arxiv.org/abs/1401.0586}{{\ttfamily 1401.0586}}].

\bibitem{Shakura:1972te}
N.I.~Shakura and R.A.~Sunyaev, \emph{{Black holes in binary systems.
  Observational appearance}}, {\emph{Astron. Astrophys.} {\bfseries 24} (1973)
  337}.

\bibitem{novikov1973astrophysics}
I.D.~Novikov and K.S.~Thorne, \emph{Astrophysics of black holes}, {\emph{Black
  holes (Les astres occlus)} {\bfseries 1} (1973) 343}.

\bibitem{lynden1974evolution}
D.~Lynden-Bell and J.E.~Pringle, \emph{The evolution of viscous discs and the
  origin of the nebular variables}, {\emph{Monthly Notices of the Royal
  Astronomical Society} {\bfseries 168} (1974) 603}.

\bibitem{Narayan:1994xi}
R.~Narayan and I.-s.~Yi, \emph{{Advection dominated accretion: A Selfsimilar
  solution}}, \href{https://doi.org/10.1086/187381}{\emph{Astrophys. J. Lett.}
  {\bfseries 428} (1994) L13}
  [\href{https://arxiv.org/abs/astro-ph/9403052}{{\ttfamily
  astro-ph/9403052}}].

\bibitem{Liu:2022cph}
B.F.~Liu and E.~Qiao, \emph{{Accretion around black holes: The geometry and
  spectra}},  \href{https://arxiv.org/abs/2201.06198}{{\ttfamily 2201.06198}}.

\bibitem{Krolik_2002}
J.H.~Krolik and J.F.~Hawley, \emph{Where is the inner edge of an accretion disk
  around a black hole?}, \href{https://doi.org/10.1086/340760}{\emph{The
  Astrophysical Journal} {\bfseries 573} (2002) 754}.

\bibitem{Arshakian:2004kh}
T.G.~Arshakian, \emph{{Direct evidence of the receding `torus' around central
  nuclei of powerful radio sources}},
  \href{https://doi.org/10.1051/0004-6361:20042341}{\emph{Astron. Astrophys.}
  {\bfseries 436} (2005) 817}
  [\href{https://arxiv.org/abs/astro-ph/0411636}{{\ttfamily
  astro-ph/0411636}}].

\bibitem{Chandrasekhar1985}
S.~Chandrasekhar, \emph{{The mathematical theory of black holes}}, Oxford
  university press (1998).

\bibitem{frank2002accretion}
J.~Frank, A.R.~King and D.~Raine, \emph{Accretion power in astrophysics},
  Cambridge university press (2002).

\bibitem{Capozziello:2011et}
S.~Capozziello and M.~De~Laurentis, \emph{{Extended Theories of Gravity}},
  \href{https://doi.org/10.1016/j.physrep.2011.09.003}{\emph{Phys. Rept.}
  {\bfseries 509} (2011) 167}
  [\href{https://arxiv.org/abs/1108.6266}{{\ttfamily 1108.6266}}].

\bibitem{Battista2023}
E.~Battista, \emph{{Quantum Schwarzschild geometry in effective field theory
  models of gravity}},
  \href{https://doi.org/10.1103/PhysRevD.109.026004}{\emph{Phys. Rev. D}
  {\bfseries 109} (2024) 026004}
  [\href{https://arxiv.org/abs/2312.00450}{{\ttfamily 2312.00450}}].

\bibitem{Casadio:2025lkn}
R.~Casadio, C.N.~Souza and R.~da~Rocha, \emph{{Quantum gravitational
  corrections at third-order curvature, acoustic analog black holes and their
  quasinormal modes}},  \href{https://arxiv.org/abs/2506.10847}{{\ttfamily
  2506.10847}}.

\bibitem{Wang:2025fmz}
Z.-L.~Wang and E.~Battista, \emph{{Dynamical features and shadows of quantum
  Schwarzschild black hole in effective field theories of gravity}},
  \href{https://doi.org/10.1140/epjc/s10052-025-13833-7}{\emph{Eur. Phys. J. C}
  {\bfseries 85} (2025) 304}
  [\href{https://arxiv.org/abs/2501.14516}{{\ttfamily 2501.14516}}].

\bibitem{Li:2025zdt}
H.~Li and X.~Zhang, \emph{{Particle Deflections around Microscopic Loop Quantum
  Black Holes with Rigorous Quantum Parameters}},
  \href{https://arxiv.org/abs/2503.20251}{{\ttfamily 2503.20251}}.

\bibitem{Tan:2024jkj}
B.~Tan, \emph{{Thermodynamics of high order correction for Schwarzschild-AdS
  black hole in non-commutative geometry}},
  \href{https://doi.org/10.1016/j.nuclphysb.2025.116868}{\emph{Nucl. Phys. B}
  {\bfseries 1014} (2025) 116868}
  [\href{https://arxiv.org/abs/2410.16799}{{\ttfamily 2410.16799}}].

\bibitem{Wang:2022ews}
Z.-L.~Wang, \emph{{Geodesic congruences in modified Schwarzschild black
  holes}}, \href{https://doi.org/10.1140/epjc/s10052-022-10843-7}{\emph{Eur.
  Phys. J. C} {\bfseries 82} (2022) 901}.

\bibitem{Cao:2024kht}
L.-M.~Cao, L.-Y.~Li, X.-Y.~Liu and Y.-S.~Zhou, \emph{{Appearance of de Sitter
  black holes and strong cosmic censorship}},
  \href{https://doi.org/10.1103/PhysRevD.109.084021}{\emph{Phys. Rev. D}
  {\bfseries 109} (2024) 084021}
  [\href{https://arxiv.org/abs/2401.15408}{{\ttfamily 2401.15408}}].

\bibitem{Wang:2023rjl}
Z.-L.~Wang, \emph{{Shadows and rings of a de Sitter\textendash{}Schwarzschild
  black hole}},
  \href{https://doi.org/10.1140/epjp/s13360-023-04756-x}{\emph{Eur. Phys. J.
  Plus} {\bfseries 138} (2023) 1131}
  [\href{https://arxiv.org/abs/2307.12361}{{\ttfamily 2307.12361}}].

\bibitem{Li:2024abk}
X.-Q.~Li, H.-P.~Yan, X.-J.~Yue, S.-W.~Zhou and Q.~Xu, \emph{{Geodesic
  structure, shadow and optical appearance of black hole immersed in
  Chaplygin-like dark fluid}},
  \href{https://doi.org/10.1088/1475-7516/2024/05/048}{\emph{JCAP} {\bfseries
  05} (2024) 048} [\href{https://arxiv.org/abs/2401.18066}{{\ttfamily
  2401.18066}}].

\bibitem{Gogoi:2025rcn}
D.J.~Gogoi, P.~Hazarika, J.~Bora and R.~Changmai, \emph{{Thermodynamics of
  Deformed AdS-Schwarzschild Black Holes in the Presence of Thermal
  Fluctuations}}, \href{https://doi.org/10.1002/prop.70004}{\emph{Fortsch.
  Phys.} {\bfseries 73} (2025) e70004}
  [\href{https://arxiv.org/abs/2501.15629}{{\ttfamily 2501.15629}}].

\bibitem{Lim:2025cne}
K.-G.~Lim and X.Y.~Chew, \emph{{Shadow of the scalar hairy black hole with
  inverted Higgs potential}},
  \href{https://doi.org/10.1103/PhysRevD.111.084085}{\emph{Phys. Rev. D}
  {\bfseries 111} (2025) 084085}
  [\href{https://arxiv.org/abs/2501.07029}{{\ttfamily 2501.07029}}].

\bibitem{Nakahara2003}
M.~Nakahara, \emph{Geometry, topology and physics}, CRC press (2018).

\bibitem{Carroll2004}
S.M.~Carroll, \emph{{Spacetime and Geometry}: {An Introduction to General
  Relativity}}, Cambridge University Press (7, 2019),
  \href{https://doi.org/10.1017/9781108770385}{10.1017/9781108770385}.

\end{thebibliography}\endgroup
\end{document}